\documentclass[conference]{IEEEtran}
\IEEEoverridecommandlockouts
\let\labelindent\relax

\usepackage{amsmath}

\usepackage{nicefrac}
\usepackage{siunitx}
\usepackage{array,framed}
\usepackage{booktabs}
\usepackage{fix-cm}
\usepackage{
  color,
  float,
  epsfig,
  wrapfig,
  graphics,
  graphicx,
  subcaption,
  ragged2e
}

\usepackage{bm}
\usepackage{textcomp,amssymb}
\usepackage{setspace}
\usepackage{amsthm}
\usepackage{latexsym,fancyhdr,url}
\usepackage{enumerate}
\usepackage{enumitem}
\usepackage{algorithm2e}
\usepackage{algpseudocode}
\usepackage{graphics}
\usepackage{xparse} 
\usepackage{xspace}
\usepackage{multirow}
\usepackage{csvsimple}
\usepackage{balance}

\usepackage{
  tikz,
  pgfplots,
  pgfplotstable
}

\usepackage{hyperref}

\usetikzlibrary{
  shapes.geometric,
  arrows,
  external,
  pgfplots.groupplots,
  matrix
}
\pgfplotsset{compat=1.18} 

\theoremstyle{plain}

\theoremstyle{definition}
\newtheorem{definition}{Definition}
\usepackage{mathtools}
\usepackage{cite}
\usepackage{amsmath,amssymb,amsfonts}
\usepackage{graphicx}
\usepackage{textcomp}
\usepackage{xcolor}
\def\BibTeX{{\rm B\kern-.05em{\sc i\kern-.025em b}\kern-.08em
    T\kern-.1667em\lower.7ex\hbox{E}\kern-.125emX}}


\begin{document}

\title{Taipan: A Query-free Multiple Sensitive Attribute Inference Attack Solely from Auxiliary Graphs}


\author{\IEEEauthorblockN{Ying Song}
\IEEEauthorblockA{\textit{University of Pittsburgh}\\
yis121@pitt.edu}
\and
\IEEEauthorblockN{Balaji Palanisamy}
\IEEEauthorblockA{\textit{University of Pittsburgh}\\
bpalan@pitt.edu}}


\maketitle

\begin{abstract}
Graph-structured data underpin a wide spectrum of modern applications, yet their multiple sensitive attributes are not isolated but deeply coupled with graph topology. This coupling facilitates intersectional privacy leakage through attribute inference attacks (AIAs). Existing AIAs predominantly assume adversaries with two capabilities: (I) access to either auxiliary data sampled from the same distribution as the target or partial training records, and (II) the ability to repeatedly query the victim model. In reality, however, these assumptions are often impractical due to stringent data protection regulations, prohibitive query budgets, and heightened detection risks. Moreover, as prior attacks are designed to infer a single sensitive attribute at a time, they fail to simultaneously recover multiple attributes and thus underestimate overall privacy risks arising from cross-attribute dependencies. More critically, they obscure a systemic blind spot: \textbf{auxiliary graphs alone are sufficient to expose multiple sensitive attributes in the target graph.} This threat persists even under distribution shifts, enabling stealthy offline attacks without any model interactions. To investigate this unexplored vulnerability, we introduce \textbf{Taipan, the first query-free framework for multiple sensitive attribute inference attacks on graphs (G-MSAIAs).} Taipan incorporates \emph{Hierarchical Attack Knowledge Routing} to capture intricate inter-attribute correlations, and \emph{Prompt-guided Attack Prototype Refinement} to mitigate distribution shifts between auxiliary and target graphs. Furthermore, as current metrics are limited to quantifying single-attribute leakage, we propose a systematic evaluation framework tailored to G-MSAIAs. Extensive experiments on diverse real-world graph datasets demonstrate that Taipan consistently achieves strong attack performance across same-, similar-, and out-of-distribution settings, and remains effective under partial label coverage or even rigorous privacy guarantees. Our findings underscore the urgent need for more robust multi-attribute privacy-preserving graph publishing methods and data-sharing practices.

\end{abstract}


\section{Introduction}
Graph-structured data have become the pivotal information infrastructure of modern interconnected systems, driving essential services in social networking \cite{pokec}, financial transaction \cite{NIFTY}, and healthcare collaborations \cite{healthcare_survey}. To foster science or unlock commercial value, graphs are disseminated through an expanding array of channels. Organizations actively release curated graph datasets to the public via open repositories such as Stanford SNAP \cite{snapnets}, Network Data Repository \cite{net_data}, Microsoft Open Academic Graph \cite{oag} and Amazon-sponsored Graph Challenge \cite{graphchallengereference}. Meanwhile, data owners, insiders or authorized recipients distribute graphs under within- or cross-institutional data sharing agreements. Beyond institutional releases, thriving data brokerage markets commodify graph-structured data at scale.
At the individual level, users routinely exchange personal data for access to digital services or platforms, inadvertently contributing to graphs in the wild.
Further compounding this landscape, the proliferation of web-scraping and recurrent large-scale relational data breaches, e.g., the LinkedIn breach exposed approximately 165 million accounts \cite{bbckey}, creates an unintended but vast reservoir of auxiliary graphs. Together, these channels ensure that auxiliary graphs encoding sensitive attributes (e.g., age, health status, criminal records) are pervasive and readily accessible, with or without the data owner's consent. 


While regulations such as GDPR \cite{gdpr} mandate the sanitization of sensitive attributes before publishing and sharing, two intertwined forms of coupling create a fundamental compliance–protection gap, rendering sensitive attributes vulnerable to attribute inference attacks (AIAs). First, multiple sensitive attributes are not isolated; they exhibit inherent cross-attribute dependencies in the feature space such that the disclosure of one attribute facilitates the inference of others \cite{multi_corr_social_net,aia_link_attr}. Second, graph topology further amplifies these dependencies through homophily, where entities sharing similar sensitive attributes tend to cluster together \cite{say_no}. This topology-attribute coupling gives rise to cascading privacy leakage: recovering one entity's sensitive profile propagates privacy risks to its associated neighbors, extending the attack surface beyond individual targets. For instance, in social networks, an adversary can infer multiple missing sensitive attributes of target users and their neighbors by exploiting cross-attribute dependencies and homophily from auxiliary graphs \cite{aia_link_attr}. Such unauthorized inference exposes individuals to surveillance, blackmail, and identity theft, thereby undermining the security and trustworthiness of social platforms. Furthermore, this attack substantially lowers the attack barrier. Auxiliary graphs are publicly accessible or readily obtainable through the channels outlined above, enabling offline and query-free AIAs without leaving any digital traces on victim models. Since current defense mechanisms primarily focus on model-centric safeguards, they are fundamentally ineffective against privacy leakage originating from the ubiquity of auxiliary graphs. Notably, this paper focuses exclusively on homophilic graphs, leaving the exploration of heterophilic ones for future work. 
\noindent\underline{\textbf{Limitations of Prior Work}} Despite the gravity of this systemic vulnerability, existing AIAs in the graph domain fail to capture it. Some utilize node embeddings generated from graph neural networks (GNNs) together with auxiliary graphs to train supervised attack models \cite{AIA_graph_emb}, requiring direct model access that is infeasible under real-world deployments. Others operate as data imputation methods to infer missing sensitive attribute values given the remaining records without modeling cross-attribute dependencies or homophily, limiting their effectiveness \cite{AIA_gnn}. A separate line of work leverages social signals, such as user-generated content, social traces, and behavior records to infer user sensitive attributes \cite{social_net_survey, aia_social_friend_behavior, aia_link_attr, multi_corr_social_net, ego_social_net}, but they are designed exclusively for social networks, require multi-source side information and partially disclosed sensitive attributes of target graphs, and achieve limited success rates.  


Beyond the graph domain, AIAs predominantly adopt a query-based paradigm to repeatedly probe the victim model. The key idea is that victim models tend to produce more confident posterior probabilities or more accurate predictions when queried with the true sensitive attribute value of a target record \cite{are_ur_sa_private}. This paradigm implicitly assumes that sensitive attributes are retained during model training and exposed through repeated probing. However, this assumption breaks down in practice for three reasons. First, GDPR \cite{gdpr} requires model owners to sanitize sensitive attributes before training and apply privacy-preserving techniques during or after training, rendering sensitive attributes no longer recoverable through repeated probing. Second, repeated queries generate anomalous access patterns that are easily detected by anomaly detection systems and incur prohibitive query budgets for sensitive attributes with multiple categorical values. Third, these attacks often require auxiliary data sampled from the same distribution as the target, which is impractical due to stringent data regulations. 

Existing AIAs are also limited in their scope and generalizability. They typically target a small subset of training records whose sensitive information is encoded in the victim model, offering limited coverage of non-training records. While imputation-based attacks may succeed when training and non-training records follow the same or highly similar distributions, their effectiveness degrades substantially when confronted with out-of-distribution (OOD) records. In addition, most existing studies focus on a single sensitive attribute, and only a few works consider multi-attribute leakage in tabular data. However, they adopt simplistic extensions, either conducting the same AIA for each attribute in parallel \cite{disparate_priv_vulnerability} or sequentially \cite{are_ur_sa_private}, and are therefore only applicable to small-scale datasets with few sensitive attributes. Most importantly, these AIAs cannot be trivially extended to graphs, where cross-attribute dependencies and topology-attribute homophily render independent per-attribute inference fundamentally insufficient and demand a holistic framework that jointly models these coupled signals to infer multiple sensitive attributes.





\noindent\underline{\textbf{Research Questions}} The aforementioned research gap motivates us to investigate the following three research questions:
\begin{enumerate}[leftmargin=*]
    \item \textit{Is it possible to design a unified attack framework that simultaneously infers multiple sensitive attributes on graphs?}
    \item \textit{Can this framework generalize to the OOD setting while remaining effective under strong privacy guarantees?}
    \item \textit{How to systematically evaluate its attack performance?}
\end{enumerate}



\noindent\underline{\textbf{Challenges}} We accordingly identify five key challenges to address the above research questions. 
\begin{enumerate}[leftmargin=*]
\item \textbf{``Information Vacuum''}: Without query responses as informative proxies, even conducting a query-free AIA for a single sensitive attribute on unseen graphs becomes highly challenging. This difficulty is further exacerbated when simultaneously inferring multiple sensitive attributes.
\item \textbf{Intertwined Correlations}: Multiple sensitive attributes exhibit intricate cross-attribute correlations that are further amplified by homophily. Training a unified attack model is therefore non-trivial: easily inferred sensitive attributes tend to dominate the learning process, overshadowing the nuanced patterns of underrepresented ones. Moreover, conflicting or negatively correlated attributes introduce contradictory learning signals into the shared latent space, ultimately degrading overall inference accuracy.
\item \textbf{Out-of-distribution (OOD) Transfer}: OOD auxiliary graphs compound negative transfer under multi-attribute settings, as per-attribute distributional misalignments accumulate and collectively degrade overall attack performance.
\item \textbf{Privacy Guarantees}: In practice, graphs are often published under strong privacy guarantees, which further amplify distribution shifts between auxiliary and target graphs.
\item \textbf{Comprehensive Evaluation}: Existing single-attribute evaluations fail to capture the coupled privacy risks inherent to multiple sensitive attribute inference attacks on graphs (G-MSAIAs), necessitating a dedicated evaluation framework.
\end{enumerate}

\noindent\underline{\textbf{Solutions and Contributions}} To address these challenges, we propose a query-free \textbf{\underline{t}}ransfer-b\textbf{\underline{a}}sed mult\textbf{\underline{ip}}le sensitive \textbf{\underline{a}}ttribute i\textbf{\underline{n}}ference attack framework named \textbf{Taipan}. This unified framework comprises two sequential modules: \emph{Hierarchical Attack Knowledge Routing} and \emph{Prompt-guided Attack Prototype Refinement}. Inspired by multi-task learning (MTL), we formulate the per-attribute inference 
as an individual attack task. Unlike traditional shared-bottom MTL architectures, we borrow ideas from the Multi-gate Mixture-of-Experts (MMoE) framework \cite{mmoe} to flexibly capture intricate cross-attribute dependencies and homophily while suppressing task interference and performance degradation. Specifically, we first profile attack tasks on the auxiliary graph and construct an attack hierarchy tree. Guided by this hierarchy, we design a hierarchical MMoE-like architecture that promotes positive transfer among similar attack tasks while mitigating negative transfer among conflicting tasks. To capture both shared and task-specific knowledge, we introduce learnable pretext tokens as task identifiers \cite{multigprompt}. During the attack transfer, we freeze the pre-trained Taipan and only tune the lightweight pretext token parameters, which act as prompts to extract structural and semantic knowledge from the auxiliary graph and adapt it to the target graph. Since we strictly assume that sensitive attributes in the target graph are masked or sanitized under rigorous privacy safeguards, this realistic setting naturally formulates attack transfer as an unsupervised domain adaptation problem, where the attack model should align knowledge learned from the auxiliary graph with sparse or distorted signals in the ill-distributed target graph. We initially employ pseudo labeling using the pre-trained Taipan and only preserve high-confidence nodes and then reuse auxiliary samples during adaptation to further alleviate data scarcity. Additionally, we adaptively adjust attack prototypes to exploit the rich information from filtered target nodes, enabling smooth and reliable distribution alignment. Finally, for comprehensive evaluation, we introduce three categories of metrics to systematically assess attack utility, task deviation, and semantic knowledge preservation, respectively.



Notably, we distinguish between conventional ``attack transferability'' and our proposed ``attack transfer''. The former refers to reusing adversarial examples across models, whereas the latter denotes transferring attack knowledge from an auxiliary graph via a pre-trained attack model to downstream attribute inference tasks. This notion resembles standard transfer learning, but serves for malicious purposes. To the best of our knowledge, this is the first study to investigate attack transfer in the graph domain, and even in general tabular settings. 

\noindent\underline{\textbf{Summary}}: In summary, our contributions include:
\begin{enumerate}[leftmargin=*]
    \item We reveal the new vulnerability stemming from ubiquitous auxiliary graphs and design the first query-free attack framework with two novel modules for G-MSAIAs.
    \item We pioneer the investigation of multiple ``attack transfer'' using auxiliary graphs with full and partial sensitive labels.
    \item We introduce a comprehensive and systematic evaluation framework to quantify G-MSAIAs.
    \item Extensive experiments demonstrate the effectiveness and efficiency of Taipan under diverse distribution settings.
    \item Our empirical results show that Taipan remains effective under strong privacy guarantees, highlighting an urgent need for more robust multi-attribute privacy-preserving graph publishing methods and for a re-evaluation of data-sharing practices and regulatory policies, as privacy harm can occur without explicit data misuse or model abuse.
\end{enumerate}

\section{Background and Related Work}
\label{sec:background}

\subsection{Notations}
Given an undirected attributed graph $\mathcal{G}=(\mathcal{V}, \mathcal{E}, \mathcal{X})$, $\mathcal{V}$ denotes the node set with $|\mathcal{V}|$ nodes, and $\mathcal{E}\subseteq\mathcal{V}\times\mathcal{V}$ denotes the edge set with $|\mathcal{E}|$ edges. The graph topology is represented by the adjacency matrix $\mathcal{A}\in\mathbb{R}^{|\mathcal{V}|\times|\mathcal{V}|}$, where $\mathcal{A}_{u,v}=1$ if and only if the edge $(u,v)\in\mathcal{E}$, otherwise $\mathcal{A}_{u,v}=0$. Each node $v\in\mathcal{V}$ is associated with a feature vector $\mathcal{X}_v=\{\mathcal{S}_v,X_v\}\in\mathbb{R}^{d}$ and a label $\mathcal{Y}_v$, where $\mathcal{S}_v$ denotes sensitive attributes (e.g., gender and age), $X$ denotes non-sensitive attributes, and $d$ represents the feature dimensionality that concatenates both sensitive and non-sensitive attributes. 

\subsection{Graph Neural Networks} 
Standard GNNs aggregate information for each node $v\in\mathcal{V}$ from
its neighborhood $\mathcal{N}(v)$ and iteratively update its representation $Z_v$. This message passing process is formulated as: 
\begin{equation}
    Z_{v}^{L}=UPD^{L}(Z_{v}^{L-1},AGG^{L-1}(\{Z_{u}^{L-1}:u\in\mathcal{N}(v)\}))
\end{equation}
where $L$ denotes the number of GNN layers and $Z_{v}^{0}=\mathcal{X}_v$ represents the initial feature vector of node $v$. The operators $UPD$ and $AGG$ are differentiable aggregation and update functions, respectively, which can be instantiated to construct various GNN architectures. For node classification tasks, the final node embedding $Z_{v}^{L}$ is mapped to the predicted label $\hat{\mathcal{Y}}_v$ through a linear classifier $f_{\theta}(\cdot)$ followed by a softmax function, where $\theta$ denotes the trainable parameters of the linear head. Throughout this paper, GNNs serve solely as graph encoders within the attack framework rather than victim models. Our attack is strictly query-free and requires no model interactions.

\subsection{Attribute Inference Attack}
Roughly speaking, an AIA aims to infer the sensitive attributes of targeted entities using accessible auxiliary data. Existing AIAs can be broadly categorized into the following two types based on the underlying data structure.
\begin{table*}[htbp]
\centering 
\caption{\centering Attack Comparison. $\surd$ represents ``access'' or ``applicable'', $\times$ represents ``no access'' or ``not applicable''. Dist.:Distribution; Impt.: Imputation; SA: Sensitive Attribute; Sim.: Similar; Diff.: Different.}
\label{attack_comp}
\resizebox{\linewidth}{!}{
\begin{tabular}{c||ccc|cc|ccc|c}
\toprule\bottomrule
\multirow{2}{*}{\textbf{Attack Name}} & \multicolumn{3}{c|}{\textbf{Attack Target}}                                                & \multicolumn{2}{c|}{\textbf{Attack Participant}} & \multicolumn{3}{c|}{\textbf{Attack Knowledge and Adversary's Capabilities}}                                                                                                                                                                                                                  & \multirow{2}{*}{\textbf{Attack Mechanism}}                                        \\ \cline{2-9}
                                      & \textbf{Phase} & \textbf{Type} & \textbf{Inference}                                        & \textbf{Model  Owner}     & \textbf{Attacker}     & \textbf{Data Access}                                                                                                                                                                              & \multicolumn{1}{c}{\textbf{Model Access}} & \textbf{Model Outputs}                                                      &                                                                                   \\ \bottomrule
\textbf{FJRMIA\cite{first_aia}}                       & Training       & Tabular       & Single                                                    & $\surd$                  & $\surd$               & \begin{tabular}[c]{@{}c@{}}Full Non-SAs in Training Data\\ Marginal Prior of SAs and Non-SAs\\ Possible Values of SAs and Non-SAs\\ Training Data’s Ground-truth Labels\end{tabular}              & Black Box                                  & \begin{tabular}[c]{@{}c@{}}Confusion Matrix\\ Confidence Score\end{tabular} & \begin{tabular}[c]{@{}c@{}}Query-based,\\ Posterior-maximized\end{tabular}        \\ \hline
\textbf{CS/LOMIA\cite{are_ur_sa_private}}                        & Training       & Tabular       & \begin{tabular}[c]{@{}c@{}}Single\\ Parallel\end{tabular} & $\surd$                  & $\surd$               & \begin{tabular}[c]{@{}c@{}}Full Non-SAs in Training Data\\ Possible Values of SAs and Non-SAs\\ Training Data’s Ground-truth Labels\\ (Auxiliary Data with Same/Similar Dist.)\end{tabular}                                                   & Black Box                                  & \begin{tabular}[c]{@{}c@{}}Confidence Score\\ (Label-only)\end{tabular}                                                             & \begin{tabular}[c]{@{}c@{}}Query-based,\\ Posterior-maximized \\ (Attack Model Trained)\end{tabular}        \\ \hline
\textbf{Sens. Val. Inf. \cite{aia_imputation}}              & Training       & Tabular       & Single                                                    & $\surd$                  & $\surd$               & \begin{tabular}[c]{@{}c@{}}Full Non-SAs in Training Data \\ Possible Values of SAs and Non-SAs\\ Auxiliary Graph with Same/Similar Dist. \\ and Similar Shadow Model\end{tabular}             & White/Black                            & Confidence Score                                                            & \begin{tabular}[c]{@{}c@{}}Neuron Impt.-exploited\\ Query-based\end{tabular} \\ \hline
\textbf{SDMIA\cite{least_info_aia_disparate}}                        & Training       & Tabular       & Single                                                    & $\surd$                  & $\surd$               & Possible Values of SAs and Non-SAs                                                                                                                                                                & Black Box                                  & Label-only                                                                  & \begin{tabular}[c]{@{}c@{}}Query-based,\\ Attack Model Trained\end{tabular}       \\ \hline
\textbf{Disparity Inf.\cite{disparate_priv_vulnerability}}               & Training       & Tabular       & \begin{tabular}[c]{@{}c@{}}Single\\ Sequential\end{tabular}   & $\surd$                  & $\surd$               & \begin{tabular}[c]{@{}c@{}}Full Non-SAs in Training Data\\ Possible Values of SAs and Non-SAs\end{tabular}                                                                                        & Black Box                                  & Confidence Score                                                            & \begin{tabular}[c]{@{}c@{}}Query-based, \\ Disparity-exploited\end{tabular}       \\ \bottomrule
\textbf{Duddu et al.\cite{AIA_graph_emb}}                 & Training       & Graph         & Single                                                    & $\surd$                  & $\surd$               & \begin{tabular}[c]{@{}c@{}}Auxiliary Graph with the Same Dist.\\ Node Embeddings of All Graphs\end{tabular}                                                                & Grey Box                                   & Node Embeddings                                                             & \begin{tabular}[c]{@{}c@{}}Query-based, \\ Attack Model Trained\end{tabular}      \\ \hline
\textbf{FP\&FP-MA\cite{AIA_gnn}}                    & Training       & Graph         & \begin{tabular}[c]{@{}c@{}}Single/\\ Missing\end{tabular} & $\surd$                  & $\surd$               & \begin{tabular}[c]{@{}c@{}}Full Non-SAs and Partial SAs \\ Auxiliary Graph with Same/Similar Dist. \\ and Similar Shadow Model\end{tabular}                             & Black Box                                  & Confidence Score                                                            & \begin{tabular}[c]{@{}c@{}}Query-based, \\ Posterior-maximized\end{tabular}       \\ \hline

\textbf{ReAI\cite{aia_link_attr}}                    & All       & Graph         & \begin{tabular}[c]{@{}c@{}}Pairwise\end{tabular} & $\bm{\times}$                  & $\surd$               & \begin{tabular}[c]{@{}c@{}}Full Non-SAs and Partial SAs \\ Possible Values of SAs and Non-SAs \\ Pairwise Attribute Relevance Prior\end{tabular}                             & $\bm{\times}$                                  & $\bm{\times}$                                                            & \begin{tabular}[c]{@{}c@{}}Social Network-only, \\ Graph-based Propagation\end{tabular}       \\ \hline

\textbf{WAPITI\cite{ego_social_net}}                    & All       & Graph         & \begin{tabular}[c]{@{}c@{}}Multiple/\\ Missing\end{tabular} & $\bm{\times}$                 & $\surd$               & \begin{tabular}[c]{@{}c@{}}Full Non-SAs and Partial SAs  \\ Possible Values of SAs and Non-SAs\\ Weighted Social Links \end{tabular}                             & $\bm{\times}$                                  & $\bm{\times}$                                                            & \begin{tabular}[c]{@{}c@{}}Social Network-only, \\ Posterior-maximized\end{tabular}       \\ \hline

\textbf{DeepAttr\cite{mul_sens_emb}}                    & All       & Graph         & \begin{tabular}[c]{@{}c@{}}Multiple/\\ Missing\end{tabular} & $\bm{\times}$                  & $\surd$               & \begin{tabular}[c]{@{}c@{}}Full Non-SAs and Partial SAs \\ Possible Values of SAs and Non-SAs\end{tabular}                             & $\bm{\times}$                                  & $\bm{\times}$                                                            & \begin{tabular}[c]{@{}c@{}}Social Network-only, \\ Attack Model Trained\end{tabular}       \\ \hline

\textbf{G-MSAIA (Ours)}                 & \textbf{All}            & \textbf{Graph}         & \textbf{Multiple}                                                  & $\bm{\times}$                 & $\surd$               & \begin{tabular}[c]{@{}c@{}}\textbf{Auxiliary Graph with Same/Sim./Diff. Dist.}\\ Possible Values of SAs\end{tabular}                                                                                     & $\bm{\times}$                              & $\bm{\times}$                                                               & \begin{tabular}[c]{@{}c@{}}\textbf{Query-free}, \\ \textbf{Transfer-based}\end{tabular}            \\ \toprule\bottomrule
\end{tabular}}
\end{table*}

\subsubsection{Attribute Inference Attacks on Tabular Data} FJRMIA \cite{first_aia} establishes the foundational AIA paradigm by repeatedly querying a victim model with auxiliary data and perturbed sensitive attribute values, identifying the value that yields the highest posterior probability. Building upon this paradigm, CSMIA \cite{are_ur_sa_private} removes the reliance on confusion matrices or marginal priors. For a target record in the training dataset, CSMIA queries the victim model with all possible sensitive attribute values while keeping non-sensitive attributes fixed, and selects the value that maximizes the model’s confidence score when the prediction is correct. 
Unlike CSMIA, LOMIA supports label-only queries. It constructs an attack dataset by querying the victim model with all possible sensitive attribute values of a target record and retaining only those queries that yield correct predictions. These correctly classified records are then used to train an attack model that exploits correlations between sensitive and non-sensitive attributes. Although both CSMIA and LOMIA can be extended to infer multiple sensitive attributes, such extensions simply repeat single-attribute inference in parallel, thereby incurring prohibitive query budgets and high detection risks. Jayaraman et al. \cite{aia_imputation} propose a sensitive value inference attack under the white-box setting, where auxiliary data are utilized to identify neurons that are most correlated with sensitive attributes. For each target record, adversaries infer the sensitive value if the weighted sum of the activations of these neurons exceeds a pre-specified threshold. SDMIA \cite{least_info_aia_disparate} assumes only minimal knowledge of the possible values of sensitive and non-sensitive attributes and employs random queries to construct an attack dataset by retaining instances whose predictions differ across all candidate sensitive values. Recently, Kabir et al. \cite{disparate_priv_vulnerability} identify disparate vulnerability across demographic subgroups by analyzing angular differences in confidence score distributions, and exploit such disparities to conduct AIAs targeting the most vulnerable records. While this disparity-based approach can be extended to multiple sensitive attributes, it infers them sequentially, resulting in prohibitively high query budgets.

\subsubsection{Attribute Inference Attack on Graphs} AIAs against GNNs remain underexplored. Duddu et al. \cite{AIA_graph_emb} propose the first AIA framework against GNNs, which assumes that adversaries can obtain node embeddings of an auxiliary graph with their corresponding sensitive attributes. However, this assumption is unrealistic in real-world deployments due to stringent security and privacy regulations. FP and FP-MA \cite{AIA_gnn} exploit the feature propagation mechanism inherent in GNNs to perform AIAs, but they are fundamentally data imputation methods that recover partially missing sensitive attributes rather than jointly inferring multiple sensitive attributes. A separate line of work focuses on social networks, encompassing content-based, social link-based, user behavior-based, and multi-source attacks \cite{social_net_survey}. Among these, social link-based attacks most closely align with our setting, as they exploit user features and graph topology to infer sensitive attributes. ReAI \cite{aia_link_attr} analyzes pairwise correlations among sensitive attributes and combines them with topological structures to recover missing attribute values via random walks with restart. WAPITI \cite{ego_social_net} leverages user profiles and local neighborhoods of target nodes to infer sensitive attributes through a weighted Bayesian model. DeepAttr \cite{mul_sens_emb} jointly infers multiple user attributes by exploiting social network structure and a small set of disclosed attributes to predict missing demographics across the graph. However, these methods require partial sensitive attributes of target users and treat AIAs as imputation or label propagation problems. Moreover, they are designed exclusively for social networks, limiting their applicability to general graph settings.

For clarity, we summarize these representative AIAs in Table \ref{attack_comp}. \textit{Overall, no unified attack framework currently exists for simultaneously inferring multiple sensitive attributes on general graphs.} Existing extensions either simply repeat single-attribute inference in parallel or sequentially, or rely on data imputation methods, which are insufficient to address the challenges posed by complex graph-structured data.

\section{Attack Framework Design}
In this section, we formalize the threat model and present Taipan, the first query-free attack framework for G-MSAIAs.

\subsection{Threat Model}

\subsubsection{Adversary’s Goal} The adversary aims to infer multiple sensitive attributes simultaneously on a target graph in a query-free manner, without interacting with any victim models. We formally define G-MSAIAs and transfer mechanism below.

\definition{(Multiple Sensitive Attribute Inference Attacks on Graphs (G-MSAIAs))}: Given a target graph $\mathcal{G}_{t}$, sensitive attributes $\mathcal{S}_{t} \in \mathbb{R}^{|\mathcal{V}_{t}|\times s}$ are masked due to privacy concerns, and only non-sensitive attributes $X_{t} \in \mathbb{R}^{|\mathcal{V}_{t}|\times (d-s)}$, which may be further processed by privacy-preserving techniques, are available, where $s$ denotes the number of sensitive attributes. The adversary's objective is to construct an attack model $\mathcal{F}$ to infer $\mathcal{S}_{t}$ of the target nodes $\mathcal{V}_{t}$ in $\mathcal{G}_{t}$.

\definition{(Attack Transfer Mechanism)}: Given a target graph $\mathcal{G}_{t}$ without access to sensitive labels $\mathcal{S}_{t}$, and an auxiliary graph $\mathcal{G}_{a}$ with sensitive labels $\mathcal{S}_{a}$, the adversary first pre-trains an attack model $\mathcal{F}_{pre}(\mathcal{G}_{a}, \mathcal{S}_{a})$, and then adapts it to the unlabeled $\mathcal{G}_{t}$ to obtain a tuned attack model $\mathcal{F}_{tun}(\mathcal{G}_{t})$, which is subsequently used to recover $\mathcal{S}_{t}$ of $\mathcal{V}_{t}$ in $\mathcal{G}_{t}$. 




\subsubsection{Adversary’s Knowledge and Capabilities}

The adversary can be any party seeking to infer sensitive attributes of the target nodes, such as service providers, insiders, advertisers, data brokers or cybercriminals. We consider a strictly query-free threat model, where the adversary has no access to any victim GNNs, including their parameters, training procedures, internal representations, or query interfaces. Instead, the adversary recovers multiple sensitive attributes from cross-attribute dependencies and homophilic patterns of auxiliary graphs. 

This attack fundamentally differs from that of conventional AIAs, which \textit{typically assume that the adversary can exploit auxiliary data and repeatedly query victim models to probe the real values of sensitive attributes}. However, such assumptions can be misaligned with practical deployment settings for two main reasons. First, to comply with stringent data protection regulations such as GDPR \cite{gdpr}, sensitive attributes are commonly masked and sensitive correlations are mitigated prior to model training, rendering attacks that rely on manipulating sensitive attributes ineffective. Second, repeated queries are often constrained by access controls, rate limiting, and model or system monitoring mechanisms, which restrict the feasibility of query-based attacks. Moreover, even under strong attack capabilities and rich prior knowledge, recent studies \cite{disparate_priv_vulnerability, aia_imputation} show that such attacks struggle to achieve consistently high attack performance. In contrast, \textit{our attack only requires auxiliary graphs, a condition far easier to satisfy in practice}. Such graphs are readily accessible through diverse channels: numerous graph datasets are widely released for research (e.g., Stanford SNAP, Network Data Repository, Microsoft Open Academic Graph) or commercial purposes (e.g., Amazon Graph Challenge), or distributed under data-sharing agreements to facilitate downstream tasks. Beyond institutional releases, auxiliary graphs are traded within data brokerage markets, harvested via web scraping and cross-platform linkage, or exposed through data breaches. The ubiquity of these graphs substantially lowers the attack barrier, making G-MSAIAs a practical and pervasive threat.
Different from existing AIAs requiring auxiliary data drawn from the same distribution as the target ones, our threat model relaxes this restrictive assumption: the adversary can access auxiliary graphs that may differ from the target in underlying distributions and prediction tasks. Following standard practice in prior AIAs \cite{aia_imputation, least_info_aia_disparate, are_ur_sa_private, AIA_gnn, disparate_priv_vulnerability}, we assume that the adversary knows the possible values of multiple sensitive attributes (e.g., gender $\in\{\text{male, female}\}$) in the target graph, not the true sensitive values. We strictly assume that the adversary obtains only non-sensitive attributes and topology of the target graph. Since real-world graphs might have label scarcity issues, we further test our attack under varying levels of sensitive label coverage in the auxiliary graph in Section \ref{coverage}. Meanwhile, as privacy-preserving techniques are often imposed before graph publishing or sharing, we additionally evaluate the attack under a rigorous privacy regime, namely differential privacy \cite{privdpr}, as detailed in Section \ref{sub:defense}.





\subsubsection{Attack Scenarios}  

We consider three attack scenarios based on the availability and alignment of auxiliary graphs with the target graph. These scenarios do not alter the adversary’s capabilities defined in the threat model, but are used to demonstrate the feasibility and generality of the proposed attack under progressively relaxed assumptions on data availability.

\noindent\textbf{Same-Distribution Auxiliary Graph.} Following common practice in prior AIAs, we first consider an idealized setting where the auxiliary graph is drawn from the same underlying distribution as the target graph. This setting establishes an upper-bound reference for G-MSAIAs' attack performance.

\noindent\textbf{Out-of-Distribution (OOD) Auxiliary Graph.} We then relax the assumption and consider auxiliary graphs in the same application domain but collected from different data sources, such as social networks scraped from different geographic regions or platforms. These auxiliary graphs are generally more accessible than same-distribution data and exhibit distributional shifts as well as mismatched feature dimensionalities. 

\noindent\textbf{Task-Shifted Auxiliary Graph.} We further relax the assumption and consider auxiliary graphs drawn from the similar broad application domain as the target graph. These two graphs possess related yet distinct prediction tasks and exhibit mismatched structural and feature representations. This weak-aligned scenario captures realistic deployments where only loosely related auxiliary graphs are 
accessible. 
For instance, the target graph may represent a social network like Facebook for modeling social communities, while the auxiliary graph is derived from LinkedIn to capture professional relationships. While both graphs belong to the similar broad domain and share overlapping sensitive attributes, e.g., basic demographics, they exhibit substantial task and distribution shifts.

Taken together, these scenarios span the realistic spectrum  of auxiliary graph availability, from idealized same-distribution access to task- and distribution-shifted settings.

\subsection{Framework Design}

\begin{figure}
    \centering
    \includegraphics[width=1\linewidth]{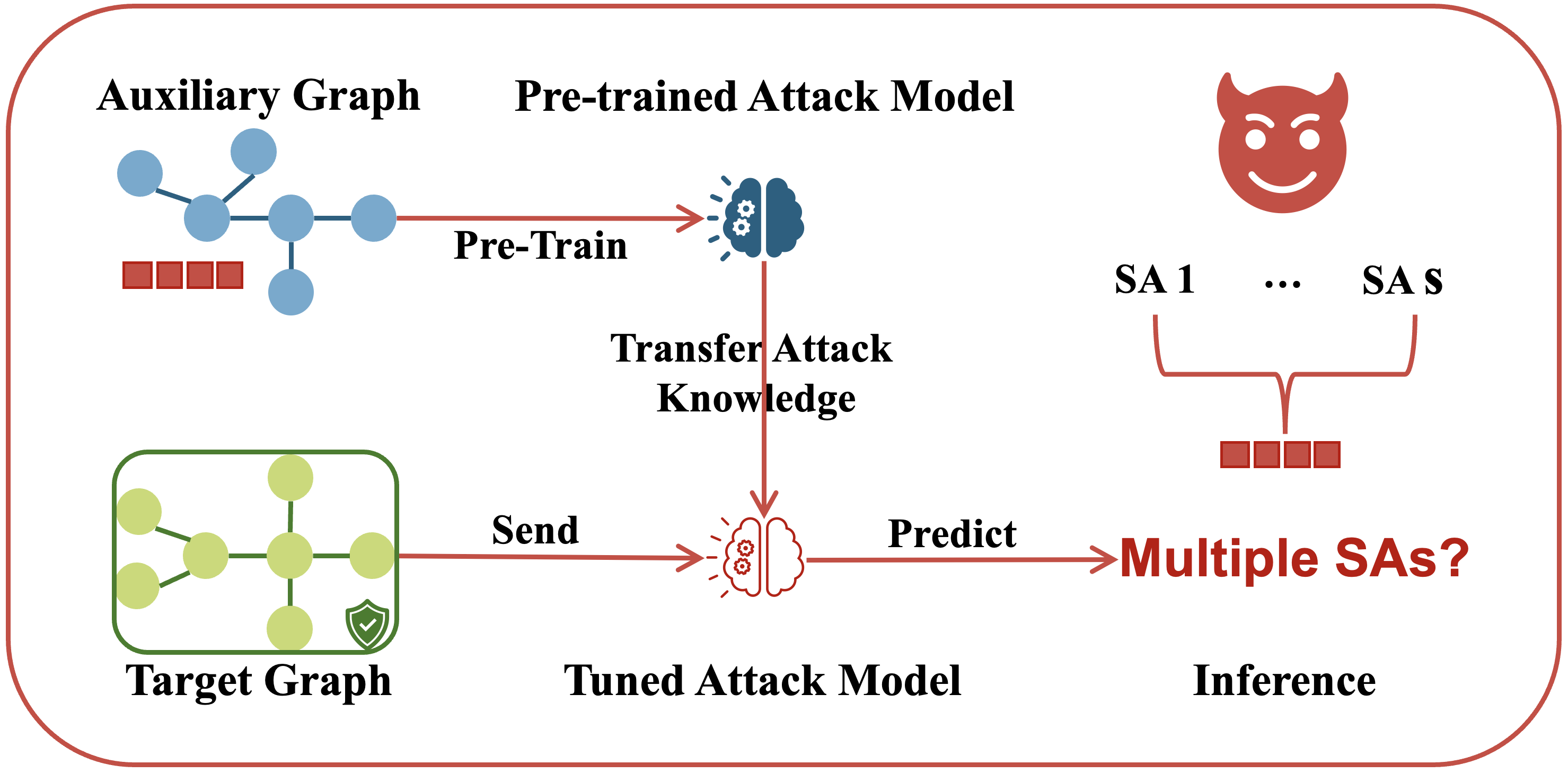}
    \caption{\centering The Overview of Taipan for G-MSAIAs. The stacked squares represent multiple sensitive attributes and the green shield indicates the privacy-preserving techniques.}
    \label{fig:taipan}
\end{figure}

In this section, we introduce Taipan, the first attack framework for G-MSAIAs and provide its overview in Figure \ref{fig:taipan}. 
We then present a systematic evaluation framework to measure multiple sensitive information leakage.


\subsubsection{Design Rationale}
G-MSAIAs aim to jointly infer multiple sensitive attributes, which naturally aligns with a graph multi-task learning paradigm. This paradigm exploits shared and complementary information among multiple related tasks to guarantee or even improve performance through multitask synergy while reducing computational and storage costs \cite{mtl_survey}. 

Under the realistic query-free setting, G-MSAIAs can be reformulated as a ``multi-task attack transfer" problem: the adversary pre-trains an attack model on an auxiliary graph and transfers it to infer multiple sensitive attributes on an unseen target graph. Although all attack tasks share the same auxiliary graph and target graph's non-sensitive attributes and topology, these two graphs may differ substantially in semantics and prediction difficulty. Furthermore, cross-attribute correlations can be positive, negative, or negligible, and homophily further amplifies non-trivial dependencies through topology, leading to complex and potentially conflicting optimization dynamics.

As a result, constructing a unified attack model for all attack tasks is inherently challenging: dominant tasks may overwhelm the optimization process, while conflicting tasks can induce gradient interference, causing optimization difficulty, slower convergence, and poor performance on weakly-related tasks or even across all tasks. Additionally, to comply with security and privacy regulations, sensitive attributes are typically removed prior to graph publishing and model training. Consequently, the adversary has no access to sensitive labels in the target graph and must infer multiple sensitive attributes from weak, indirect or even distorted signals alone, which further exacerbates the difficulty of query-free G-MSAIAs.

These challenges call for an attack framework that can flexibly balance shared and task-specific attack knowledge, explicitly model diverse task correlations, and remain parameter-efficient under realistic query-free settings. The Multi-gate Mixture-of-Experts (MMoE \cite{mmoe}) paradigm has emerged as a promising solution for flexibly handling complex task correlations while mitigating model collapse and performance degradation. It constructs a suite of shared expert networks to handle multiple tasks and utilizes a task-specific multi-gate mechanism to dynamically modulate these experts by assigning different weights. This parameter-efficient architecture substantially reduces the cumulative computational costs and storage requirements compared to independent single-task training. Nonetheless, MMoE cannot be trivially adapted to G-MSAIAs for several reasons. First, to the best of our knowledge, no unified graph MMoE framework exists. Since graphs exhibit complex topological structures and rich node features, existing MMoE frameworks tailored for tabular data cannot be directly extended to graphs. Second, G-MSAIAs necessitate transferring attack knowledge from a labeled auxiliary graph to an unlabeled target graph, which may follow a different distribution from the auxiliary graph under realistic attack settings. However, classical MMoE frameworks cannot address such unsupervised domain adaptation. Third, despite its parameter efficiency, naively fine-tuning a pre-trained MMoE attack model incurs additional computational and storage overheads that are not feasible in low-resource attack scenarios. In addition, updating all parameters might corrupt the general attack knowledge, leading to poor adaptation and generalization.

To address the above challenges, we propose Taipan, a novel attack framework for G-MSAIAs. Taipan consists of two core modules:  \emph{Hierarchical Attack Knowledge Routing} and \emph{Prompt-guided Attack Prototype Refinement}. We detail the design of Taipan in the following sections.

\subsubsection{Hierarchical Attack Knowledge Routing}
This module is designed to adaptively route shared and task-specific attack signals within a unified pre-trained model, enabling each inference task to extract transferable knowledge while mitigating negative interference among multiple sensitive attributes. It involves two stages: (1) profiling attack tasks to characterize and exploit diverse sensitive correlations; (2) assigning task identifiers to support task-aware attack knowledge routing and downstream sensitive attribute inference.

\noindent\textbf{Attack Profiling}
G-MSAIAs first require profiling the complex correlations among multiple attack tasks. Existing multi-task learning frameworks typically quantify such correlations either through the inter-task similarity of task gradients on shared prompts \cite{hierar_mtl} or via learnable task covariance matrices \cite{mtl_survey}. Despite their success, these approaches often incur high optimization difficulty and substantial computational overhead. 
To alleviate this issue, we adopt a simple yet effective strategy. Specifically, we compute pairwise cosine similarities among sensitive attributes in the auxiliary graph and then apply agglomerative hierarchical clustering \cite{Day_Edelsbrunner_1984} on their inter-similarity matrix to profile attack tasks. This profiling produces an attack hierarchy that captures fine-grained shared attack signals, enabling positive transfer among similar tasks while mitigating negative transfer among conflicting tasks. 

\begin{figure}[htbp] 
    \centering
    \includegraphics[width=1\linewidth]{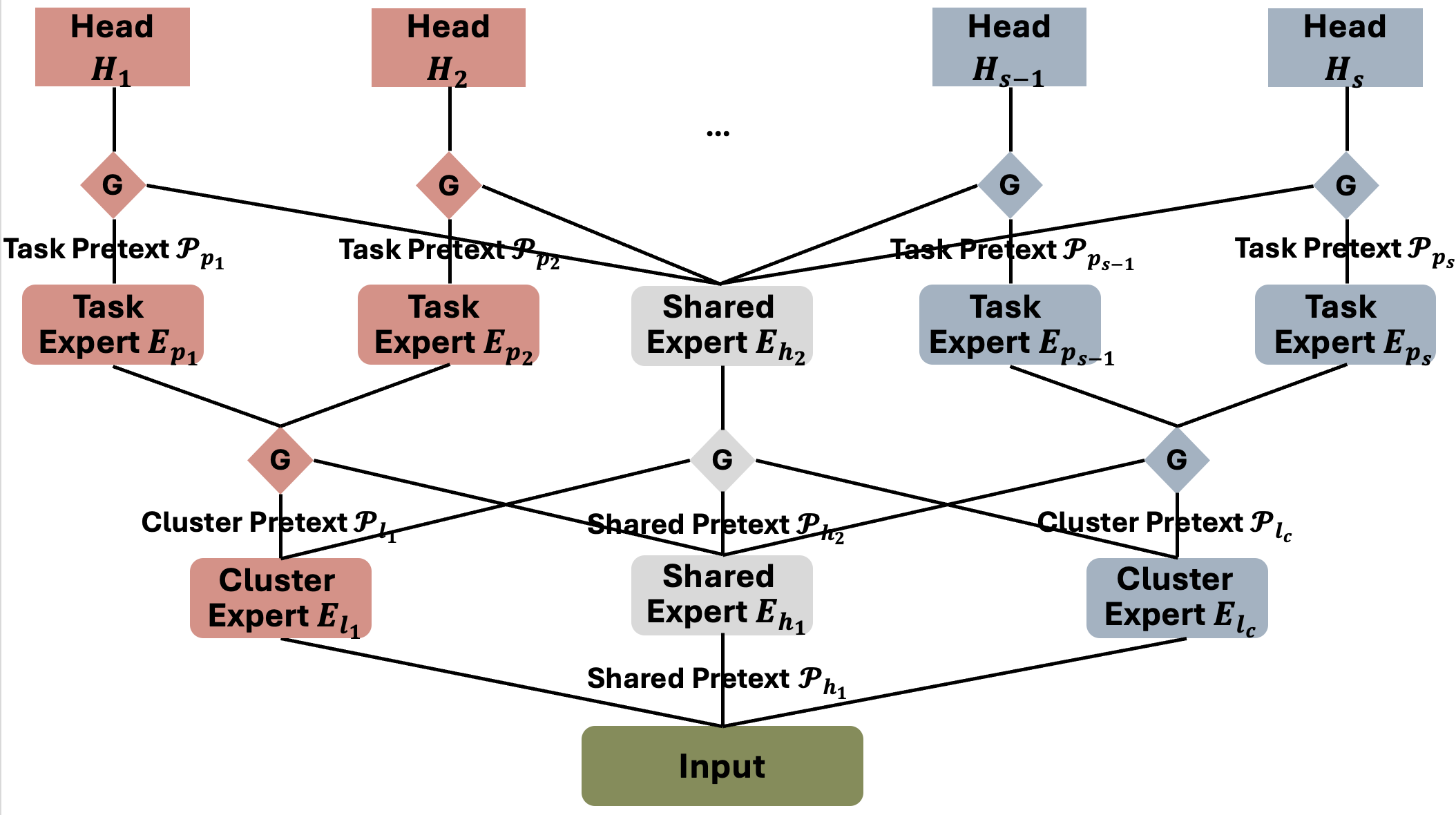} 
    \caption{Hierarchical Attack Knowledge Routing}
    \label{fig:mmoe}
\end{figure}

Guided by the attack hierarchy, we construct an MMoE-like multi-task attack framework with a suite of expert networks, gating networks, and task heads, as shown in Figure \ref{fig:mmoe}. The framework governs how experts are shared and how gates route attack knowledge across hierarchical clusters and inference tasks. When sensitive attributes exhibit no meaningful correlations, the hierarchy becomes uninformative, and the framework naturally degenerates to a flat multi-task architecture with a shared expert alongside task-specific experts, gates, and heads. 

\noindent\textbf{Task Identifier Assignment}
The framework's primary objective is to extract both shared and task-specific attack knowledge to assist downstream G-MSAIAs. While the attack hierarchy alleviates negative transfer among conflicting tasks, the shared encoder may struggle to disentangle fine-grained, task-specific semantics from global shared representations when routing experts across diverse tasks. We therefore introduce attack task identifiers to enable more task-aware attack knowledge routing.


Inspired by MULTIGPrompt \cite{multigprompt}, we instantiate attack task identifiers using pretext tokens, which are learnable vectors that can be injected into pre-training graphs \cite{gppt, allinone} or integrated into GNN architectures \cite{multigprompt}. Unlike MULTIGPrompt, which employs composite pretext tokens to modulate the input, hidden, and output layers of a shared encoder, we assign pretext tokens to each expert of the attack hierarchy to explicitly construct a hierarchical attack path. This design enables pretext tokens to function as task instructions that guide both shared and task-specific attack knowledge extraction, while inducing well-structured computational pathways tailored to different inference tasks. As a result, hierarchical task-aware routing effectively reduces gradient interference and mitigates negative transfer, while enhancing transferability and robustness across multiple attack tasks. In addition, pretext tokens serve as controllable semantic anchors, facilitating efficient and reliable adaptation to downstream sensitive attribute inference.

Formally, given a full hierarchical attack path which integrates an attack hierarchy $\mathcal{T}$ with $\kappa$ hierarchy and $c$ clusters (for clarity of illustration, we omit subclusters, i.e., $\kappa=2$. Extensions to deeper hierarchies are straightforward), we instantiate three types of experts—shared experts $E_{h}$, cluster-shared experts $E_{l}$ and task-specific experts $E_{p}$— and place them at their corresponding levels in the path. Each expert is paired with a corresponding gating network, denoted by $G_{h}$, $G_{l}$, and $G_{p}$, together with task-specific heads $H$. We further initialize a pretext token set $\mathcal{P}=\{\mathcal{P}_{h},\mathcal{P}_{l}, \mathcal{P}_{p}\}$ and attach them to their respective expert along the path, as illustrated in Figure~\ref{fig:mmoe}. The output of each attack task $\mathcal{S}_{a_{i}}, i\in\{1,2,\cdots,s\}$ can be formulated as follows: 
\begin{align}
    \mathcal{G}_{a}^*&= (\mathcal{V}_a, \mathcal{E}_a, X_a^*\odot \mathcal{P}_{h_1}) \notag\\
    Z_{\mathcal{C}_j}&=G_{l_j}(\mathcal{G}_{a}^*)^T\{E_{h_1}(\mathcal{G}_{a}^*)\odot \mathcal{P}_{h_2},E_{l_j}(\mathcal{G}_{a}^*)\odot \mathcal{P}_{l_j}\} \notag
    \\
    Z_{h_1}&=G_{h}(\mathcal{G}_{a}^*)^T\{E_{h_1}(\mathcal{G}_{a}^*)\odot \mathcal{P}_{h_2}, E_{l_1}(\mathcal{G}_{a}^*)\odot \mathcal{P}_{l_1}, \notag\\
    &\phantom{=\;\;}\dots,E_{l_c}(\mathcal{G}_{a}^*)\odot \mathcal{P}_{l_c}\} \notag \\
    Z_{\mathcal{S}_{a_{i}}}&=G_{p_i}(Z_{\mathcal{C}_j})^T\{E_{p_i}(Z_{\mathcal{C}_j})\odot \mathcal{P}_{p_i},E_{h_2}(Z_{h_1})\}  \notag \\
    \hat{\mathcal{S}}_{a_{i}}&=H_{i}(Z_{\mathcal{S}_{a_{i}}})
\end{align}
where $\mathcal{C}_j \text{ denotes a cluster in $\mathcal{T}$, }j\in\{1,2,\cdots,c\}$ and there exists a valid path from $\mathcal{C}_j$ to $\mathcal{S}_{a_{i}}$. $\mathcal{P}_{h}$ represents shared pretext tokens, and the remaining tokens are defined analogously. Each expert network $E=GNN(\cdot)$ is a GNN encoder, each gating network $G=softmax(\sigma(MLP(\cdot)))$ is a $2$-layer MLP, and each task head $H=MLP(\sigma(\cdot))$ is a $1$-layer MLP, where $\sigma$ is a specified activation function, such as ELU \cite{elu}. Since auxiliary  node labels $\mathcal{Y}_{a}$ also contain informative signals, we augment the non-sensitive node features as $X_{a}^*=X_{a}\oplus\mathcal{Y}_{a}$.

Finally, the entire framework is optimized by:
\begin{equation}
    \mathcal{L}_{att}=\sum_{i=1}^{s}w_i\cdot l(\mathcal{S}_{a_{i}}, \hat{\mathcal{S}}_{a_{i}})
    \label{pre_loss}
\end{equation}
where $l(\cdot)$ denotes the attack loss (e.g., cross-entropy). The task weight $w_i=dCor\left(\mathcal{S}_{a_{i}}, X^{*}_{a}\oplus(\mathcal{S}_{a}\backslash \mathcal{S}_{a_{i}})\right)$ is measured by the distance correlation ($dCor$) \cite{dcor} between each sensitive attribute and the remaining node features. We choose $dCor$ because it measures statistical dependence in arbitrary dimensions without exhaustive pairwise feature comparisons.

\subsubsection{Prompt-guided Attack Prototype Refinement} During attack transfer, we first freeze the pre-trained attack model and adapt it to the target graph by only tuning lightweight pretext token parameters. These pretext tokens act as graph prompts, substantially reducing computational overhead while mitigating task interference and negative transfer. Notably, this graph prompting fundamentally differs from the existing node-level reformulation methods that inject learnable vectors or subgraphs into pre-training graphs \cite{allinone, gppt, dagprompt}. Since the target graph is unlabeled, downstream G-MSAIAs can be naturally formulated as an unsupervised domain adaptation (UDA) problem. However, prior multi-task prompt tuning frameworks are primarily designed for supervised domain adaptation \cite{mtl_prompt_rec, hierar_mtl, multigprompt, mtl_survey, gmtl_replayandforgetfree}, leaving multi-task unsupervised domain adaptation largely unexplored, particularly in the graph domain. Additionally, existing unsupervised domain adaptation methods \cite{guda_smooth, guda_unsupervised} tailored for tabular data cannot be trivially extended to multi-task graph prompting. Motivated by these limitations, we propose a simple yet effective multi-task prompt tuning attack framework for unsupervised graph domain adaptation. The framework comprises two submodules: \emph{High-confidence Tuning} to mitigate data scarcity and catastrophic forgetting, and \emph{Adaptive Prototype Adjustment} to adaptively align semantic and structural signals across graphs.

\noindent\textbf{High-confidence Tuning} 
When target labels are unavailable, pseudo-labeling combined with data augmentation is a widely adopted strategy for mitigating distribution shift in unsupervised domain adaptation \cite{guda_unsupervised}. In the unlabeled target graph, we first generate pseudo-labels $\tilde{\mathcal{S}}_{t}$ using the well-calibrated pre-trained attack model and only preserve those nodes whose confidence scores exceed a pre-defined threshold $\gamma$. However, when the number of retained nodes is small or the target graph is limited in size, this few-shot scenario may lead to overfitting on noisy pseudo-labels. To alleviate data scarcity, we jointly fine-tune the attack model using both the auxiliary graph and the retained pseudo-labeled nodes. This strategy further aligns with auxiliary  attack semantics while avoiding catastrophic forgetting, thereby enabling stable and robust adaptation. Formally, its objective can be formulated as follows: 
\begin{align}
    \mathcal{L}_{cnf}=\mathcal{L}_{att}+\sum_{i=1}^{s}\sum_{j=1}^{m}w_i\cdot l(\tilde{\mathcal{S}}_{{t}_{ij}}, \hat{\mathcal{S}}_{{t}_{ij}})\mathbb{I}(\max P(\hat{\mathcal{S}}_{{t}_{ij}})\ge \gamma)
\end{align}
where $m\leq|\mathcal{V}_{t}|$ is the number of target nodes in the target graph, $\mathbb{I}(\cdot)$ represents the indicator function , $P(\cdot)$ denotes the predicted probability and $\gamma$ is the pre-defined threshold.


\noindent\textbf{Adaptive Prototype Adjustment}
A remaining challenge lies in the filtered nodes. As the pre-trained attack model cannot provide reliable confidence estimates for these nodes, semantic misalignment may persist between the auxiliary  and target graphs. Thus, we introduce the \emph{Adaptive Prototype Adjustment} submodule to align prototypes across graphs. 

We first initialize class prototypes for each attack task using node embeddings from the auxiliary graph, which serve as reliable semantic anchors. Formally, the prototype $\mathrm{C}$ for class $k$ of attack task $\mathcal{S}_i$ at iteration $t$ is defined as:
\begin{equation}
    \mathrm{C}_{\mathcal{S}_i,k}^{t}=\frac{\sum_{j=1}^{m+|\mathcal{V}_a|}\mathbb{I}(\mathcal{S}_i=k)Z_{\mathcal{S}_i}}{\sum_{j=1}^{m+|\mathcal{V}_a|}\mathbb{I}(\mathcal{S}_i=k)}
\end{equation}

We further incorporate exponential moving average (EMA) \cite{ema} to enable smooth domain adaptation. 
\begin{equation}
    \mathrm{C}_{\mathcal{S}_i,k}^{t*}=\alpha\mathrm{C}_{\mathcal{S}_i,k}^{t-1}+(1-\alpha)\mathrm{C}_{\mathcal{S}_i,k}^{t}
\end{equation}
where the momentum $\alpha\in[0,1]$ controls the update rate.

For the filtered nodes, we define the following loss:
\begin{equation}
    \mathcal{L}_{fil}=-\sum_{i=1}^{s}\sum_{j=1}^{m_{fil}}P(\hat{\mathcal{S}}_{ij})\log\{P(\hat{\mathcal{S}}_{ij})\}
\end{equation}
where $m_{fil}$ denotes the number of filtered target nodes, $P(\hat{\mathcal{S}}_{ij})=softmax \left(cos(Z_{fil_{\mathcal{S}_{ij}}},\mathrm{C}_{\mathcal{S}_{ij}})/\tau \right)$ is the prediction for each attack task and $\tau$ is the temperature coefficient.

To align two graphs' distributions while mitigating catastrophic forgetting, we define the overall adaptation loss as:
\begin{equation}
    \mathcal{L}_{prm}=\mathcal{L}_{cnf}+\mathcal{L}_{fil}
\end{equation}

Specifically, when auxiliary  and target graphs originate from different distributions with mismatched feature dimensionalities, we apply principal component analysis (PCA) to project both graphs into a unified feature space. Our goal is to demonstrate the feasibility of G-MSAIAs under diverse auxiliary data settings, more advanced mismatched UDA techniques are left for future work.









\subsection{Evaluation Metrics}
Existing evaluation metrics for AIAs are primarily designed for a single sensitive attribute and therefore fail to capture multi-attribute privacy leakage. To bridge this gap and answer our third research question, we propose a systematic evaluation framework for G-MSAIAs. We further empirically demonstrate its necessity in the following section. For clarity, we provide mathematical definitions for all metrics that are specific to the multi-attribute attack setting.

\subsubsection{Confidence-based Metrics} 
\hfill

\noindent\textbf{Average AUC (AA) and F1 (AF).} We compute the standard AUC and F1 scores for each sensitive attribute and report their averages to provide a global assessment of attack confidence and discriminative capability in a unified way. 

\noindent\textbf{Task Deviation AUC (TDA) and F1 (TDF).} To quantify the performance gap across attack tasks, for each sensitive attribute $\mathcal{S}_i$, we define TDA as $\max\limits_{\mathcal{S}_i}\text{AUC}_{\mathcal{S}_i}-\min\limits_{\mathcal{S}_i}\text{AUC}_{\mathcal{S}_i}$, and TDF is defined analogously. Larger TDA/TDF indicates greater disparity and less stability across different attack tasks.

\subsubsection{Correctness-based Metrics} 
\hfill

\noindent\textbf{Hamming Distance (HD)} HD quantifies the average fraction of incorrectly inferred sensitive attributes across all nodes. Mathematically, let $\mathcal{S}_v\text{ and }\hat{\mathcal{S}}_v\in\{0,1\}^{s}$, HD is defined as $\text{HD} = \frac{1}{|\mathcal{V}|}\sum_{v\in\mathcal{V}}\frac{1}{s}\sum_{i=1}^{s}\mathbb{I}(\hat{\mathcal{S}}_{v,i}\neq\mathcal{S}_{v,i})$.

\noindent\textbf{Subset Accuracy (SuA)} While HD captures marginal privacy leakage, it overlooks joint attack success across multiple sensitive attributes. To address this limitation, we introduce SuA, a strict ``all-or-nothing'' metric that measures the probability of simultaneous G-MSAIAs, defined as $\text{SuA}=\frac{1}{|\mathcal{V}|}\sum_{v\in\mathcal{V}}\frac{1}{s}\mathbb{I}(\hat{\mathcal{S}}_{v,1}=\mathcal{S}_{v,1},\dots,\hat{\mathcal{S}}_{v,s}=\mathcal{S}_{v,s})$. An attack is counted as successful under SuA only if \textit{all sensitive attributes of a specific node are correctly inferred}. However, SuA alone can be misleading: a low SuA does not necessarily imply effective privacy protection, as partial leakage across attributes may still be substantial. In this sense, HD and SuA are complementary: HD reflects marginal leakage, while SuA captures joint compromise. Their combination provides a more comprehensive assessment of sensitive information erosion. 

\subsubsection{Semantics-based Metrics} Multiple sensitive attributes often exhibit complex cross-attribute correlations, however, existing metrics largely ignore this facet and fail to assess whether an attack captures the underlying semantic dependencies of the target graph.

\noindent\textbf{Semantic Difference (SD)} SD measures the discrepancy in feature–label alignment between inferred and ground-truth sensitive attributes, defined as $dCor(\hat{S}, X)-dCor(S, X)$. Lower SD values indicate that the attack preserves more alignment, whereas higher SD suggests that the attack may overly rely on biased feature signals. 

\noindent\textbf{Label Consistency (LC)} LC evaluates whether inferred sensitive attributes retain the dependency structure of the ground-truth, denoted as $dCor(\hat{S},S)$. Higher LC indicates that the attack effectively captures internal correlations among multiple sensitive attributes while lower LC suggests weak or distorted dependency modeling.

\section{Experiments}
\label{sec:exp}
In this section, we conduct extensive experiments on four real-world graph datasets to systematically evaluate the attack performance of Taipan. Specifically, we investigate the following six research questions: 

\noindent\textbf{Q1}: Why is the comprehensive evaluation framework necessary for G-MSAIAs?

\noindent\textbf{Q2}: How effective is Taipan in performing G-MSAIAs?

\noindent\textbf{Q3}: How well does Taipan generalize across GNN encoders?

\noindent\textbf{Q4}: What factors can influence the performance of Taipan?

\noindent\textbf{Q5}: Which nodes are more/less vulnerable to Taipan?

\noindent\textbf{Q6}: How robust is Taipan against defense mechanisms?



\subsection{Experiment Setups}
\label{exp_setup}

\noindent\textit{\textbf{Datasets.}}
We evaluate Taipan on four real-world benchmark graph datasets: German, Credit, Pokec-n and Pokec-z. Graph statistics are presented in Table \ref{data_stat} in Appendix \ref{sec_data_stat}. 

\noindent\textit{\textbf{GNN encoders.}} We consider three representative GNNs: GCN \cite{gcn}, GraphSAGE \cite{sage} and GIN \cite{gin}. Their original performance is summarized in Table \ref{data_fair} in Appendix \ref{sec_ori_perf}, where GCN encoders exhibit larger prediction biases toward majority classes compared to GraphSAGE and GIN. 

\noindent\textit{\textbf{Baselines.}} Taipan is the first attack framework for G-MSAIAs. Existing AIAs in the graph domain either rely on access to node embeddings or repeated queries to victim GNNs \cite{AIA_graph_emb, AIA_gnn}, or are confined to social networks and require partially disclosed sensitive attributes of target users \cite{aia_link_attr, ego_social_net, mul_sens_emb}. Therefore, they are not directly comparable. Instead, we consider the following baselines: (1) Frequency Prior (FrePr.): Sensitive attributes are inferred by sampling from Bernoulli distributions based on class frequency. 
Despite its simplicity, this baseline can achieve deceptively high accuracy by always predicting the majority class on imbalanced datasets.
(2) Single Prediction (SingP.): We separately train an attack model for each sensitive attribute, and then aggregate the best-performing models. SingP. serves as an ablation without attack hierarchy and the ``attack transfer'' paradigm. (3) Pre-training (PreTr.): We directly apply pre-trained Taipan to infer multiple sensitive attributes on the target graph. This baseline also acts as an ablation without \emph{Prompt-guided Attack Prototype Refinement}.

\noindent\textit{\textbf{Implementation.}} Experiments are conducted on three Nvidia A$100$ GPUs. Please see Appendix \ref{sec_imp_det} for more details.

\subsection{Attack Performance (Q1-Q3)}
\subsubsection{Necessity of Comprehensive Evaluation (Q1)} 
The attack results in Table \ref{att_perf} highlight the limitations of classical single-attribute metrics. Under class imbalance on German, even FrePr. can yield deceptively high average AUC and F1 scores, but their task deviations are substantial, which indicates that poor performance on certain attack tasks can be obscured by strong performance on others. Correctness-based metrics further reveal these shortcomings. For instance, while FrePr. attains acceptable HD, its SuA drops sharply, suggesting weak joint inference across multiple sensitive attributes. Similarly, SingP. exhibits low SuA, as independently trained attack models may behave inconsistently on the same node samples, leading to severe performance degradation in the multi-attribute setting. In contrast, SuA can better capture inference synergy across multiple sensitive attributes. The competitive SuA achieved by pre-trained and full Taipan supports this observation, as Taipan is designed to model task correlations and mitigate negative transfer. Notably, low SuA alone does not imply strong privacy protection when HD still indicates substantial partial leakage, such as SingP. on German. Finally, the semantics-based metrics assess whether inferred attributes align with sensitive attributes in auxiliary graphs. Consistent with earlier findings, SD reveals that GCN tends to induce higher bias, while LC shows that Taipan can effectively capture intersectional correlations among multiple sensitive attributes.

\subsubsection{Effectiveness (Q2)}
From Table \ref{att_perf}, we observe consistent patterns of Taipan across diverse graph datasets. Both Taipan and its pre-trained variant (PreTr.) outperform frequency-based sampling and achieve performance comparable to or exceeding SingP. across most evaluation metrics. Specifically, full Taipan generally maintains superior performance over PreTr., highlighting the effectiveness of \emph{Prompt-guided Attack Prototype Refinement}. We note that while SingP. attains higher AA and AF in some cases, it yields unstable and unfair attack performance, as reflected by TDA and TDF, and significantly lags behind Taipan in correctness and semantics-based metrics. For instance, on Pokec-n with GraphSAGE as the encoder, SingP. exceeds Taipan by $4.57\%$ in AA, yet performs worse in AF and incurs substantially larger task deviations. Overall, Taipan not only achieves comparable or higher AA/AF than SingP. but also demonstrates superior stability and fairness while more effectively preserving intersectional sensitive correlations. We further visualize node embeddings of the inferred multiple sensitive attributes along with their joint confidence score distributions to illustrate the effectiveness of Taipan. Due to space limitations, details are provided in Appendix \ref{sec_viz}.

\subsubsection{Flexibility (Q3)}
As shown in Table \ref{att_perf}, Taipan and PreTr. exhibit robust performance across all GNN encoders. In general, attack performance degrades with GIN as the encoder. The only exception occurs on the dense Credit graph. Unlike GCN and GraphSAGE, which employ smoothing ``low-pass'' filters, GIN relies on sum-based aggregation and is therefore more sensitive to structural noise, particularly on graphs with low average degree. This behavior is consistent with the lower original performance of GIN on such graphs, suggesting a fundamental mismatch between its aggregation mechanism and the underlying graph structure. Furthermore, the additional learnable parameters introduced by pretext tokens may amplify noise, impairing transferable knowledge extraction and semantic alignment, as reflected by SD and LC. 

\begin{table}[!htbp]
\caption{\centering The Attack Performance of Taipan — Same-distribution Transfer, where the auxiliary and target graphs both come from the same graph dataset without overlapping. Bold: Best result; Underline: Second best result (close to the best).}
\label{att_perf}
\resizebox{\linewidth}{!}{
\begin{tabular}{c||c|c|cccccccc}
\toprule\bottomrule
\textbf{Dataset}                   & \textbf{GNN}          & \textbf{Baseline} & \textbf{AA ($\uparrow$)} & \textbf{AF ($\uparrow$)} & \textbf{TDA ($\downarrow$)} & \textbf{TDF ($\downarrow$)} & \textbf{HD ($\downarrow$)} & \textbf{SuA ($\uparrow$)} & \textbf{SD ($|\downarrow|$)} & \textbf{LC ($\uparrow$)} \\ \bottomrule

\multirow{10}{*}{\textbf{German}}  & -                & FrePr.  & 54.21 & 60.25 & 13.08 & 58.70 & 36.33 & 26.00 & -2.28 & 20.37 \\ \cline{2-11}
                                   & \multirow{3}{*}{GCN}       & SingP. & \underline{66.50} & 42.50 & 13.46 & 88.05 & 33.27 & 14.00 & \textbf{0.48}  & 34.36 \\
                                   &                            & PreTr. & 66.34 & \underline{73.62} & \underline{\textbf{10.45}} & \underline{40.24} & \textbf{20.33} & \underline{\textbf{54.00}} & \underline{7.22}  & \underline{42.77} \\
                                   &                            & Taipan & \textbf{70.01} & \textbf{76.10} & \textbf{7.58}  & \textbf{31.44} & \underline{\textbf{21.33}} & \textbf{57.00} & 9.88  & \textbf{46.35} \\ \cline{2-11}
                                   & \multirow{3}{*}{GraphSAGE} & SingP. & 69.47 & 54.72 & \textbf{8.28}  & 55.69 & 35.20 & 35.20 & \textbf{-1.80} & 32.43 \\
                                   &                            & PreTr. & \underline{69.37} & \underline{70.49} & 19.95 & \underline{53.75} & \underline{19.67} & \underline{53.00} & 6.26  & \underline{56.15} \\
                                   &                            & Taipan & \textbf{72.62} & \textbf{77.11} & \underline{16.99} & \textbf{41.41} & \textbf{16.33} & \textbf{60.00} & \underline{\textbf{3.38}}  & \textbf{64.40} \\ \cline{2-11}
                                   & \multirow{3}{*}{GIN}       & SingP. & 55.64 & 31.49 & 22.18 & 89.50 & 34.80 & 12.00 & \textbf{-8.37} & 6.89  \\
                                   &                            & PreTr. & \underline{\textbf{66.39}} & \textbf{73.38} & \underline{21.16} & \underline{33.26} & \textbf{25.00} & \textbf{49.00} & \underline{\textbf{9.39}}  & \underline{57.20} \\
                                   &                            & Taipan & \textbf{67.30} & \underline{\textbf{72.28}} & \textbf{5.70}  & \textbf{28.12} & \underline{30.67} & \underline{39.00} & 18.91 & \textbf{62.58} \\ \hline
\multirow{10}{*}{\textbf{Credit}}  & -                & FrePr.  & 49.97 & 72.31 & 0.30  & 36.30 & 33.47 & 41.63 & -5.50 & 1.37  \\ \cline{2-11}
                                   & \multirow{3}{*}{GCN}       & SingP. & \textbf{62.80} & \textbf{82.63} & 21.14 & \textbf{24.48} & 27.60 & 46.16 & \textbf{4.70}  & 9.12  \\
                                   &                            & PreTr. & 57.18 & \underline{\textbf{80.94}} & \textbf{5.13}  & 28.16 & \textbf{24.30} & \textbf{53.67} & \underline{\textbf{5.03}}  & \textbf{19.34} \\
                                   &                            & Taipan & \underline{\textbf{60.02}} & \underline{80.16} & \underline{\textbf{6.95}}  & \underline{28.53} & \underline{\textbf{26.13}} & \underline{\textbf{51.37}} & 19.38 & \underline{15.91} \\ \cline{2-11}
                                   & \multirow{3}{*}{GraphSAGE} & SingP. & \textbf{69.69} & \textbf{78.99} & 25.02 & \underline{\textbf{32.86}} & 25.98 & 50.35 & 14.82 & 11.79 \\
                                   &                            & PreTr. & 60.45 & 77.96 & \underline{\textbf{5.54}}  & 34.77 & \underline{\textbf{25.40}} & \underline{\textbf{52.40}} & \textbf{9.08}  & \underline{16.24} \\
                                   &                            & Taipan & \underline{61.86} & \underline{\textbf{78.91}} & \textbf{4.87}  & \textbf{31.66} & \textbf{24.97} & \textbf{53.77} & \underline{\textbf{9.17}}  & \textbf{19.62} \\ \cline{2-11}
                                   & \multirow{3}{*}{GIN}       & SingP. & \textbf{74.71} & 71.50 & 28.62 & 47.78 & 27.02 & 49.58 & 16.30 & 15.84 \\
                                   &                            & PreTr. & \underline{66.69} & \underline{\textbf{81.30}} & \underline{\textbf{8.58}}  & \underline{\textbf{28.00}} & \underline{\textbf{22.78}} & \underline{\textbf{56.90}} & \underline{\textbf{1.52}}  & \underline{\textbf{26.24}} \\
                                   &                            & Taipan & 65.67 & \textbf{82.53} & \textbf{6.20}  & \textbf{26.12} & \textbf{22.23} & \textbf{57.47} & \textbf{0.74}  & \textbf{26.68} \\ \hline
\multirow{10}{*}{\textbf{Pokec-n}} & -                & FrePr.  & 48.40 & 28.30 & 4.60  & 23.50 & 41.67 & 21.70 & -6.41 & 8.01  \\ \cline{2-11}
                                   & \multirow{3}{*}{GCN}       & SingP. & \textbf{81.06} & 57.14 & 21.14 & 32.62 & 22.18 & 46.22 & \textbf{-0.04} & 44.37 \\
                                   &                            & PreTr. & 72.34 & \underline{61.01} & \underline{15.13} & \underline{27.91} & \textbf{20.76} & \textbf{50.11} & \underline{2.82}  & \textbf{48.23} \\
                                   &                            & Taipan & \underline{74.85} & \textbf{64.39} & \textbf{10.80} & \textbf{20.46} & \underline{\textbf{21.74}} & \underline{46.93} & 4.85  & \underline{\textbf{47.76}} \\ \cline{2-11}
                                   & \multirow{3}{*}{GraphSAGE} & SingP. & \textbf{81.76} & 63.39 & 29.84 & 40.75 & 20.82 & 48.69 & 4.55  & 56.71 \\
                                   &                            & PreTr. & 77.12 & \underline{\textbf{66.68}} & \underline{\textbf{21.06}} & \textbf{32.15} & \underline{\textbf{18.45}} & \underline{\textbf{54.20}} & \underline{\textbf{3.23}}  & \underline{\textbf{59.74}} \\
                                   &                            & Taipan & \underline{77.19} & \textbf{67.08} & \textbf{20.91} & \underline{\textbf{32.21}} & \textbf{17.42} & \textbf{56.70} & \textbf{1.50}  & \textbf{61.67} \\ \cline{2-11}
                                   & \multirow{3}{*}{GIN}       & SingP. & \textbf{77.48} & 55.25 & 19.64 & 31.62 & 24.01 & 42.31 & \textbf{-0.17} & 41.21 \\
                                   &                            & PreTr. & 69.74 & \underline{\textbf{57.88}} & \textbf{10.17} & \textbf{21.11} & \underline{\textbf{21.89}} & \underline{\textbf{47.05}} & \underline{\textbf{0.23}}  & \underline{\textbf{44.24}} \\
                                   &                            & Taipan & \underline{70.89} & \textbf{58.96} & \underline{\textbf{11.52}} & \underline{25.25} & \textbf{21.67} & \textbf{47.27} & 1.45  & \textbf{44.89} \\ \hline
\multirow{10}{*}{\textbf{Pokec-z}} & -                & FrePr.  & 50.28 & 37.07 & 4.50  & 25.89 & 42.94 & 18.40 & -9.31 & 6.45  \\ \cline{2-11}
                                   & \multirow{3}{*}{GCN}       & SingP. & \underline{\textbf{80.18}} & 59.15 & \textbf{22.85} & \underline{\textbf{34.17}} & 23.53 & 44.55 & \textbf{0.49}  & 43.43 \\ 
                                   &                            & PreTr. & 76.53 & \underline{67.78} & 28.28 & 41.21 & \textbf{18.70} & \textbf{53.16} & \underline{\textbf{2.44}}  & \textbf{61.83} \\
                                   &                            & Taipan & \textbf{80.28} & \textbf{73.14} & \underline{\textbf{25.51}} & \textbf{31.96} & \underline{\textbf{19.51}} & \underline{\textbf{52.87}} & 4.53  & \underline{\textbf{61.05}} \\ \cline{2-11}
                                   & \multirow{3}{*}{GraphSAGE} & SingP. & \textbf{82.59} & 65.88 & 30.57 & 38.04 & 20.91 & 49.15 & \textbf{5.11}  & 56.41 \\
                                   &                            & PreTr. & 78.69 & \underline{\textbf{70.68}} & \underline{\textbf{21.90}} & \underline{\textbf{31.46}} & \underline{\textbf{18.44}} & \underline{\textbf{54.92}} & 5.95  & \underline{\textbf{60.81}} \\
                                   &                            & Taipan & \underline{\textbf{80.49}} & \textbf{72.70} & \textbf{20.19} & \textbf{29.15} & \textbf{17.66} & \textbf{56.28} & \underline{\textbf{5.86}}  & \textbf{61.90} \\ \cline{2-11}
                                   & \multirow{3}{*}{GIN}       & SingP. & \textbf{78.06} & 59.96 & 20.76 & 28.16 & 25.38 & 40.91 & \textbf{0.96}  & 41.35 \\
                                   &                            & PreTr. & 74.28 & \underline{\textbf{67.39}} & \textbf{9.43}  & \underline{\textbf{17.41}} & \underline{\textbf{22.65}} & \underline{46.25} & \underline{\textbf{-1.11}} & \underline{47.49} \\
                                   &                            & Taipan & \underline{\textbf{76.96}} & \textbf{69.87} & \underline{\textbf{11.99}} & \textbf{16.57} & \textbf{20.64} & \textbf{50.05} & 3.77  & \textbf{51.92}

                     \\ \toprule\bottomrule
\end{tabular}}
\end{table}


\subsubsection{Ablation Studies (Q4)}
To systematically assess the contribution of each component in Taipan, we conduct a series of ablation studies on German with GCN as the encoder. In \emph{Hierarchical Attack Knowledge Routing}, we separately remove the attack hierarchy, pretext tokens, and gating mechanism, along with their combinations. In \emph{Prompt-guided Attack Prototype Refinement}, we ablate the source-based prototype adjustment. As target-based adaptation constitutes the core design of this module, removing it entirely reduces the framework to the pre-trained Taipan. As shown in Table \ref{att_perf_ablation}, removing any component results in a clear degradation in attack performance, with the extent of degradation reflecting its relative importance. Particularly, pretext tokens contribute the most to G-MSAIAs, followed by the attack hierarchy. The combination of pretext tokens and gating mechanism yields the strongest gains for partial sensitive attribute inference while the combination of pretext tokens and attack hierarchy follows behind. Source-based prototype adjustment further provides a performance gain in \emph{Prompt-guided Attack Prototype Refinement}. Overall, \emph{Hierarchical Attack Knowledge Routing} plays a central role in Taipan for effective G-MSAIAs.

\begin{table}[h]
\caption{\centering The Ablation Studies of Taipan for G-MSAIAs}
\label{att_perf_ablation}
\resizebox{\linewidth}{!}{
\begin{tabular}{c||c|cccccccc}
 \toprule\bottomrule
\textbf{Stage}                          & \textbf{Variant}    & \textbf{AA ($\uparrow$)} & \textbf{AF ($\uparrow$)} & \textbf{TDA ($\downarrow$)} & \textbf{TDF ($\downarrow$)} & \textbf{HD ($\downarrow$)} & \textbf{SuA ($\uparrow$)} & \textbf{SD ($|\downarrow|$)} & \textbf{LC ($\uparrow$)}  \\ \bottomrule
\textbf{Full}                           & \textbf{Taipan}     & \textbf{70.01} & \textbf{76.10} & \underline{\textbf{7.58}}  & \textbf{31.44} & 21.33 & \textbf{57.00} & 9.88  & \underline{\textbf{46.35}} \\ \hline
\multirow{7}{*}{\textbf{Pre-training}}  & w.o.hierarchy       & 64.65 & 67.35 & 21.29 & 54.77 & 25.33 & 49.00 & 5.02  & 44.99 \\
                                        & w.o.pretexts        & 58.06 & 71.53 & 12.84 & 46.41 & 25.00 & 54.00 & 3.73  & 39.42 \\
                                        & w.o.gating          & \underline{\textbf{67.32}} & 70.50 & 21.71 & \underline{\textbf{35.34}} & 28.33 & 44.00 & 15.64 & \textbf{51.77} \\
                                        & w.o. hier.+pret.    & 54.72 & 67.57 & 11.71 & 54.35 & 25.33 & 55.00 & \textbf{3.19}  & 37.91 \\
                                        & w.o. hiery+gate.    & 63.40 & 68.38 & \underline{\textbf{7.53}}  & 52.20 & 24.67 & \underline{\textbf{56.00}} & \underline{\textbf{3.29}}  & 38.08 \\
                                        & w.o. pret+gate      & 58.91 & 66.39 & 18.83 & 61.89 & 22.67 & 52.00 & 15.93 & 36.90 \\
                                        & w.o.all             & 52.44 & 61.94 & \textbf{4.87}  & 71.95 & 26.67 & \underline{\textbf{56.00}} & \underline{\textbf{-3.59}} & 25.01 \\ \hline
\multirow{2}{*}{\textbf{Prompt-tuning}} & w.o. source adjust. & \underline{\textbf{67.92}} & \underline{\textbf{74.95}} & 14.23 & \underline{\textbf{35.59}} & \underline{\textbf{20.67}} & 54.00 & 11.42 & \underline{\textbf{46.58}} \\
                                        & w.o.all             & 66.34 & 73.62 & 10.45 & 40.24 & \textbf{20.33} & 54.00 & 7.22  & 42.77             \\ \toprule\bottomrule
\end{tabular}}
\end{table}


\subsection{Impact Factors (Q4 and Q5)}

\subsubsection{Mismatched Attack Transfer (Q4)} Previously, we assumed the auxiliary graph is drawn from the same distribution as the target graph. We now consider a more challenging setting where the adversary can only access mismatched auxiliary graphs, including both similar-distribution and OOD cases. Specifically, we study two transfer scenarios: (1) Pokec-n as the auxiliary graph and Pokec-z as the target graph, which originate from the same social network but are sampled from different provinces and exhibit mismatched feature dimensionalities; and (2) Credit as the auxiliary graph and German as the target graph, which stem from different data sources with distinct learning objectives. From Table \ref{att_perf_within}, Taipan and PreTr. consistently outperform FrePr. and SingP. across diverse GNN encoders and transfer configurations. Consistent with earlier observations, GIN yields inferior attack performance in the Pokec-n-to-Pokec-z transfer. While the Credit-to-German transfer introduces larger distribution shifts and results in lower attack performance than Pokec-n-to-Pokec-z, Taipan remains effective under this stringent setting. 

\begin{table}[!htbp]
\caption{\centering The Attack Performance — Mismatched Similar-distribution and Out-of-distribution Transfers}
\label{att_perf_within}
\resizebox{\linewidth}{!}{
\begin{tabular}{c||c|c|cccccccc}
\toprule\bottomrule
\textbf{Dataset}                          & \textbf{GNN}               & \textbf{Baseline} & \textbf{AA ($\uparrow$)} & \textbf{AF ($\uparrow$)} & \textbf{TDA ($\downarrow$)} & \textbf{TDF ($\downarrow$)} & \textbf{HD ($\downarrow$)} & \textbf{SuA ($\uparrow$)} & \textbf{SD ($|\downarrow|$)} & \textbf{LC ($\uparrow$)} \\ \bottomrule
\multirow{9}{*}{\textbf{Pokec-n to Pokec-z}} & -                & FrePr.  & 50.63 & 34.05 & 2.19  & 58.62 & 41.38 & 18.89 & -7.50 & 5.68 \\ \cline{2-11}

& \multirow{3}{*}{GCN}       & SingP. & \textbf{71.37} & 43.15 & 35.58 & \underline{49.14} & \textbf{29.10} & \textbf{34.92} & \textbf{-2.01}  & 33.67 \\
                                          &                            & PreTr. & 62.22 & \underline{44.87} & \underline{\textbf{24.50}} & 51.88 & \underline{\textbf{29.18}} & \underline{\textbf{34.08}} & -3.75  & \textbf{37.51} \\
                                          &                            & Taipan & \underline{64.33} & \textbf{50.61} & \textbf{23.66} & \textbf{41.87} & \underline{29.83} & \underline{32.04} & \underline{\textbf{2.77}}   & \underline{\textbf{36.14}} \\ \cline{2-11} 
                                          & \multirow{3}{*}{GraphSAGE} & SingP. & \textbf{73.01} & 50.77 & 36.87 & 48.06 & \textbf{27.27} & \textbf{37.36} & \underline{\textbf{-0.77}}  & \textbf{43.40} \\
                                          &                            & PreTr. & 64.64 & \underline{52.16} & \underline{29.43} & \underline{42.74} & 29.21 & 33.89 & -1.80  & \underline{\textbf{40.62}} \\
                                          &                            & Taipan & \underline{65.79} & \textbf{55.29} & \textbf{25.77} & \textbf{36.34} & \underline{\textbf{28.89}} & \underline{\textbf{34.86}} & \textbf{0.53}   & \underline{40.35} \\ \cline{2-11} 
                                          & \multirow{3}{*}{GIN}       & SingP. & \textbf{66.62} & \underline{47.33} & 23.48 & \underline{30.56} & 32.10 & 30.07 & \textbf{-0.16}  & 26.92 \\
                                          &                            & PreTr. & 58.02 & 36.21 & \underline{17.04} & 54.77 & \textbf{30.28} & \textbf{33.79} & \underline{-4.41}  & \underline{\textbf{29.96}} \\
                                          &                            & Taipan & \underline{\textbf{64.22}} & \textbf{54.71} & \textbf{13.32} & \textbf{23.70} & \underline{\textbf{30.67}} & \underline{\textbf{32.72}} & 7.78   & \textbf{30.93} \\ \hline
\multirow{9}{*}{\textbf{Credit to German}}   & -                & FrePr.  & 46.11 & 66.18 & 6.89  & 60.50 & 39.50 & 35.00 & -11.63 & 12.91 \\ \cline{2-11}

& \multirow{3}{*}{GCN}       & SingP. & 44.25 & 69.11 & 17.56 & 40.30 & 39.10 & 33.00 & \underline{43.49}  & \underline{19.82} \\
                                          &                            & PreTr. & \underline{47.20} & \textbf{75.34} & \underline{5.60}  & \textbf{31.93} & \underline{\textbf{34.00}} & \underline{40.00} & 45.62  & \textbf{23.15} \\
                                          &                            & Taipan & \textbf{50.75} & \underline{\textbf{74.22}} & \textbf{1.51}  & \underline{\textbf{34.16}} & \textbf{32.00} & \textbf{44.00} & \textbf{31.37}  & 5.88  \\ \cline{2-11} 
                                          & \multirow{3}{*}{GraphSAGE} & SingP. & \underline{\textbf{50.74}} & \textbf{71.06} & 3.28  & 34.37 & \textbf{35.50} & \textbf{38.60} & \textbf{-8.62}  & 5.90  \\
                                          &                            & PreTr. & 48.74 & \underline{\textbf{68.29}} & \textbf{1.03}  & \underline{26.84} & \underline{\textbf{40.50}} & \underline{\textbf{37.00}} & \underline{11.96}  & \underline{\textbf{29.35}} \\
                                          &                            & Taipan & \textbf{51.93} & 66.30 & \underline{\textbf{2.96}}  & \textbf{3.57}  & 44.50 & 33.00 & 35.42  & \textbf{29.36} \\ \cline{2-11} 
                                          & \multirow{3}{*}{GIN}       & SingP. & \underline{\textbf{51.66}} & \textbf{78.44} & 4.69  & \underline{\textbf{15.27}} & \textbf{34.70} & \textbf{46.00} & \textbf{-10.37} & 4.08  \\
                                          &                            & PreTr. & 46.91 & 68.17 & \textbf{0.23}  & 40.19 & \underline{\textbf{37.50}} & \underline{38.00} & \underline{\textbf{10.84}}  & \underline{9.27}  \\
                                          &                            & Taipan & \textbf{54.28} & \underline{\textbf{70.91}} & \underline{\textbf{1.25}}  & \textbf{9.46}  & 42.00 & 34.00 & 43.01  & \textbf{13.97} \\ \toprule\bottomrule
\end{tabular}} 
\end{table}

\subsubsection{Adaptation Steps (Q4)}
\begin{figure*}[htbp] 
\centering
    \begin{subfigure}{0.5\textwidth} 
        \centering
        \includegraphics[width=\linewidth]{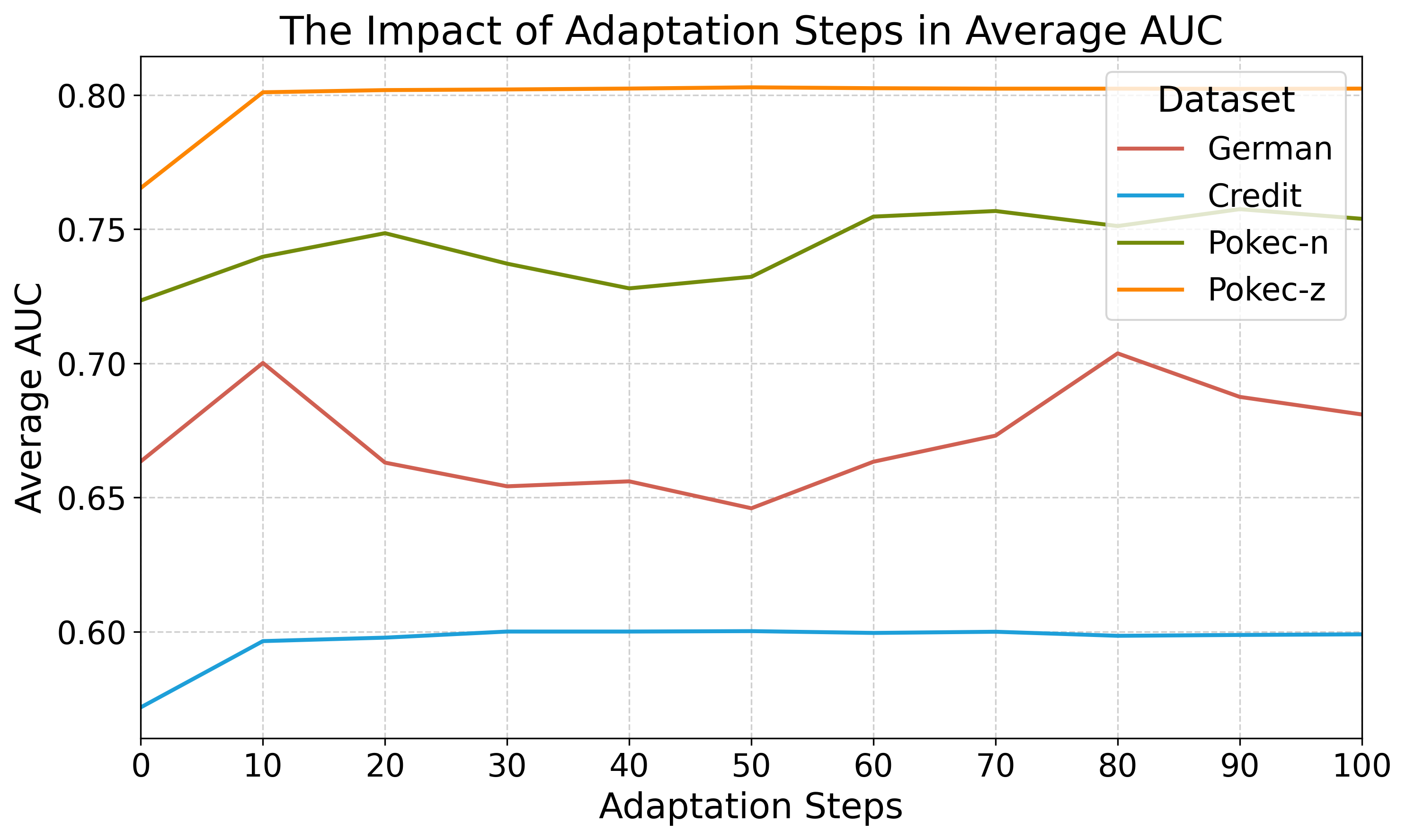} 
    \end{subfigure}\hfill 
    \begin{subfigure}{0.5\textwidth}
        \centering
        \includegraphics[width=\linewidth]{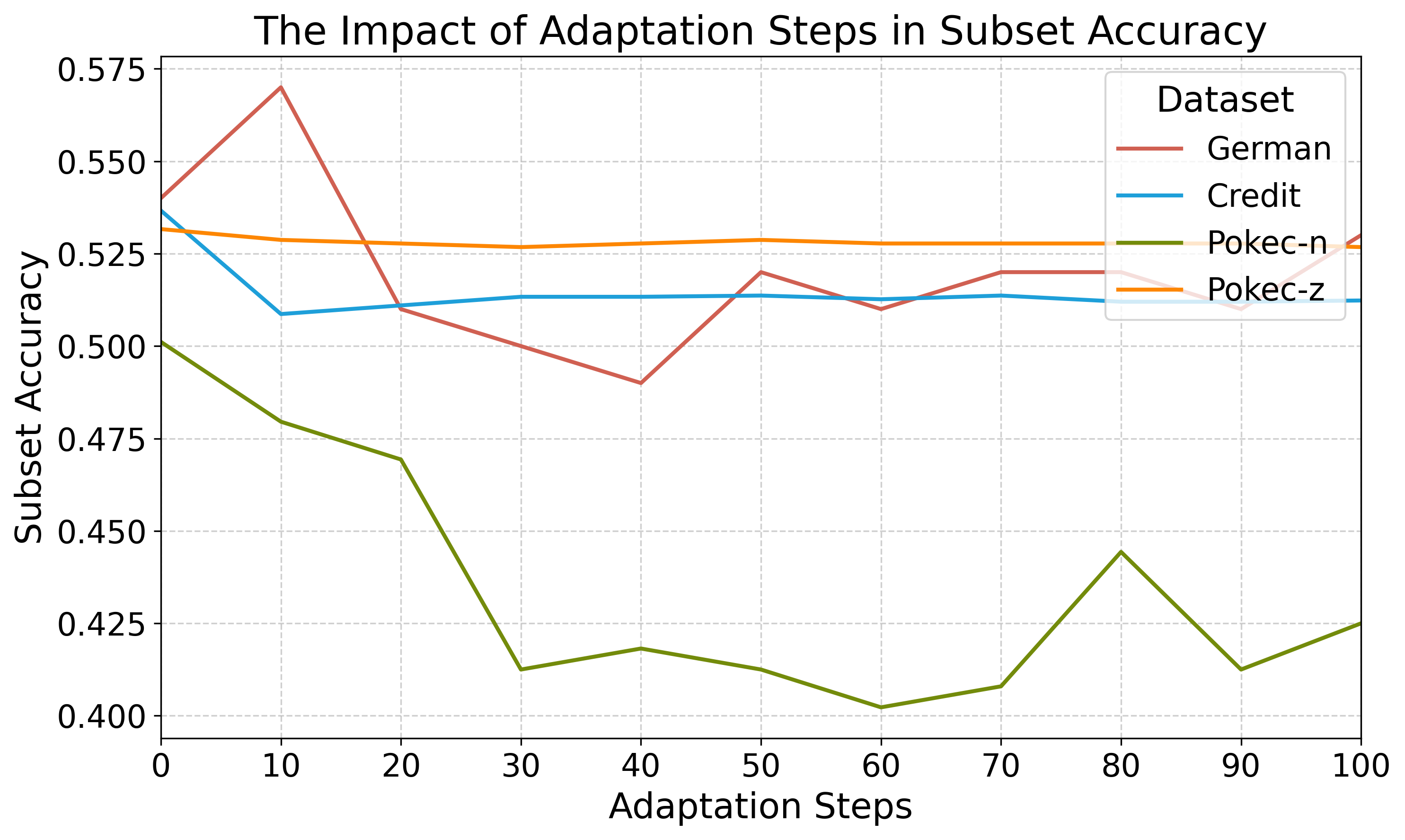} 
    \end{subfigure}
    
    

    
    \caption{Impact of Adaptation Steps}
    \label{fig:adaptation_trends}
\end{figure*}

We assess the impact of adaptation steps in Taipan for G-MSAIAs and present the main results in Figure \ref{fig:adaptation_trends}. Additional details on semantic-based metrics are provided in Figure \ref{fig:adaptation_trends_rest} in Appendix \ref{sec_adpt}. We observe a consistent trend across graphs: confidence and semantic-based metrics improve rapidly within a small number of adaptation steps, followed by oscillation or mild degradation as adaptation continues. This trend indicates that most transferable attack semantic signals are captured during early adaptation, while additional steps provide limited marginal benefits or induce instability. Meanwhile, correctness-based metrics exhibit only slight degradation within small adaptation steps, as reflected by the narrow y-axis range. From a practical standpoint, these observations suggest that a small number of adaptation steps is sufficient to achieve strong attack performance, offering a favorable trade-off between effectiveness and computational overhead. We further observe that the relative performance of different graphs varies substantially across all metrics. Such variations stem from a combination of intrinsic graph characteristics (e.g., node degree distribution, homophily, and data imbalance) and task-level factors, including inter-attribute correlations and learning difficulty. For instance, although Credit exhibits relatively weaker confidence-based performance, it attains comparable SuA due to its limited number of sensitive attributes and simple task correlations. 

\subsubsection{Node Degree and Node Homophily (Q4 and Q5)}
\begin{figure*}[htbp]
    \centering
    \begin{subfigure}{0.5\textwidth}
        \includegraphics[width=\linewidth]{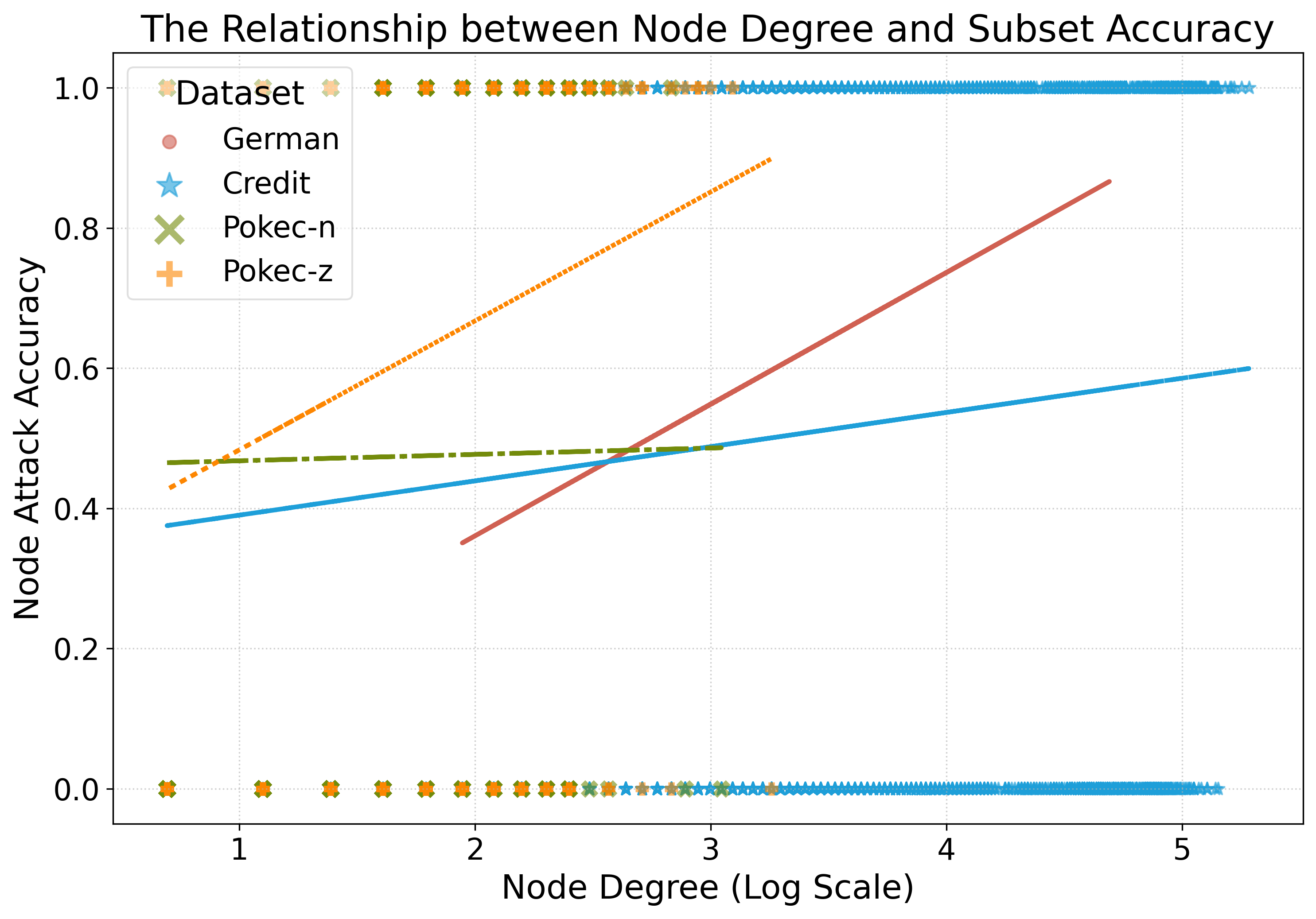}
    \end{subfigure}\hfill
    \begin{subfigure}{0.5\textwidth}
        \includegraphics[width=\linewidth]{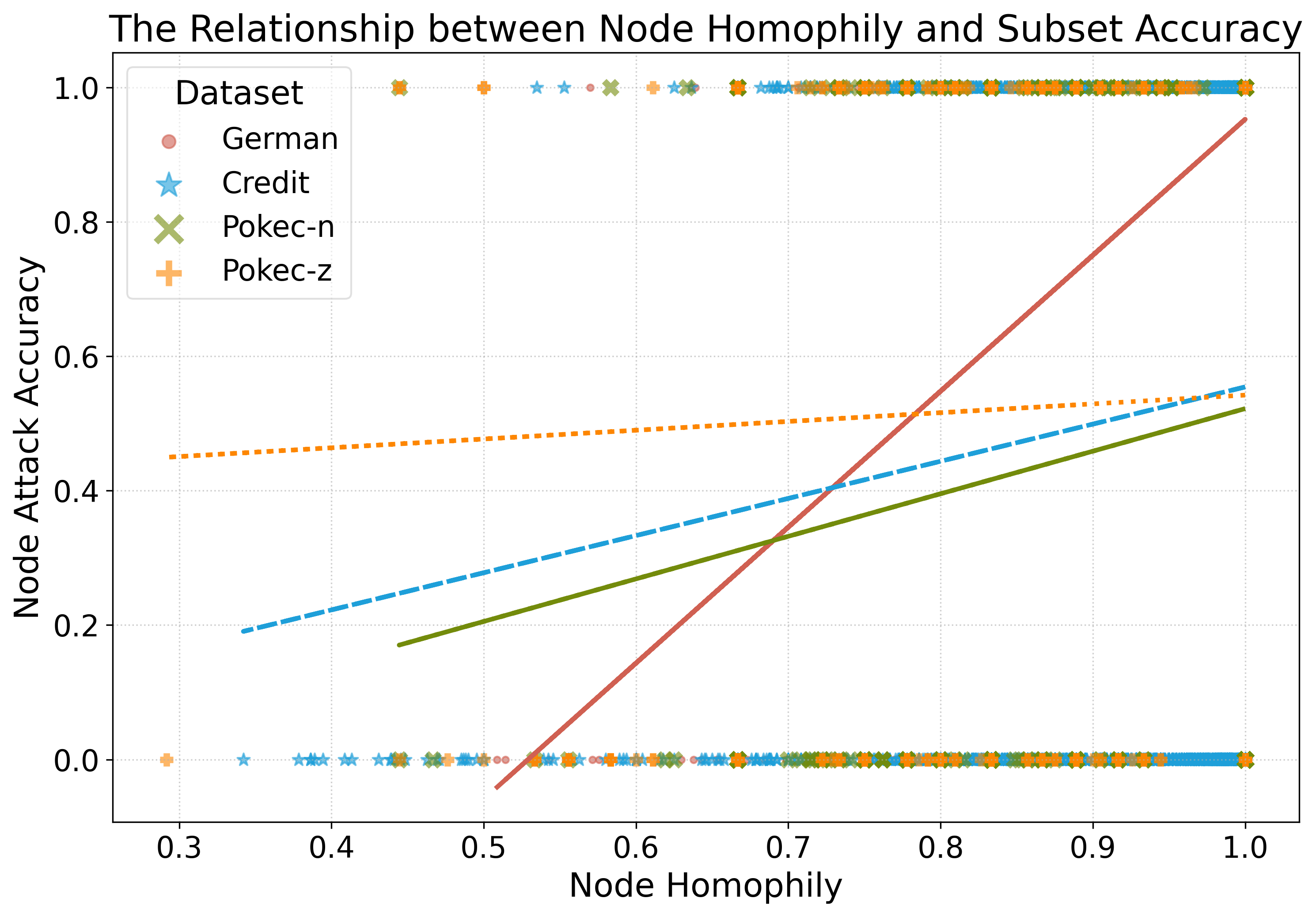}
    \end{subfigure}
    
    
    \caption{Impacts of Node Degree and Node Homophily}
    \label{deg_homo}
\end{figure*}

Building upon our prior analysis of impact factors across node, group, task-levels, we investigate the two node-level factors, namely node degree and homophily, to identify the most vulnerable nodes and simultaneously examine their corresponding impacts on attack performance. We visualize the relationship between node degree, node homophily, and per-node subset accuracy in Figure \ref{deg_homo}. Notably, these scattered points exhibit degree-dependent patterns but not linear trends. Since we evaluate subset accuracy at the node level, this ``all-or-nothing'' metric is inherently discrete, yielding values of either $0$ or $1$ for individual nodes. For brevity, we defer the details of per-node attack accuracy defined over the range $[0,1]$ in Figure \ref{deg_homo_rest} in Appendix \ref{sec_deg_homo}. We observe that node degree exhibits a strong positive correlation with subset accuracy, particularly on German and Pokec-z. Intuitively, nodes with higher degree aggregate information from larger neighborhoods, thereby providing richer and more stable structural signals for G-MSAIAs. 
We also observe a clear positive relationship between node homophily and attack success across all graphs. When neighboring nodes tend to share the same sensitive attributes, the message-passing mechanism reinforces consistent semantic and structural signals, reducing noise and amplifying confidence in attribute inference. 
In conclusion, these observations indicate that nodes embedded in dense and homophilic regions are more vulnerable to G-MSAIAs. 

\subsubsection{Sensitive Label Coverage in Auxiliary Graphs (Q4 and Q5)}
\label{coverage}

Our main threat model assumes that nodes in the auxiliary graph are fully labeled with sensitive attributes. While this assumption follows the standard setting in the AIA literature \cite{AIA_graph_emb, first_aia,least_info_aia_disparate,are_ur_sa_private}, in reality, the adversary may obtain auxiliary nodes without any available labels, posing additional challenges to Taipan. In this subsection, we relax this assumption and evaluate Taipan's performance under partial label coverage. We consider two strategies: GNN information flow (Flow) and pseudo-label augmentation (Pseudo). Flow trains exclusively on labeled nodes per attribute and ignores unlabeled nodes. Due to message passing, these ignored nodes still contribute to representation learning. Notably, under the transductive GNN setting (i.e., all nodes participate in message passing while supervised loss is computed only on labeled nodes), this experiment also serves as a graph size reduction study. Pseudo adopts the aforementioned high-confidence pseudo-labeling mechanism for unlabeled nodes. After pre-training on the labeled subset, it generates pseudo-labels for the unlabeled auxiliary nodes whose confidence exceeds the pre-fixed threshold, augments the labeled set, and retrains. We vary the coverage rate $r \in \{10\%, 25\%, 50\%, 75\%, 100\%\}$. For fair comparison, we leverage the same hyperparameters, please check Appendix \ref{sec_imp_det} for more details.

\begin{figure*}[htbp]
    \centering
    \includegraphics[width=0.99\linewidth]{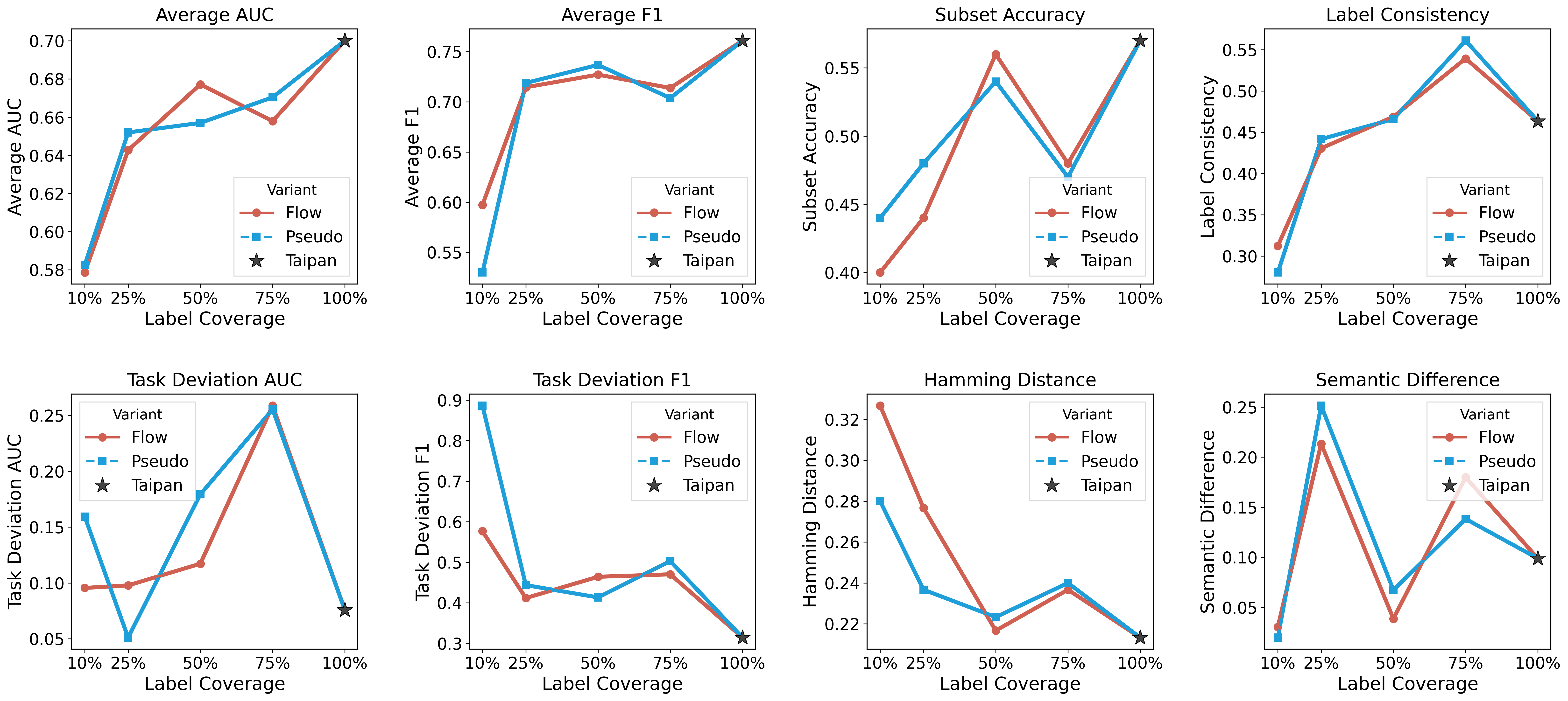}
    \caption{Impacts of Label Coverage in Auxiliary Graphs}
    \label{label_coverage}
\end{figure*}

Figure \ref{label_coverage} shows that Taipan's attack performance degrades under partial label coverage. Even with only $10\%$ of sensitive labels available, Taipan remains competitive with SingP. trained on full labels, which confirms substantial attack signals persist under severe label scarcity. Pseudo achieves asymmetric gains at low coverage and converges to Flow when the coverage increases. Compared to Flow, Pseudo exhibits much larger TDF with $10\%$ coverage while still attaining $44.00\%$ in SuA. 
Under extremely low coverage, pseudo-labels for the most underrepresented attribute may become unreliable, thereby widening the attribute-level performance gap. Additionally, we observe non-monotonic trends across metrics when $r \in \{50\%,75\%\}$. Except for the full-label Taipan, most metrics improve around $r\text{=}50\%$ while LC peaks at $75\%$ coverage. We attribute this phenomenon to implicit regularization. Moderate sensitive label scarcity forces the model to extract feature- and structural-driven dependencies propagated through message passing while extra supervision at $r\text{=}75\%$ induces overfitting to dominated attack tasks. With full coverage, all attack tasks receive sufficient supervision, effectively mitigating task interference. We leave in-depth investigation for future work. 



\subsection{Defense (Q6)}
We now utilize two representative defenses to assess their effectiveness against Taipan, which also serve as case studies for Taipan under privacy-preserving graph publishing.

\subsubsection{Differential Privacy}
\label{sub:defense}
We first adopt the de facto standard privacy-preserving mechanism--differential privacy (DP)--to provide rigorous privacy guarantees. Even though DP is primarily designed to protect individual membership information by injecting noise into either data or training process, it has also been widely employed to defend against AIAs \cite{aia_imputation, feasibility_aia}. Since Taipan operates in a query-free setting, DP-protected GNN training is not applicable. Instead, we resort to DP graph synthesis. However, existing DP graph synthesis methods mainly focus on edge-level DP, which protects the presence or absence of individual edges \cite{privgraph, linkdp}. Enforcing node-level DP is substantially more challenging, as modifying a single node induces high sensitivity due to the removal of all incident edges, particularly for high-degree nodes.

To assess the impact of DP on Taipan, we leverage the state-of-the-art DP graph synthesis method, PrivDPR \cite{privdpr}, which achieves $(\epsilon,\delta)$-node DP via deep PageRank. Following the default configuration of PrivDPR, we fix $\delta$ as $1e\text{-}5$ and vary the privacy budget $\epsilon \in \{0.1, 1, 10\}$ and evaluate Taipan under a stringent  node-DP guarantee. Our objective is not to preserve a favorable privacy–utility trade-off, but rather to stress-test Taipan under rigorous privacy protection. 

From Table \ref{att_perf_defense}, PrivDPR fails to effectively defend against Taipan. On small-scale datasets like German, PrivDPR can mitigate the overall privacy leakage. One plausible explanation is that smaller graphs possess lower structural redundancy, thus, injecting DP noise more easily obfuscates the limited number of critical structural paths, leading to a noticeable drop in attack performance. However, Taipan still recovers partial sensitive information. Under the multi-attribute setting, this injection cannot simultaneously disrupt all attack tasks, informative signals for easier-to-attack sensitive attributes are preserved. On larger and more homophilic graphs, Taipan maintains attack effectiveness across diverse privacy budgets, its performance in certain metrics even closely resembles that of Taipan without any defense. For instance, on Credit, Taipan obtains an AF of $79.50\%$ even when $\epsilon\text{=}0.1$. This slight degradation suggests that Taipan under a rigorous DP protection still reveals substantial privacy leakage. Furthermore, we observe a flat trend between DP guarantee strength and attack mitigation. Since PrivDPR mainly generates node-DP synthetic graphs via edge pruning, these pruned edges might not carry critical attack signals while the remaining ones continue to propagate sufficient across-attribute dependencies through GNN message passing for Taipan to recover sensitive attributes. 


To investigate the underlying mechanism behind this leakage, we resort to a standard $2$-layer GCN without nonlinear activations to obtain node embeddings isolated from confounding effects of nonlinear feature transformations. We then leverage $dCor$ to measure sensitive correlations between embeddings $Z$ and multiple sensitive attributes and report $dCor$ on the original (ODC=$dCor(Z_a, \mathcal{S})$) and on the synthetic target graphs (DDC=$dCor(\tilde{Z}_t, \mathcal{S})$) in Table \ref{att_perf_defense}. Our results show that although strict node-DP largely reduces sensitive correlations, the residual ones remain sufficiently high for G-MSAIAs. This empirical evidence supports the robustness of Taipan: traceable footprints persist in DP-protected graphs and can be exploited by Taipan for multiple sensitive attribute inference. 


\subsubsection{Fair Learning}
Following Dibbo et al. \cite{least_info_aia_disparate}, we leverage fair graph learning to mitigate AIAs. To our knowledge, MAPPING \cite{mapping} is the only existing method that explicitly handles multiple sensitive attributes prior to GNN training. It shows that fairness interventions can effectively remove sensitive correlations between multiple sensitive attributes and model predictions, thereby reducing sensitive information leakage. Accordingly, we employ MAPPING to generate debiased synthetic graphs with reduced privacy leakage.  

We follow the default configuration of MAPPING and fix its fairness-relevant coefficients to $\lambda_2=3.5\text{e-}5$, $\lambda_3=50$, $\lambda_2=1\text{e-}4$ and $r_p=0.72$ across all graphs. Compared to strict node-DP, attack performance on MAPPING-generated graphs is consistently higher for both average single-attribute and joint multi-attribute inferences. 
Correspondingly, sensitive correlations also increase. The only exception occurs for DDC on Pokec-z, where the sensitive correlation under node-DP is slightly higher than that under MAPPING. A plausible explanation is that PrivDPR primarily enforces node-level privacy by aggressively pruning edges, which may overly suppress informative structural signals required for G-MSAIAs.
Moreover, PrivDPR does not explicitly remove sensitive traces from node features, resulting in relatively larger DDC. In contrast, MAPPING jointly mitigates sensitive signals in both node features and graph topology, achieving a better trade-off between fairness and utility. The enhanced synthesis quality further translates into stronger attack performance.

\begin{table}[htbp]
\caption{\centering The Defense against Taipan for G-MSAIAs}
\label{att_perf_defense}
\resizebox{\linewidth}{!}{
\begin{tabular}{c||c|c|cccccccc|cc}
\toprule\bottomrule
\textbf{Defense}     & $\epsilon$             & \textbf{Dataset} & \textbf{AA ($\uparrow$)} & \textbf{AF ($\uparrow$)} & \textbf{TDA ($\downarrow$)} & \textbf{TDF ($\downarrow$)} & \textbf{HD ($\downarrow$)} & \textbf{SuA ($\uparrow$)} & \textbf{SD ($|\downarrow|$)} & \textbf{LC ($\uparrow$)} & \textbf{ODC ($\downarrow$)} & \textbf{DDC ($\downarrow$)} \\ \bottomrule
\multirow{12}{*}{\textbf{PrivDPR}} 
 & \multirow{4}{*}{0.1} 
   & German   & 52.82 & 60.38 & \textbf{3.22}  & 55.60 & 36.33 & 37.00 & \textbf{12.57} & 14.79 & 24.00 & 9.68  \\
 & & Credit   & \textbf{51.31} & 79.50 & 3.25  & 30.81 & \textbf{27.52} & \textbf{47.87} & 13.70 & \textbf{6.38}  & 6.43  & 4.71  \\
 & & Pokec-n  & 65.81 & 52.99 & 18.24 & \textbf{28.75} & 28.98 & 36.14 & -0.90 & \textbf{36.35} & 11.56 & 6.95  \\
 & & Pokec-z  & 74.39 & 64.74 & \textbf{34.19} & 46.95 & \textbf{22.23} & \textbf{46.35} & 2.69  & \textbf{57.27} & 33.09 & 16.93 \\
\cline{2-13}
 & \multirow{4}{*}{1.0} 
   & German   & 58.18 & 64.76 & 9.15  & 54.27 & \textbf{27.00} & \textbf{47.00} & 14.66 & \textbf{23.27} & 24.00 & \textbf{8.76}  \\
 & & Credit   & 50.83 & 80.52 & 1.30  & 29.09 & 27.73 & 46.67 & \textbf{2.02}  & 3.48  & 6.43  & 3.92  \\
 & & Pokec-n  & 65.92 & \textbf{53.14} & 14.96 & 29.61 & 28.75 & 34.77 & \textbf{-0.76} & 35.49 & 11.56 & 7.46  \\
 & & Pokec-z  & \textbf{75.07} & \textbf{67.54} & 34.34 & 39.92 & 23.76 & 42.06 & 3.53  & 56.18 & 33.09 & \textbf{16.55} \\
\cline{2-13}
 & \multirow{4}{*}{10.0} 
   & German   & \textbf{58.82} & \textbf{65.89} & 8.00  & \textbf{41.87} & 30.33 & 37.00 & 14.55 & 19.47 & 24.00 & 10.73 \\
 & & Credit   & 50.20 & \textbf{81.76} & \textbf{0.85}  & \textbf{26.48} & 27.80 & 46.17 & 7.76  & 2.85  & 6.43  & \textbf{3.27}  \\
 & & Pokec-n  & \textbf{65.93} & 52.28 & \textbf{10.72} & 30.59 & \textbf{28.26} & \textbf{37.05} & -1.49 & 34.99 & 11.56 & \textbf{6.78}  \\
 & & Pokec-z  & 74.27 & 66.41 & 34.32 & \textbf{39.72} & 24.57 & 41.19 & \textbf{1.99}  & 54.90 & 33.09 & 17.01 \\
\midrule
\multirow{4}{*}{\textbf{FairGNN}} 
 & \multirow{4}{*}{--} 
   & German   & \textbf{64.94} & \textbf{70.96} & 9.88  & 46.31 & \textbf{26.67} & \textbf{49.00} & 8.71  & \textbf{37.15} & 24.00 & 10.13 \\
 & & Credit   & \textbf{54.16} & 79.78 & 4.26  & 29.83 & 27.92 & 46.93 & 21.48 & \textbf{6.79}  & 6.43  & 5.94  \\
 & & Pokec-n  & \textbf{71.43} & \textbf{55.80} & 22.20 & 46.78 & \textbf{21.93} & \textbf{48.18} & 5.40  & \textbf{49.81} & 11.56 & 11.13 \\
 & & Pokec-z  & \textbf{79.53} & \textbf{72.74} & \textbf{27.50} & \textbf{33.08} & \textbf{20.55} & \textbf{50.83} & 4.56  & \textbf{60.01} & 33.09 & \textbf{14.51}      \\ \toprule\bottomrule
\end{tabular}}
\end{table}



\subsection{Limitations and Future Work}

\noindent\textbf{Extreme Domain Shifts} Recent advances in transfer learning under extreme domain shifts suggest a promising direction for extending Taipan beyond same- or similar-domain graphs.

\noindent\textbf{Scalability \& Heterophily} 
While evaluated on large-scale graphs, scaling to very large graphs remains an open challenge. We also leave heterophilic MSAIAs for future work.

\noindent\textbf{Cross-task Interaction} We identified several key impact factors for G-MSAIAs, deeper analysis of their intricate interplays across multiple attack tasks is left for future work.

\noindent\textbf{Multi-class and Multi-task} Taipan can be naturally generalized to multi-class settings and regression-based tasks. These extensions necessitate a thorough investigation of complex interactions among multiple sensitive groups and individuals.

\section{Conclusion}
\label{sec:conclusion}
In this work, we pioneer the investigation of an inherent yet unexplored vulnerability arising solely from auxiliary graphs. We introduce a new attack paradigm for G-MSAIAs through ``multi-task attack transfer'', and propose Taipan, the first query-free attack framework that extracts shared and task-specific attack knowledge from auxiliary graphs to simultaneously infer multiple sensitive attributes on target graphs. By integrating \emph{Hierarchical Attack Knowledge Routing} with \emph{Prompt-guided Attack Prototype Refinement}, Taipan effectively mitigates distribution shifts between auxiliary and target graphs under realistic settings. Extensive experiments on real-world graph datasets demonstrate that Taipan achieves strong attack performance even under rigorous privacy safeguards. Our findings highlight an urgent need to re-examine privacy risks posed by auxiliary graphs and establish more robust multi-attribute privacy-preserving graph publishing practices.

\section*{Open Science}
In compliance with the CFP Open Science policy, we publicly release our datasets, source code, and evaluation scripts. To promote reproducibility and transparency, these materials are packaged in an anonymous repository \url{https://anonymous.4open.science/r/Taipan-2D34}. These resources are intended to facilitate further research, enable rigorous assessment of multiple sensitive information leakage, and support the development of more effective and robust privacy safeguards for graph publishing and sharing.

\section*{LLM Acknowledgements}
We thank Gemini for assistance with grammar checking and manuscript polishing. All LLM-generated suggestions were manually reviewed, verified, and revised by the authors to ensure accuracy, originality, and academic integrity.


\section*{Ethical Considerations}
This research identifies a critical yet unexplored vulnerability that adversaries can solely exploit auxiliary graphs to conduct multiple sensitive attribute inference attacks. We introduce a new query-free data-only attack paradigm and demonstrate its substantial privacy risks in same or similar-distribution and out-of-distribution settings. To mitigate privacy concerns and prevent potential data misuse, we only utilize auxiliary datasets that have been extensively adopted in prior research. All experiments are conducted on local devices in controlled environments. No personally identifiable information is disclosed, and no real-world systems or servers are compromised.


Massive datasets are increasingly published for research or commercial purposes, or are easily acquired via web-scraping or data breaches. However, existing defense mechanisms are predominantly model-centric and designed to protect against single sensitive attribute leakage, and therefore cannot effectively defend against the proposed attacks. It is imperative for researchers to develop more robust and reliable multi-attribute privacy-preserving graph publishing methods, and for practitioners to re-evaluate current data-sharing practices. Furthermore, we call upon policymakers to strengthen corresponding data protection regulations.

The techniques, models, and open-source codes provided in this paper are intended only for research purposes or for improving data privacy protection in real-world applications. 




\bibliographystyle{IEEEtranS}
\bibliography{Reference}

@misc{bbckey,
  author = {{BBC News}},
  title  = {Millions of hacked LinkedIn IDs advertised 'for sale'},
  year   = {2016},
  month  = may,
  howpublished = {\url{https://www.bbc.com/news/technology-36320322}},
  note   = {Accessed: 2026-04-28}
}

@misc{graphchallengereference,
      title={Improving the Graph Challenge Reference Implementation}, 
      author={Inna Voloshchuk and Hayden Jananthan and Chansup Byun and Jeremy Kepner},
      year={2026},
      eprint={2601.00974},
      archivePrefix={arXiv},
      primaryClass={cs.NI},
      url={https://arxiv.org/abs/2601.00974}, 
}

@inproceedings{oag,
author = {Zhang, Fanjin and Liu, Xiao and Tang, Jie and Dong, Yuxiao and Yao, Peiran and Zhang, Jie and Gu, Xiaotao and Wang, Yan and Shao, Bin and Li, Rui and Wang, Kuansan},
title = {OAG: Toward Linking Large-scale Heterogeneous Entity Graphs},
year = {2019},
isbn = {9781450362016},
publisher = {Association for Computing Machinery},
address = {New York, NY, USA},
url = {https://doi.org/10.1145/3292500.3330785},
doi = {10.1145/3292500.3330785},
booktitle = {Proceedings of the 25th ACM SIGKDD International Conference on Knowledge Discovery \& Data Mining},
pages = {2585–2595},
numpages = {11},
location = {Anchorage, AK, USA},
series = {KDD '19}
}

@article{net_data, title={The Network Data Repository with Interactive Graph Analytics and Visualization}, volume={29}, url={https://ojs.aaai.org/index.php/AAAI/article/view/9277}, DOI={10.1609/aaai.v29i1.9277}, abstractNote={ &lt;p&gt; NetworkRepository (NR) is the first interactive data repository with a web-based platform for visual interactive analytics. Unlike other data repositories (e.g., UCI ML Data Repository, and SNAP), the network data repository (networkrepository.com) allows users to not only download, but to interactively analyze and visualize such data using our web-based interactive graph analytics platform. Users can in real-time analyze, visualize, compare, and explore data along many different dimensions. The aim of NR is to make it easy to discover key insights into the data extremely fast with little effort while also providing a medium for users to share data, visualizations, and insights. Other key factors that differentiate NR from the current data repositories is the number of graph datasets, their size, and variety. While other data repositories are static, they also lack a means for users to collaboratively discuss a particular dataset, corrections, or challenges with using the data for certain applications. In contrast, NR incorporates many social and collaborative aspects that facilitate scientific research, e.g., users can discuss each graph, post observations, and visualizations. &lt;/p&gt; }, number={1}, journal={Proceedings of the AAAI Conference on Artificial Intelligence}, author={Rossi, Ryan and Ahmed, Nesreen}, year={2015}, month={Mar.} }

@misc{snapnets,
  author       = {Jure Leskovec and Andrej Krevl},
  title        = {{SNAP Datasets}: {Stanford} Large Network Dataset Collection},
  howpublished = {\url{http://snap.stanford.edu/data}},
  month        = jun,
  year         = 2014
}

@article{ema,
title={Exponential Moving Average of Weights in Deep Learning: Dynamics and Benefits},
author={Daniel Morales-Brotons and Thijs Vogels and Hadrien Hendrikx},
journal={Transactions on Machine Learning Research},
issn={2835-8856},
year={2024},
url={https://openreview.net/forum?id=2M9CUnYnBA},
note={}
}

@inproceedings{gdpr,
   title={The General Data Protection Regulation: Requirements, Architectures, and Constraints},
   url={http://dx.doi.org/10.1109/RE.2019.00036},
   DOI={10.1109/re.2019.00036},
   booktitle={2019 IEEE 27th International Requirements Engineering Conference (RE)},
   publisher={IEEE},
   author={Hjerppe, Kalle and Ruohonen, Jukka and Leppanen, Ville},
   year={2019},
   month=sep }

@inproceedings{say_no,
author = {Dai, Enyan and Wang, Suhang},
title = {Say No to the Discrimination: Learning Fair Graph Neural Networks with Limited Sensitive Attribute Information},
year = {2021},
isbn = {9781450382977},
publisher = {Association for Computing Machinery},
address = {New York, NY, USA},
url = {https://doi.org/10.1145/3437963.3441752},
doi = {10.1145/3437963.3441752},
booktitle = {Proceedings of the 14th ACM International Conference on Web Search and Data Mining},
pages = {680–688},
numpages = {9},
location = {Virtual Event, Israel},
series = {WSDM '21}
}

@article{guda_unsupervised,
  title={Unsupervised Prompting for Graph Neural Networks},
  author={Baghershahi, Peyman and Medya, Sourav},
  journal={arXiv preprint arXiv:2505.16903},
  year={2025}
}

@inproceedings{guda_smooth,
author = {Chen, Wei and Ye, Guo and Wang, Yakun and Zhang, Zhao and Zhang, Libang and Wang, Daixin and Zhang, Zhiqiang and Zhuang, Fuzhen},
title = {Smoothness really matters: a simple yet effective approach for unsupervised graph domain adaptation},
year = {2025},
isbn = {978-1-57735-897-8},
publisher = {AAAI Press},
url = {https://doi.org/10.1609/aaai.v39i15.33743},
doi = {10.1609/aaai.v39i15.33743},
booktitle = {Proceedings of the Thirty-Ninth AAAI Conference on Artificial Intelligence and Thirty-Seventh Conference on Innovative Applications of Artificial Intelligence and Fifteenth Symposium on Educational Advances in Artificial Intelligence},
articleno = {1770},
numpages = {9},
series = {AAAI'25/IAAI'25/EAAI'25}
}

@inproceedings{gmtl_replayandforgetfree,
title={Replay-and-Forget-Free Graph Class-Incremental Learning: A Task Profiling and Prompting Approach},
author={Chaoxi Niu and Guansong Pang and Ling Chen and Bing Liu},
booktitle={The Thirty-eighth Annual Conference on Neural Information Processing Systems},
year={2024},
url={https://openreview.net/forum?id=FXdMgfCDer}
}

@article{mtl_prompt_rec,
author = {Bai, Ting and Huang, Le and Yu, Yue and Yang, Cheng and Hou, Cheng and Zhao, Zhe and Shi, Chuan},
title = {Efficient Multi-task Prompt Tuning for Recommendation},
year = {2025},
issue_date = {July 2025},
publisher = {Association for Computing Machinery},
address = {New York, NY, USA},
volume = {43},
number = {4},
issn = {1046-8188},
url = {https://doi.org/10.1145/3736403},
doi = {10.1145/3736403},
journal = {ACM Trans. Inf. Syst.},
month = jul,
articleno = {104},
numpages = {21}
}

@inproceedings{dagprompt,
author = {Chen, Qin and Wang, Liang and Zheng, Bo and Song, Guojie},
title = {DAGPrompT: Pushing the Limits of Graph Prompting with a Distribution-aware Graph Prompt Tuning Approach},
year = {2025},
isbn = {9798400712746},
publisher = {Association for Computing Machinery},
address = {New York, NY, USA},
url = {https://doi.org/10.1145/3696410.3714917},
doi = {10.1145/3696410.3714917},
booktitle = {Proceedings of the ACM on Web Conference 2025},
pages = {4346–4358},
numpages = {13},
location = {Sydney NSW, Australia},
series = {WWW '25}
}

@article{elu,
  title={Fast and accurate deep network learning by exponential linear units (elus)},
  author={Clevert, Djork-Arn{\'e} and Unterthiner, Thomas and Hochreiter, Sepp},
  journal={arXiv preprint arXiv:1511.07289},
  volume={4},
  number={5},
  pages={11},
  year={2015}
}

@inproceedings{mmoe,
author = {Ma, Jiaqi and Zhao, Zhe and Yi, Xinyang and Chen, Jilin and Hong, Lichan and Chi, Ed H.},
title = {Modeling Task Relationships in Multi-task Learning with Multi-gate Mixture-of-Experts},
year = {2018},
isbn = {9781450355520},
publisher = {Association for Computing Machinery},
address = {New York, NY, USA},
url = {https://doi-org.pitt.idm.oclc.org/10.1145/3219819.3220007},
doi = {10.1145/3219819.3220007},
booktitle = {Proceedings of the 24th ACM SIGKDD International Conference on Knowledge Discovery \& Data Mining},
pages = {1930–1939},
numpages = {10},
location = {London, United Kingdom},
series = {KDD '18}
}

@misc{mtl_survey,
      title={Unleashing the Power of Multi-Task Learning: A Comprehensive Survey Spanning Traditional, Deep, and Pretrained Foundation Model Eras}, 
      author={Jun Yu and Yutong Dai and Xiaokang Liu and Jin Huang and Yishan Shen and Ke Zhang and Rong Zhou and Eashan Adhikarla and Wenxuan Ye and Yixin Liu and Zhaoming Kong and Kai Zhang and Yilong Yin and Vinod Namboodiri and Brian D. Davison and Jason H. Moore and Yong Chen},
      year={2024},
      eprint={2404.18961},
      archivePrefix={arXiv},
      primaryClass={cs.LG},
      url={https://arxiv.org/abs/2404.18961}, 
}

@INPROCEEDINGS{hierar_mtl,
  author={Liu, Yajing and Lu, Yuning and Liu, Hao and An, Yaozu and Xu, Zhuoran and Yao, Zhuokun and Zhang, Baofeng and Xiong, Zhiwei and Gui, Chenguang},
  booktitle={2023 IEEE/CVF Conference on Computer Vision and Pattern Recognition (CVPR)}, 
  title={Hierarchical Prompt Learning for Multi-Task Learning}, 
  year={2023},
  volume={},
  number={},
  pages={10888-10898},
  doi={10.1109/CVPR52729.2023.01048}}

@article{Day_Edelsbrunner_1984, title={Efficient algorithms for agglomerative hierarchical clustering methods}, volume={1}, ISSN={1432-1343}, DOI={10.1007/BF01890115}, abstractNote={Whenevern objects are characterized by a matrix of pairwise dissimilarities, they may be clustered by any of a number of sequential, agglomerative, hierarchical, nonoverlapping (SAHN) clustering methods. These SAHN clustering methods are defined by a paradigmatic algorithm that usually requires 0(n3) time, in the worst case, to cluster the objects. An improved algorithm (Anderberg 1973), while still requiring 0(n3) worst-case time, can reasonably be expected to exhibit 0(n2) expected behavior. By contrast, we describe a SAHN clustering algorithm that requires 0(n2 logn) time in the worst case. When SAHN clustering methods exhibit reasonable space distortion properties, further improvements are possible. We adapt a SAHN clustering algorithm, based on the efficient construction of nearest neighbor chains, to obtain a reasonably general SAHN clustering algorithm that requires in the worst case 0(n2) time and space.}, number={1}, journal={Journal of Classification}, author={Day, William H. E. and Edelsbrunner, Herbert}, year={1984}, month=dec, pages={7–24} }

@inproceedings{linkdp,
  title     = {Secure Deep Graph Generation with Link Differential Privacy},
  author    = {Yang, Carl and Wang, Haonan and Zhang, Ke and Chen, Liang and Sun, Lichao},
  booktitle = {Proceedings of the Thirtieth International Joint Conference on
               Artificial Intelligence, {IJCAI-21}},
  publisher = {International Joint Conferences on Artificial Intelligence Organization},
  editor    = {Zhi-Hua Zhou},
  pages     = {3271--3278},
  year      = {2021},
  month     = {8},
  note      = {Main Track},
  doi       = {10.24963/ijcai.2021/450},
  url       = {https://doi.org/10.24963/ijcai.2021/450},
}

@inproceedings {privgraph,
author = {Quan Yuan and Zhikun Zhang and Linkang Du and Min Chen and Peng Cheng and Mingyang Sun},
title = {{PrivGraph}: Differentially Private Graph Data Publication by Exploiting Community Information},
booktitle = {32nd USENIX Security Symposium (USENIX Security 23)},
year = {2023},
isbn = {978-1-939133-37-3},
address = {Anaheim, CA},
pages = {3241--3258},
url = {https://www.usenix.org/conference/usenixsecurity23/presentation/yuan-quan},
publisher = {USENIX Association},
month = aug
}

@inproceedings{privdpr,
author = {Zhang, Sen and Hu, Haibo and Ye, Qingqing and Xu, Jianliang},
title = {PrivDPR: Synthetic Graph Publishing with Deep PageRank under Differential Privacy},
year = {2025},
isbn = {9798400712456},
publisher = {Association for Computing Machinery},
address = {New York, NY, USA},
url = {https://doi.org/10.1145/3690624.3709334},
doi = {10.1145/3690624.3709334},
booktitle = {Proceedings of the 31st ACM SIGKDD Conference on Knowledge Discovery and Data Mining V.1},
pages = {1936–1947},
numpages = {12},
location = {Toronto ON, Canada},
series = {KDD '25}
}

@inproceedings {inf_gnn,
author = {Zhikun Zhang and Min Chen and Michael Backes and Yun Shen and Yang Zhang},
title = {Inference Attacks Against Graph Neural Networks},
booktitle = {31st USENIX Security Symposium (USENIX Security 22)},
year = {2022},
isbn = {978-1-939133-31-1},
address = {Boston, MA},
pages = {4543--4560},
url = {https://www.usenix.org/conference/usenixsecurity22/presentation/zhang-zhikun},
publisher = {USENIX Association},
month = aug
}

@INPROCEEDINGS{mia_gnn_shadow,
  author={Olatunji, Iyiola E. and Nejdl, Wolfgang and Khosla, Megha},
  booktitle={2021 Third IEEE International Conference on Trust, Privacy and Security in Intelligent Systems and Applications (TPS-ISA)}, 
  title={Membership Inference Attack on Graph Neural Networks}, 
  year={2021},
  volume={},
  number={},
  pages={11-20},
  doi={10.1109/TPSISA52974.2021.00002}}

@inproceedings {shadow_data,
author = {Boyang Zhang and Zheng Li and Ziqing Yang and Xinlei He and Michael Backes and Mario Fritz and Yang Zhang},
title = {{SecurityNet}: Assessing Machine Learning Vulnerabilities on Public Models},
booktitle = {33rd USENIX Security Symposium (USENIX Security 24)},
year = {2024},
isbn = {978-1-939133-44-1},
address = {Philadelphia, PA},
pages = {3873--3890},
url = {https://www.usenix.org/conference/usenixsecurity24/presentation/zhang-boyang},
publisher = {USENIX Association},
month = aug
}

@article{dcor,
  title={Measuring and testing dependence by correlation of distances},
  author={G'abor J. Sz'ekely and Maria L. Rizzo and Nail K. Bakirov},
  journal={Annals of Statistics},
  year={2007},
  volume={35},
  pages={2769-2794},
  url={https://api.semanticscholar.org/CorpusID:5661488}
}

@article{mapping, title={MAPPING: debiasing graph neural networks for fair node classification with limited sensitive information leakage}, volume={27}, ISSN={1573-1413}, DOI={10.1007/s11280-024-01312-0}, abstractNote={Despite remarkable success in diverse web-based applications, Graph Neural Networks (GNNs) inherit and further exacerbate historical discrimination and social stereotypes, which critically hinder their deployments in high-stake domains such as online clinical diagnosis, financial crediting, etc. However, existing research in fair graph learning typically favors pairwise constraints to achieve fairness but fails to cast off dimensional limitations and generalize them into multiple sensitive attributes. Besides, most studies focus on in-processing techniques to enforce and calibrate fairness, constructing a model-agnostic debiasing GNN framework at the pre-processing stage to prevent downstream misuses and improve training reliability is still largely under-explored. Furthermore, previous work tends to enhance either fairness or privacy individually but few probes into how fairness issues trigger privacy concerns and whether such concerns can be alleviated with fairness intervention. In this paper, we propose a novel model-agnostic debiasing framework named MAPPING (Masking And Pruning and Message-Passing trainING) for fair node classification, in which we adopt the distance covariance (dCov)-based fairness constraints to simultaneously reduce feature and topology biases under multiple sensitive memberships, and combine them with adversarial debiasing to confine the risks of sensitive attribute inference. Experiments on real-world datasets with different GNN variants demonstrate the effectiveness and flexibility of MAPPING. Our results show that MAPPING can achieve better trade-offs between utility and fairness, and mitigate privacy risks of sensitive information leakage. This work paves the way for a new direction in trustworthy GNNs by addressing fairness and privacy concerns simultaneously, rather than achieving fairness at the expense of privacy.}, number={6}, journal={World Wide Web}, author={Song, Ying and Palanisamy, Balaji}, year={2024}, month=nov, pages={74} }

@article{fair_aia,
  title={Do Fairness Interventions Come at the Cost of Privacy: Evaluations for Binary Classifiers},
  author={Huan Tian and Guangsheng Zhang and Bo Liu and Tianqing Zhu and Ming Ding and Wanlei Zhou},
  journal={IEEE Transactions on Dependable and Secure Computing},
  year={2025},
  volume={22},
  pages={4654-4670},
  url={https://api.semanticscholar.org/CorpusID:276903745}
}

@inproceedings{gin,
title={How Powerful are Graph Neural Networks?},
author={Keyulu Xu and Weihua Hu and Jure Leskovec and Stefanie Jegelka},
booktitle={International Conference on Learning Representations},
year={2019},
url={https://openreview.net/forum?id=ryGs6iA5Km},
}

@inproceedings{sage,
author = {Hamilton, William L. and Ying, Rex and Leskovec, Jure},
title = {Inductive representation learning on large graphs},
year = {2017},
isbn = {9781510860964},
publisher = {Curran Associates Inc.},
address = {Red Hook, NY, USA},
booktitle = {Proceedings of the 31st International Conference on Neural Information Processing Systems},
pages = {1025–1035},
numpages = {11},
location = {Long Beach, California, USA},
series = {NIPS'17}
}

@inproceedings{gcn,
title={Semi-Supervised Classification with Graph Convolutional Networks},
author={Thomas N. Kipf and Max Welling},
booktitle={International Conference on Learning Representations},
year={2017},
url={https://openreview.net/forum?id=SJU4ayYgl}
}

@inproceedings{pokec,
  title={Data analysis in public social networks},
  author={Takac, Lubos and Zabovsky, Michal},
  booktitle={International scientific conference and international workshop present day trends of innovations},
  volume={1},
  number={6},
  year={2012}
}

@InProceedings{NIFTY,
  title = 	 {Towards a unified framework for fair and stable graph representation learning},
  author =       {Agarwal, Chirag and Lakkaraju, Himabindu and Zitnik, Marinka},
  booktitle = 	 {Proceedings of the Thirty-Seventh Conference on Uncertainty in Artificial Intelligence},
  pages = 	 {2114--2124},
  year = 	 {2021},
  editor = 	 {de Campos, Cassio and Maathuis, Marloes H.},
  volume = 	 {161},
  series = 	 {Proceedings of Machine Learning Research},
  month = 	 {27--30 Jul},
  publisher =    {PMLR},
  url = 	 {https://proceedings.mlr.press/v161/agarwal21b.html}
}

@inproceedings{allinone,
  title={All in one: Multi-task prompting for graph neural networks},
  author={Sun, Xiangguo and Cheng, Hong and Li, Jia and Liu, Bo and Guan, Jihong},
  booktitle={Proceedings of the 29th ACM SIGKDD Conference on Knowledge Discovery and Data Mining},
  pages={2120--2131},
  year={2023}
}

@inproceedings{gppt,
author = {Sun, Mingchen and Zhou, Kaixiong and He, Xin and Wang, Ying and Wang, Xin},
title = {GPPT: Graph Pre-training and Prompt Tuning to Generalize Graph Neural Networks},
year = {2022},
isbn = {9781450393850},
publisher = {Association for Computing Machinery},
address = {New York, NY, USA},
url = {https://doi.org/10.1145/3534678.3539249},
doi = {10.1145/3534678.3539249},
booktitle = {Proceedings of the 28th ACM SIGKDD Conference on Knowledge Discovery and Data Mining},
pages = {1717–1727},
numpages = {11},
location = {Washington DC, USA},
series = {KDD '22}
}

@inproceedings{multigprompt,
author = {Yu, Xingtong and Zhou, Chang and Fang, Yuan and Zhang, Xinming},
title = {MultiGPrompt for Multi-Task Pre-Training and Prompting on Graphs},
year = {2024},
isbn = {9798400701719},
publisher = {Association for Computing Machinery},
address = {New York, NY, USA},
url = {https://doi.org/10.1145/3589334.3645423},
doi = {10.1145/3589334.3645423},
booktitle = {Proceedings of the ACM Web Conference 2024},
pages = {515–526},
numpages = {12},
location = {Singapore, Singapore},
series = {WWW '24}
}

@inproceedings {disparate_priv_vulnerability,
author = {Kabir, Ehsanul and Craig, Lucas and Mehnaz, Shagufta},
title = {Disparate Privacy Vulnerability: Targeted Attribute Inference Attacks and Defenses},
booktitle = {34th USENIX Security Symposium (USENIX Security 25)},
year = {2025},
url = {https://www.usenix.org/conference/usenixsecurity25/presentation/kabir},
publisher = {USENIX Association},
month = aug
}

@INPROCEEDINGS{least_info_aia_disparate,
  author={Dibbo, Sayanton V. and Chung, Dae Lim and Mehnaz, Shagufta},
  booktitle={2023 IEEE Conference on Secure and Trustworthy Machine Learning (SaTML)}, 
  title={Model Inversion Attack with Least Information and an In-depth Analysis of its Disparate Vulnerability}, 
  year={2023},
  volume={},
  number={},
  pages={119-135},
  doi={10.1109/SaTML54575.2023.00017}}

@INPROCEEDINGS{feasibility_aia,
  author={Zhao, Benjamin Zi Hao and Agrawal, Aviral and Coburn, Catisha and Asghar, Hassan Jameel and Bhaskar, Raghav and Kaafar, Mohamed Ali and Webb, Darren and Dickinson, Peter},
  booktitle={2021 IEEE European Symposium on Security and Privacy (EuroS\&P)}, 
  title={On the (In)Feasibility of Attribute Inference Attacks on Machine Learning Models}, 
  year={2021},
  volume={},
  number={},
  pages={232-251},
  doi={10.1109/EuroSP51992.2021.00025}}

@inproceedings{aia_imputation,
author = {Jayaraman, Bargav and Evans, David},
title = {Are Attribute Inference Attacks Just Imputation?},
year = {2022},
isbn = {9781450394505},
publisher = {Association for Computing Machinery},
address = {New York, NY, USA},
url = {https://doi.org/10.1145/3548606.3560663},
doi = {10.1145/3548606.3560663},
booktitle = {Proceedings of the 2022 ACM SIGSAC Conference on Computer and Communications Security},
pages = {1569–1582},
numpages = {14},
location = {Los Angeles, CA, USA},
series = {CCS '22}
}

@inproceedings{first_aia,
author = {Fredrikson, Matt and Jha, Somesh and Ristenpart, Thomas},
title = {Model Inversion Attacks that Exploit Confidence Information and Basic Countermeasures},
year = {2015},
isbn = {9781450338325},
publisher = {Association for Computing Machinery},
address = {New York, NY, USA},
url = {https://doi.org/10.1145/2810103.2813677},
doi = {10.1145/2810103.2813677},
booktitle = {Proceedings of the 22nd ACM SIGSAC Conference on Computer and Communications Security},
pages = {1322–1333},
numpages = {12},
location = {Denver, Colorado, USA},
series = {CCS '15}
}

@inproceedings {are_ur_sa_private,
author = {Shagufta Mehnaz and Sayanton V. Dibbo and Ehsanul Kabir and Ninghui Li and Elisa Bertino},
title = {Are Your Sensitive Attributes Private? Novel Model Inversion Attribute Inference Attacks on Classification Models},
booktitle = {31st USENIX Security Symposium (USENIX Security 22)},
year = {2022},
isbn = {978-1-939133-31-1},
address = {Boston, MA},
pages = {4579--4596},
url = {https://www.usenix.org/conference/usenixsecurity22/presentation/mehnaz},
publisher = {USENIX Association},
month = aug
}

@InProceedings{fair_audit_AIA,
author="Aalmoes, Jan
and Duddu, Vasisht
and Boutet, Antoine",
editor="Barhamgi, Mahmoud
and Wang, Hua
and Wang, Xin",
title="On the Alignment of Group Fairness with Attribute Privacy",
booktitle="Web Information Systems Engineering -- WISE 2024",
year="2025",
publisher="Springer Nature Singapore",
address="Singapore",
pages="333--348",
isbn="978-981-96-0567-5"
}

@inproceedings{AIA_graph_emb,
author = {Duddu, Vasisht and Boutet, Antoine and Shejwalkar, Virat},
title = {Quantifying Privacy Leakage in Graph Embedding},
year = {2021},
isbn = {9781450388405},
publisher = {Association for Computing Machinery},
address = {New York, NY, USA},
url = {https://doi.org/10.1145/3448891.3448939},
doi = {10.1145/3448891.3448939},
booktitle = {MobiQuitous 2020 - 17th EAI International Conference on Mobile and Ubiquitous Systems: Computing, Networking and Services},
pages = {76–85},
numpages = {10},
location = {Darmstadt, Germany},
series = {MobiQuitous '20}
}

@misc{AIA_gnn,
      title={Does Black-box Attribute Inference Attacks on Graph Neural Networks Constitute Privacy Risk?}, 
      author={Iyiola E. Olatunji and Anmar Hizber and Oliver Sihlovec and Megha Khosla},
      year={2023},
      eprint={2306.00578},
      archivePrefix={arXiv},
      primaryClass={cs.LG},
      url={https://arxiv.org/abs/2306.00578}, 
}

@inproceedings {aia_social_friend_behavior,
author = {Neil Zhenqiang Gong and Bin Liu},
title = {You Are Who You Know and How You Behave: Attribute Inference Attacks via Users{\textquoteright} Social Friends and Behaviors},
booktitle = {25th USENIX Security Symposium (USENIX Security 16)},
year = {2016},
isbn = {978-1-931971-32-4},
address = {Austin, TX},
pages = {979--995},
url = {https://www.usenix.org/conference/usenixsecurity16/technical-sessions/presentation/gong},
publisher = {USENIX Association},
month = aug
}

@ARTICLE{social_net_survey,
  author={Piao, Yangheran and Ye, Kai and Cui, Xiaohui},
  journal={IEEE Access}, 
  title={Privacy Inference Attack Against Users in Online Social Networks: A Literature Review}, 
  year={2021},
  volume={9},
  number={},
  pages={40417-40431},
  doi={10.1109/ACCESS.2021.3064208}}

@INPROCEEDINGS{ego_social_net,
  author={Simo, Hervais and Kreutzer, Michael},
  booktitle={2024 IEEE 23rd International Conference on Trust, Security and Privacy in Computing and Communications (TrustCom)}, 
  title={WAPITI - A Weighted Bayesian Method for Private Information Inference on Social Ego Networks}, 
  year={2024},
  volume={},
  number={},
  pages={2269-2280},
  doi={10.1109/TrustCom63139.2024.00313}}

@INPROCEEDINGS{multi_corr_social_net,
  author={Yang, Yitong and Lin, Qixiao and Mao, Jian and Liu, Lipei},
  booktitle={2021 IEEE SmartWorld, Ubiquitous Intelligence \& Computing, Advanced \& Trusted Computing, Scalable Computing \& Communications, Internet of People and Smart City Innovation (SmartWorld/SCALCOM/UIC/ATC/IOP/SCI)}, 
  title={An improved social attribute inference scheme based on multi-attribute correlation}, 
  year={2021},
  volume={},
  number={},
  pages={370-377},
  doi={10.1109/SWC50871.2021.00057}}

@ARTICLE{aia_link_attr,
  author={Mao, Jian and Tian, Wenqian and Yang, Yitong and Liu, Jianwei},
  journal={IEEE Access}, 
  title={An Efficient Social Attribute Inference Scheme Based on Social Links and Attribute Relevance}, 
  year={2019},
  volume={7},
  number={},
  pages={153074-153085},
  doi={10.1109/ACCESS.2019.2946179}}

@ARTICLE{mul_sens_emb,
  author={Luo, Xucheng and Xie, Minrui and Zhang, Kunpeng and Zhou, Fan and Zhong, Ting},
  journal={IEEE Access}, 
  title={DeepAttr: Inferring Demographic Attributes via Social Network Embedding}, 
  year={2019},
  volume={7},
  number={},
  pages={130270-130282},
  doi={10.1109/ACCESS.2019.2940752}}

@article{healthcare_survey,
  title={A systematic review of graph neural network in healthcare-based applications: Recent advances, trends, and future directions},
  author={Paul, Showmick Guha and Saha, Arpa and Hasan, Md Zahid and Noori, Sheak Rashed Haider and Moustafa, Ahmed},
  journal={IEEE access},
  volume={12},
  pages={15145--15170},
  year={2024},
  publisher={IEEE}
}

\appendix
\setcounter{section}{0}

\subsection{Data Statistics}
\label{sec_data_stat}
German and Credit are financial credit networks constructed by Agarwal et al. \cite{NIFTY}. In German, nodes represent bank clients and edges are constructed if two clients exhibit high similarity in their credit accounts. The task is to classify credit risk as high or low given three sensitive attributes—marriage status, gender, and age. In Credit, nodes denote credit card users and edges are formed based on similarity in payment behaviors. The task is to predict whether an applicant will default on an upcoming payment considering two sensitive attributes—marriage status and age. Pokec-n and Pokec-z \cite{pokec} are derived from a popular social network application in two different provinces of Slovakia, respectively. Nodes are application users and edges represent  friendship relations. The task is to predict the working field of the users with three sensitive attributes—region, gender and age. 

\begin{table}[!htbp]
\caption{\centering Detailed Graph Statistics. Imb. Ratio: \#majority/\#minority, where $1$ indicates perfect balance.}
\label{data_stat}
\resizebox{\linewidth}{!}{
\begin{tabular}{c|ccccc|ccc}
\toprule\bottomrule
\textbf{Dataset} & \textbf{\#Nodes} & \textbf{\#Edges} & \textbf{\#Attr.} & \textbf{\#SAs} & \textbf{Label}  & \textbf{Avg. Deg.} & \textbf{Homo.(\%)} &\textbf{Imb. Ratio Range}\\ \bottomrule
\textbf{German}  & 1,000            & 22,242           & 27                    & 3                     & Credit Status   & 44.48                & 60.86    &[1.15-3.68]               \\
\textbf{Credit}  & 30,000           & 1,436,858        & 13                    & 2                     & Payment Default & 95.79                & 72.55      &[1.20-10.17]             \\
\textbf{Pokec-n}  & 66,569           & 583,616          & 266                   & 3                     & Working Field   & 16.53                & 79.79   &[1.05-2.65]                \\
\textbf{Pokec-z}  & 67,796           & 617,958          & 277                   & 3                     & Working Field   & 19.23                & 77.05        &[1.02-2.77]           \\ \toprule\bottomrule
\end{tabular}}
\end{table}

\subsection{Attack Task Correlation and Hierarchy}
\begin{figure}[htbp]
    \centering
    \includegraphics[width=0.99\linewidth]{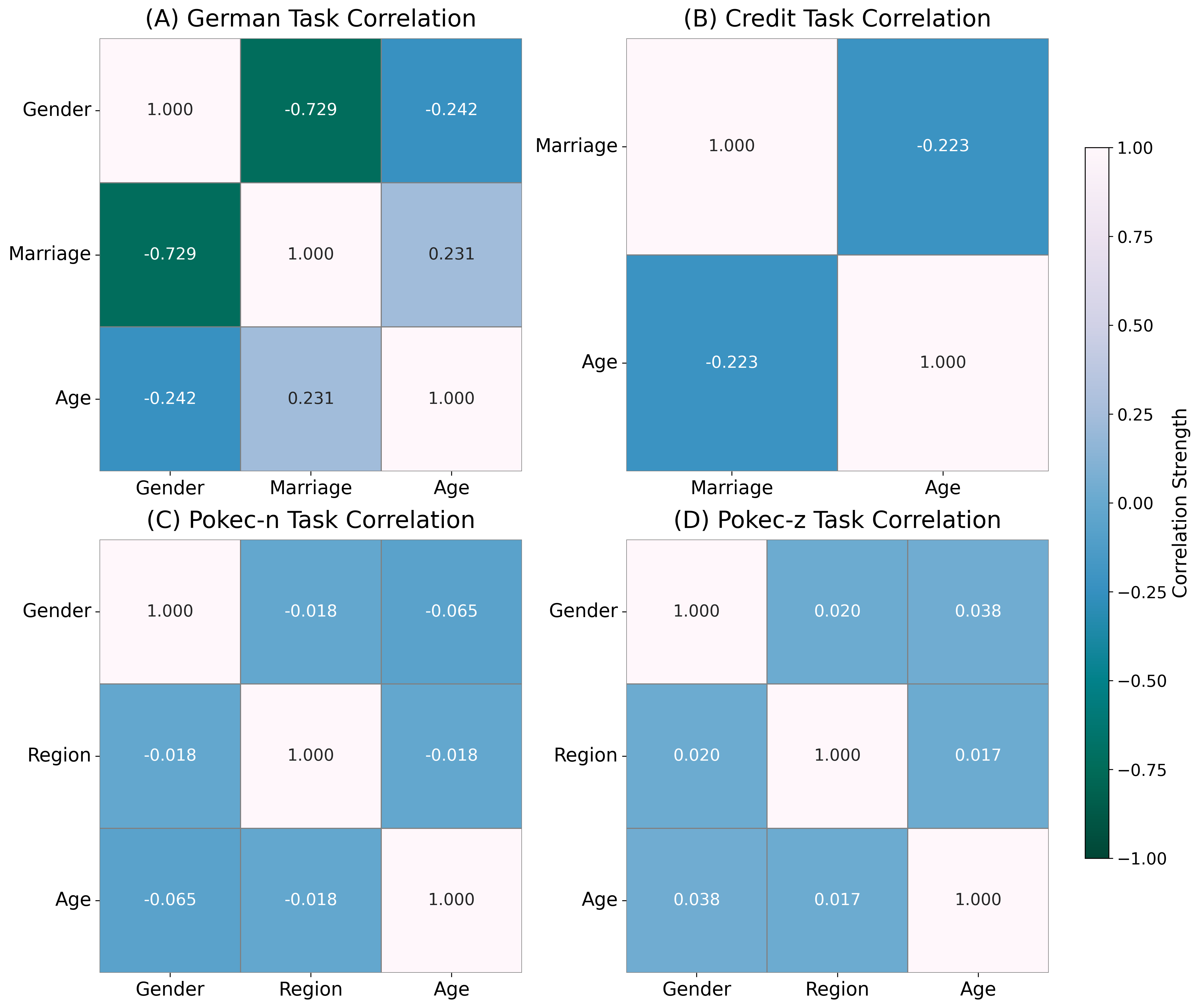}
    \caption{\centering The Attack Task Correlations among Diverse Graph Datasets}
    \label{att_corr}
\end{figure}

\begin{figure}[htbp]
    \centering
    \includegraphics[width=0.99\linewidth]{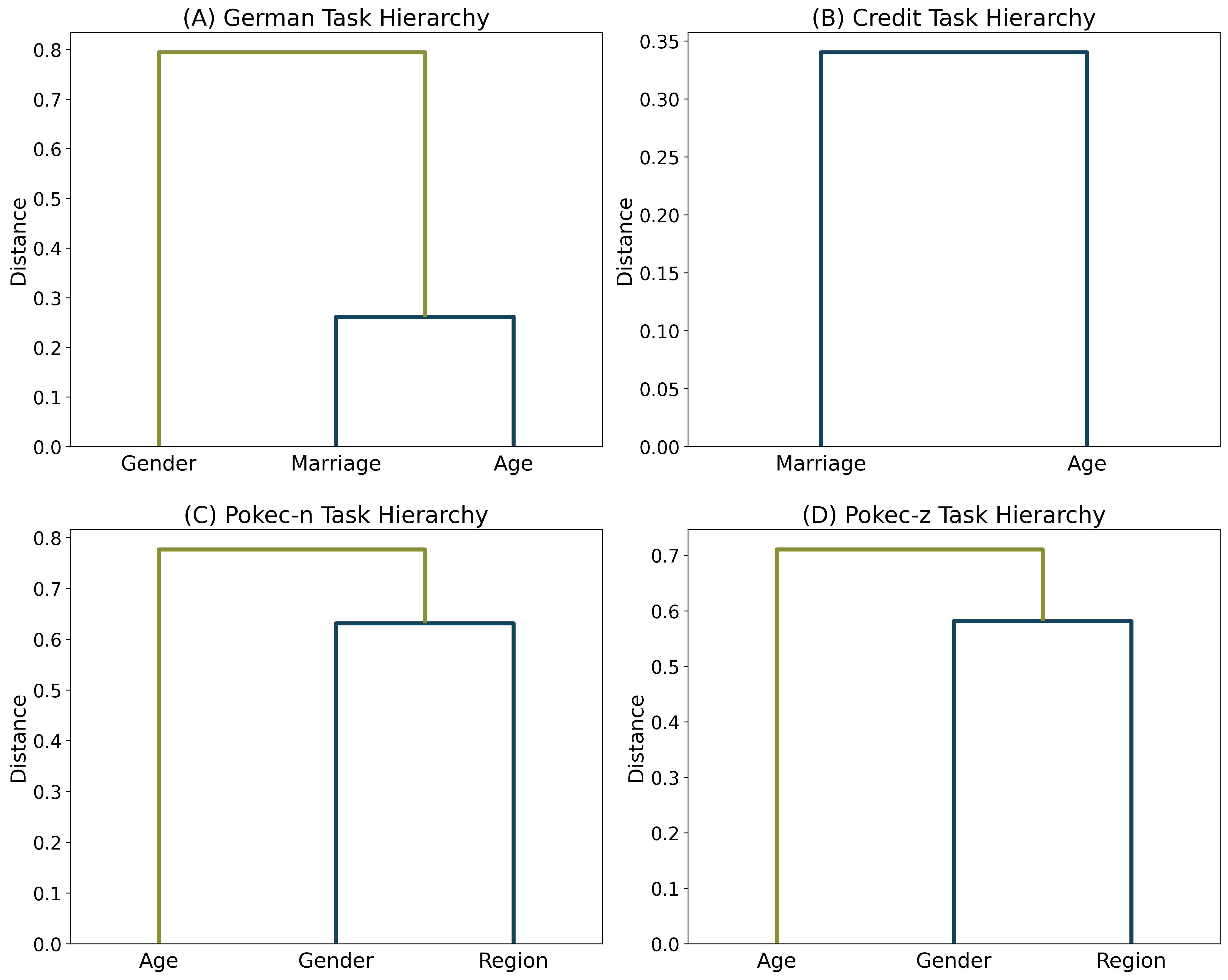}
    \caption{The Attack Hierarchy of Diverse Graph Datasets}
    \label{att_hierarchy}
\end{figure}

The attack task correlations and corresponding hierarchy of all graph datasets are illustrated in Figure \ref{att_corr} and \ref{att_hierarchy}.

\subsection{Implementation Details}
\label{sec_imp_det}
We adopt the architecture design of GCN, GraphSAGE and GIN in existing studies \cite{NIFTY,mapping}, and follow standard practice \cite{shadow_data, mia_gnn_shadow, inf_gnn} to split each graph dataset into two equal parts: one serving as the victim graph and the other as the auxiliary graph. The splitting ratio is $6/2/2$ for training, validation and testing, respectively. To maintain consistent loss scales and ensure stable optimization, we adopt the averaged form for each loss function. We provide the hyperparameter settings for different stages as follows. 

\begin{enumerate}
    \item Victim GNN: The hidden layer dimension is set to $16$ for German and Credit and $64$ for Pokec-n and Pokec-z. The learning rate is $0.01$, weight decay is $5e\text{-}4$ and dropout rate is $0.05$. The number of training epochs is $1000$ for German and Credit and $200$ for Pokec-n and Pokec-z. We use Adam as the optimizer. 
    \item Taipan: During \emph{Hierarchical Attack Knowledge Routing}, we adopt the same setting, except for learning rate $\in\{0.005,0.01\}$. During \emph{Prompt-guided Attack Prototype Refinement}, the adaptive threshold is tuned from $\{0.5, 0.6, 0.7, 0.8\}$, the number of adaptation steps is from $\{10,20,30,40,50\}$, and learning rate is from $\{0.001, 0.005, 0.01\}$. We follow the standard practice and set momentum as $0.9$ and temperature as $0.1$. 
    \item Baselines: For Rand., we utilize the empirical frequencies of sensitive attributes for Bernoulli sampling. For SingP., learning rate is tuned from $\{0.001, 0.005\}$ while other settings remain identical. 
    \item Label Coverage: we set learning rate as $0.001$, adaptation step as $10$,  threshold as $0.7$, the rest are the same as before.
\end{enumerate}

\subsection{Original Performance}
\label{sec_ori_perf} 
From Table \ref{data_fair}, we observe that GraphSAGE consistently outperforms GCN and GIN; GCN exceeds GIN on small-scale datasets, while GIN excels on large-scale datasets. Since attribute privacy is closely aligned with group fairness \cite{fair_audit_AIA, mapping}, as both aim to limit the model’s reliance on sensitive attributes, we adopt two widely used group fairness metrics to provide indirect evidence of sensitive information leakage: statistical parity difference ($\Delta$SP) and equalized odds difference ($\Delta$EO), which measure prediction rate disparity and discrepancies in true/false positive rates across sensitive groups, respectively \cite{fair_aia, fair_audit_AIA}. However, our results indicate that these conventional metrics fail to reliably capture biases hidden behind graphs. Both $\Delta$SP and $\Delta$EO are highly model-dependent and exhibit substantial variability even on the same dataset. For instance, GCN trained on German incurs significantly higher bias than other GNNs, whereas GraphSAGE trained on Pokec-z exhibits high bias for one sensitive attribute but much lower bias for the remaining attributes. These inconsistencies highlight the limitations of existing fairness metrics in the multi-attribute setting. 

Due to the alignment between attribute privacy and group fairness \cite{mapping, fair_audit_AIA}, Taipan can also serve as a fairness auditing tool to simultaneously infer multiple sensitive attributes, enabling the identification of both sensitive information leakage and structural group disparities.
\begin{table}[!htbp]
\caption{\centering Original Performance across Different Graph Datasets. The \textcolor{red}{red color} indicates the commonly used sensitive attribute (SA) for fair or private graph learning, the arrow indicates the direction of better performance and the \textbf{bold font} represents the best results.}
\label{data_fair}
\resizebox{\linewidth}{!}{
\begin{tabular}{c||c|ccc|cccccc}
\toprule\bottomrule
\multirow{2}{*}{\textbf{Dataset}} & \multirow{2}{*}{\textbf{GNN}} & \multirow{2}{*}{\textbf{AUC ($\uparrow$)}} & \multirow{2}{*}{\textbf{ACC ($\uparrow$)}} & \multirow{2}{*}{\textbf{F1 ($\uparrow$)}} & \multicolumn{2}{c}{\textbf{SA1}} & \multicolumn{2}{c}{\textbf{SA2}} & \multicolumn{2}{c}{\textbf{SA3}} \\ \cline{6-11} 
                                  &                               &                               &                               &                              & \textbf{$\Delta$SP ($\downarrow$)}     & \textbf{$\Delta$EO ($\downarrow$)}    & \textbf{$\Delta$SP ($\downarrow$)}     & \textbf{$\Delta$EO ($\downarrow$)}    & \textbf{$\Delta$SP ($\downarrow$)}     & \textbf{$\Delta$EO ($\downarrow$)}    \\ \bottomrule
\multirow{3}{*}{\textbf{German}}  & GCN                           & 66.62                         & 69.42                         & 79.48                        & \textcolor{red}{21.37}           & \textcolor{red}{13.50}          & 17.22           & 12.50          & 27.95           & 22.50          \\
                                  & GraphSAGE                     & 75.85                         & 68.93                         & 81.07                        & \textcolor{red}{3.67}            & \textcolor{red}{0.50}           & 2.48            & 0.83           & 3.73            & 2.85           \\
                                  & GIN                           & 60.12                         & 67.96                         & 80.70                        & \textcolor{red}{0.47}            & \textcolor{red}{1.50}           & 2.22            & 3.33           & 2.47            & 1.69           \\ \hline
\multirow{3}{*}{\textbf{Credit}}  & GCN                           & 71.86                         & 75.22                         & 83.93                        & 0.98            & 15.73          & \textcolor{red}{2.21}            & \textcolor{red}{13.65}          & —               & —              \\
                                  & GraphSAGE                     & 73.40                         & 79.55                         & 87.81                        & 1.50            & 0.42           & \textcolor{red}{1.11}            & \textcolor{red}{0.74}           & —               & —              \\
                                  & GIN                           & 73.73                         & 78.98                         & 87.54                        & 0.95            & 1.27           & \textcolor{red}{1.68}            & \textcolor{red}{1.54}           & —               & —              \\ \hline
\multirow{3}{*}{\textbf{Pokec-n}} & GCN                           & 60.91                         & 64.43                         & 57.24                        & 5.46            & 10.73          & \textcolor{red}{12.45}           & \textcolor{red}{1.46}           & 74.12           & 32.32          \\
                                  & GraphSAGE                     & 62.18                         & 65.34                         & 57.38                        & 1.35            & 5.90           & \textcolor{red}{7.28}            & \textcolor{red}{1.33}           & 73.07           & 32.32          \\
                                  & GIN                           & 60.36                         & 64.60                         & 56.90                        & 2.28            & 13.58          & \textcolor{red}{11.07}           & \textcolor{red}{0.67}           & 67.13           & 33.52          \\ \hline
\multirow{3}{*}{\textbf{Pokec-z}} & GCN                           & 64.92                         & 59.52                         & 51.86                        & 9.33            & 12.60          & \textcolor{red}{6.54}            & \textcolor{red}{14.87}          & 81.70           & 33.33          \\
                                  & GraphSAGE                     & 66.55                         & 59.13                         & 54.41                        & 69.80           & 101.56         & \textcolor{red}{3.00}            & \textcolor{red}{18.71}          & 79.42           & 5.74           \\
                                  & GIN                           & 65.61                         & 60.64                         & 53.46                        & 44.43           & 61.06          & \textcolor{red}{1.53}            & \textcolor{red}{20.84}          & 88.50           & 27.66          \\ \toprule\bottomrule
\end{tabular}}
\end{table}

\subsection{Visualization}
\label{sec_viz}
\begin{figure*}[htbp]
\centering
    \begin{subfigure}{0.5\textwidth} 
        \centering
        \includegraphics[width=\linewidth]{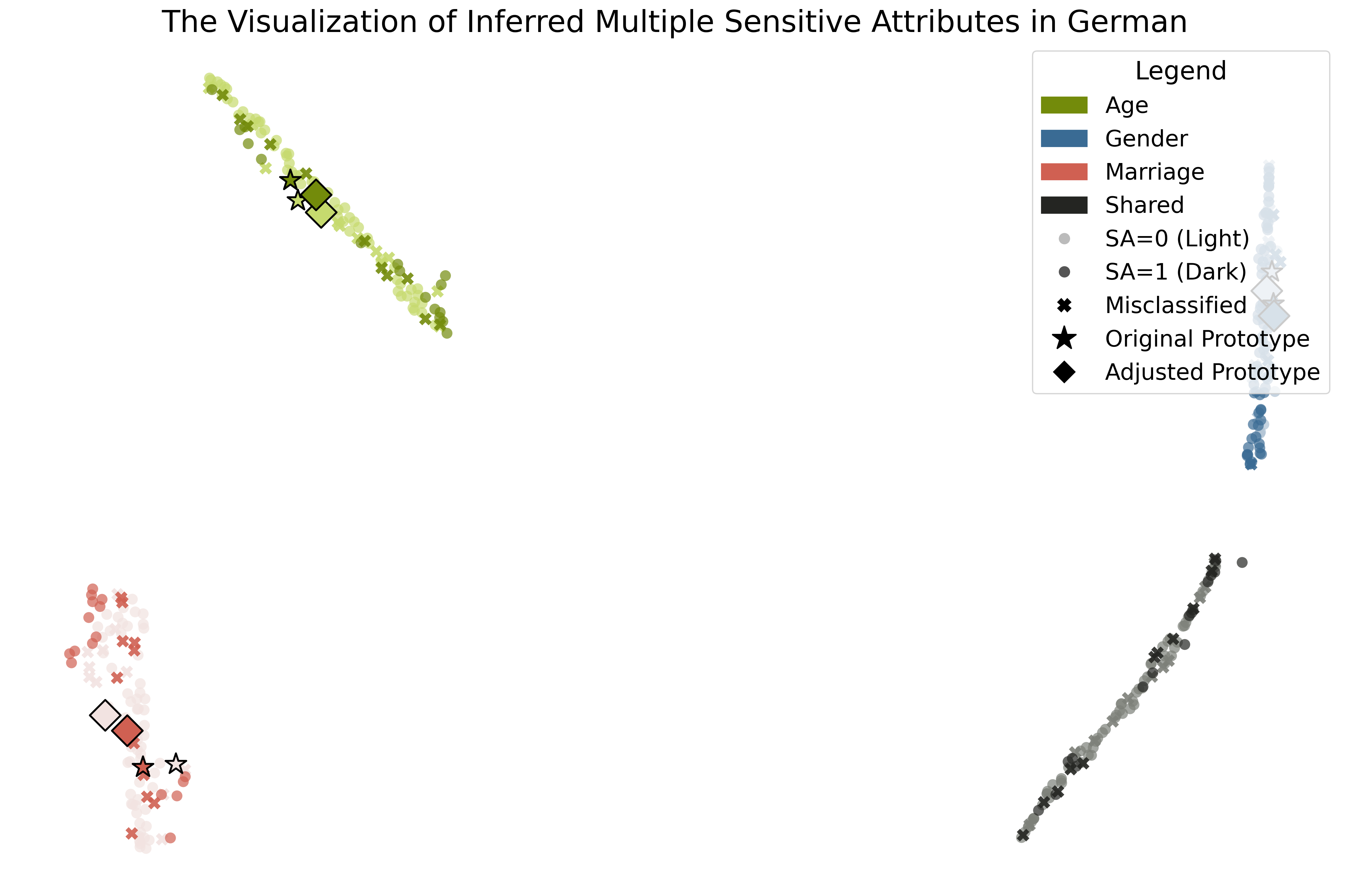} 
    \end{subfigure}\hfill 
    \begin{subfigure}{0.5\textwidth}
        \centering
        \includegraphics[width=\linewidth]{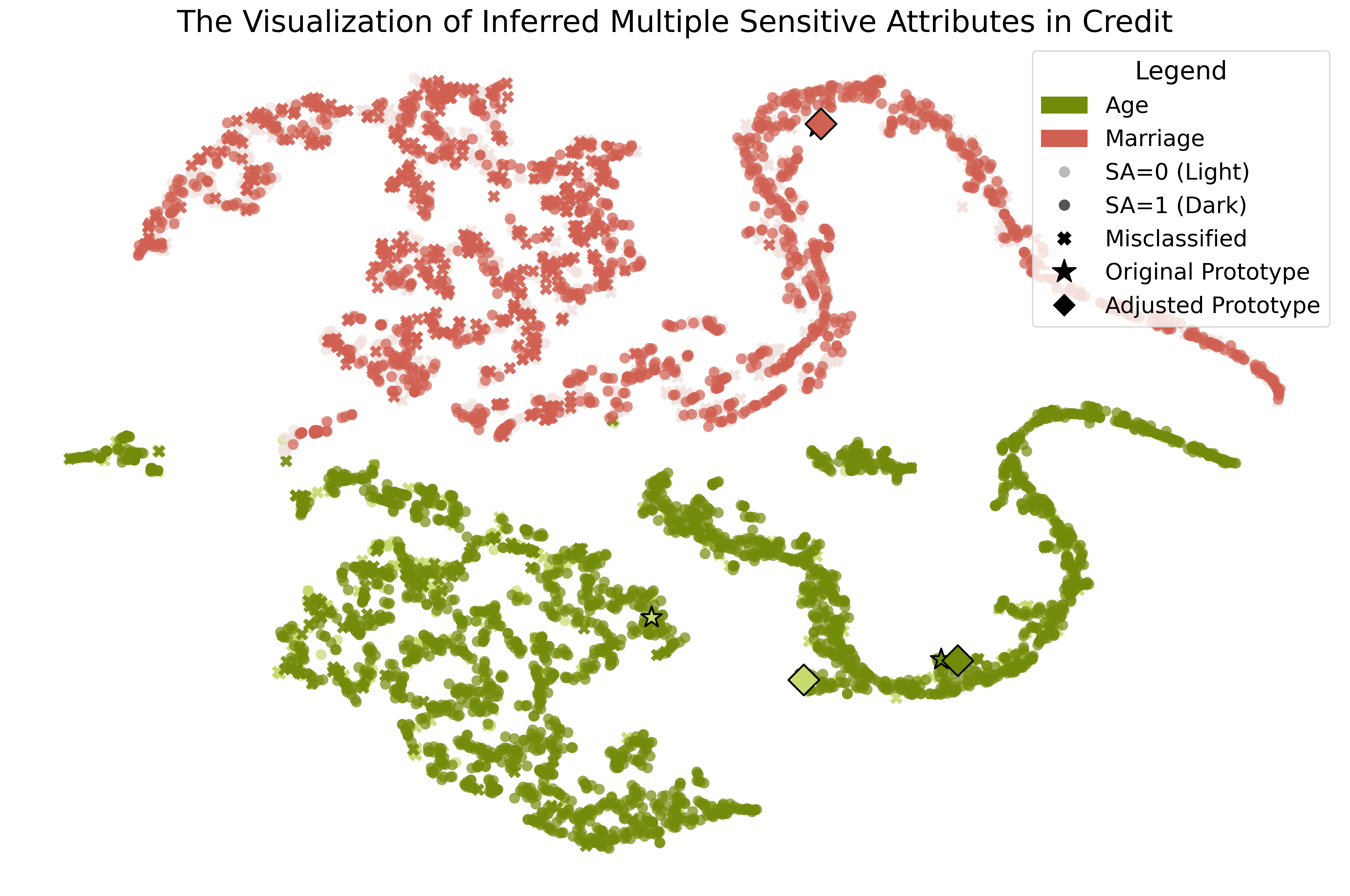} 
    \end{subfigure}
    
    
    \begin{subfigure}{0.5\textwidth}
        \centering
        \includegraphics[width=\linewidth]{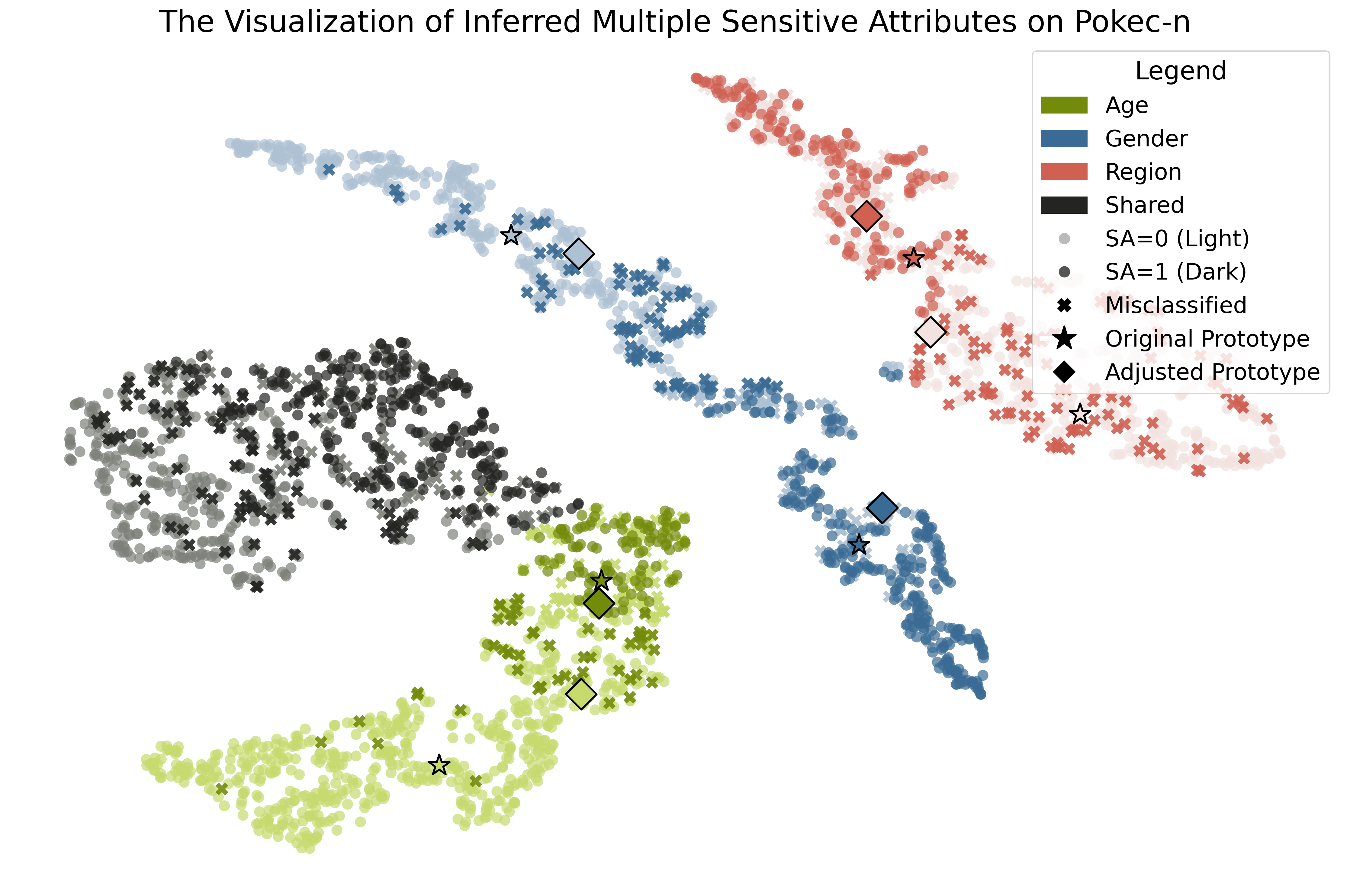} 
    \end{subfigure}\hfill
    \begin{subfigure}{0.5\textwidth}
        \centering
        \includegraphics[width=\linewidth]{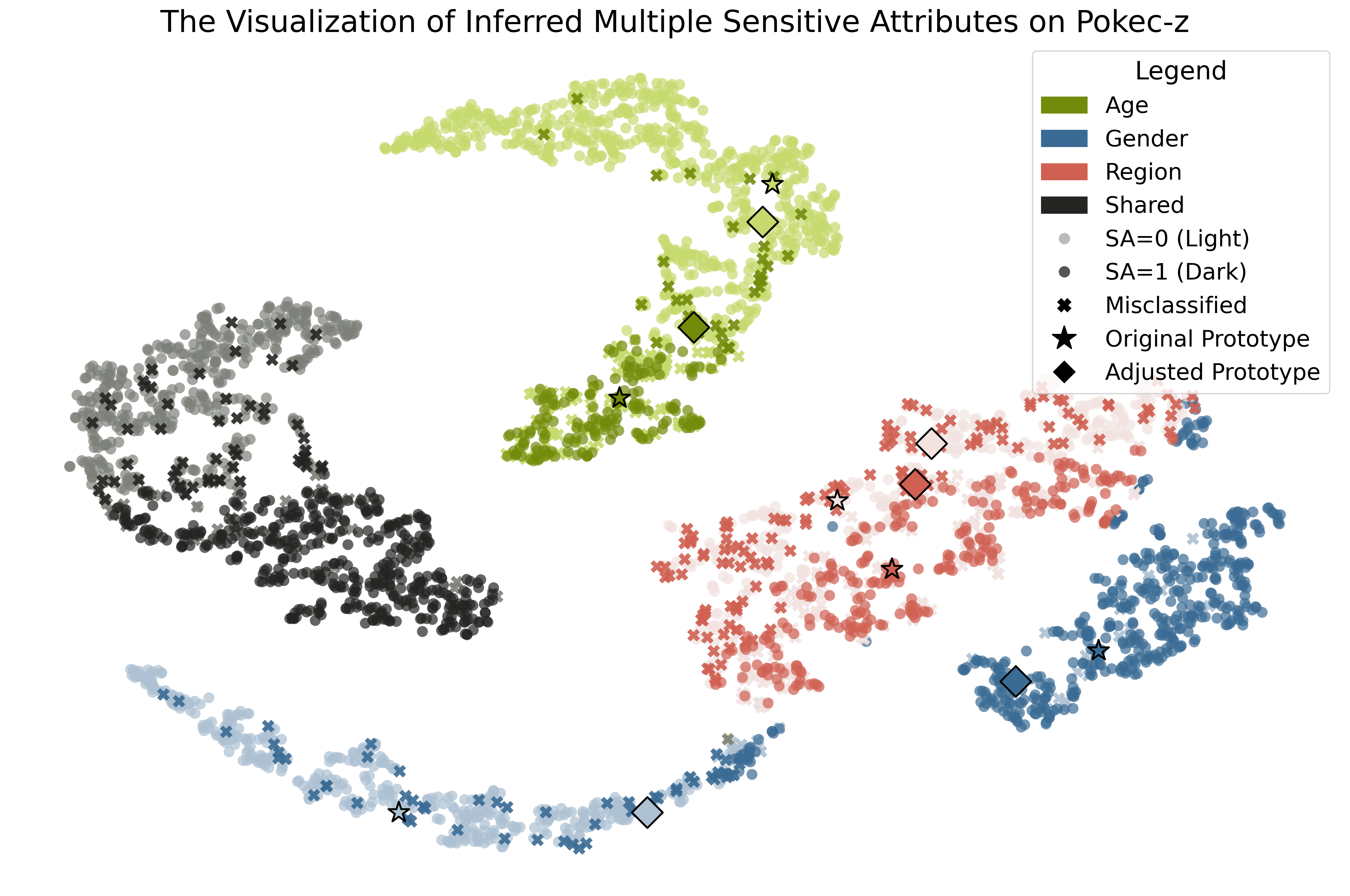} 
    \end{subfigure}
    \caption{The tSNE Visualization of Inferred Multiple Sensitive Attributes}
    \label{tsne}
\end{figure*}

\noindent\textbf{Embeddings of Multiple Sensitive Attributes} As illustrated in Figure \ref{tsne}, we utilize t-SNE to project the high-dimensional attack shared and task-specific node embeddings of multiple sensitive attributes into a two-dimensional space. We observe clear boundaries between shared and task-specific knowledge across all graphs, which indicates that Taipan successfully disentangles overlapping sensitive attributes into well-structured and identifiable clusters. Within most clusters, the original prototypes and their adjusted prototypes are located in close proximity to the corresponding cluster centers. The limited drift of the adapted prototypes from their original positions suggests that domain adaptation acts as a refinement process rather than a complete realignment, thereby avoiding overfitting to domain-specific noise. Furthermore, the embeddings on German exhibit more compact and well-separated patterns than those on other graphs, suggesting that Taipan can more effectively infer multiple sensitive attributes in this case. In addition, misclassified nodes primarily lie on the periphery or cluster boundaries. Intuitively, such boundary nodes typically have lower local homophily and generate more conflicting signals, which increasingly hinders G-MSAIAs.



\begin{figure*}[htbp]
\centering
    \begin{subfigure}{0.5\textwidth} 
        \centering
        \includegraphics[width=\linewidth]{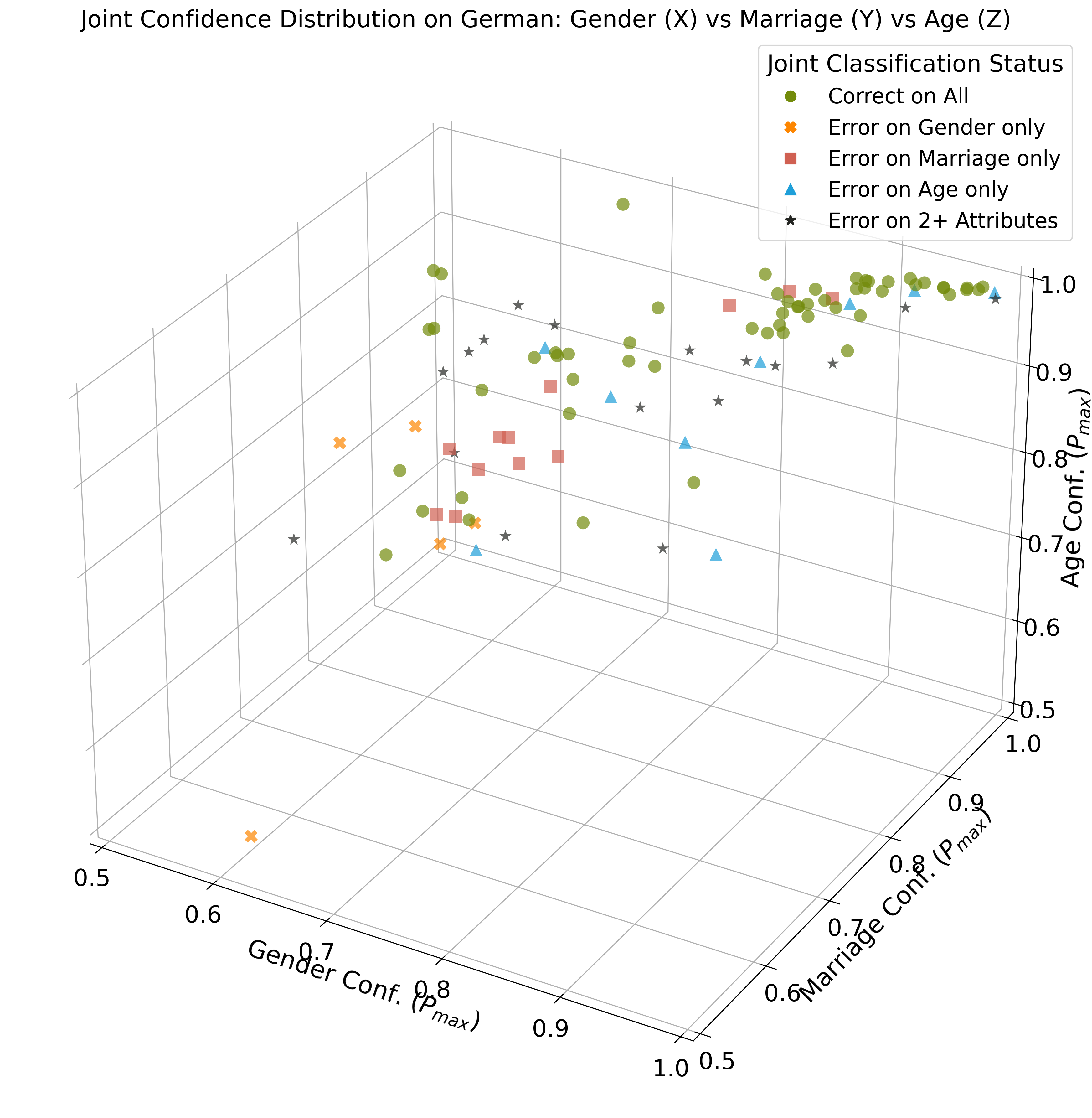} 
    \end{subfigure}\hfill 
    \begin{subfigure}{0.5\textwidth}
        \centering
        \includegraphics[width=\linewidth]{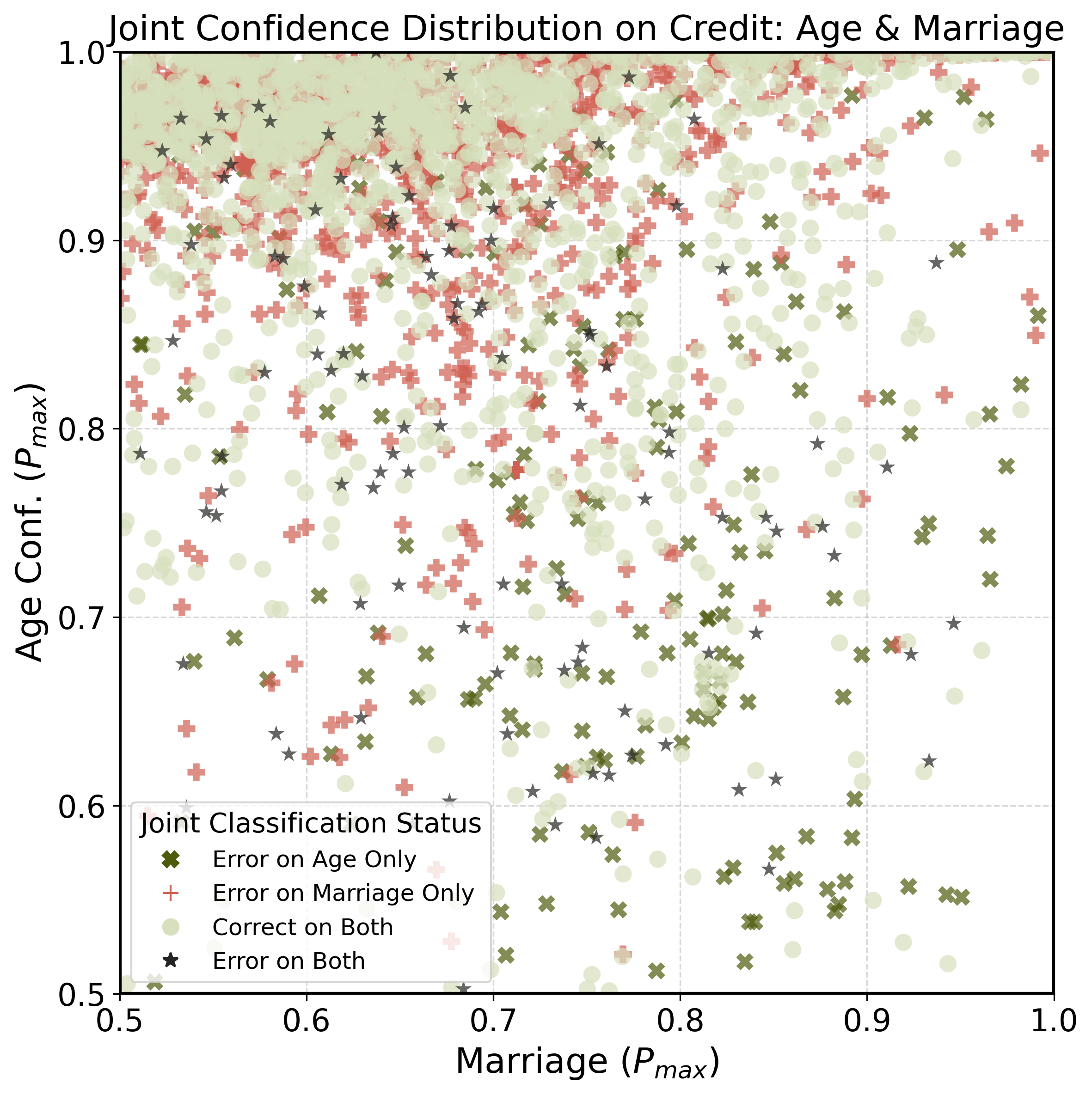} 
    \end{subfigure}
    
    
    \begin{subfigure}{0.5\textwidth}
        \centering
        \includegraphics[width=\linewidth]{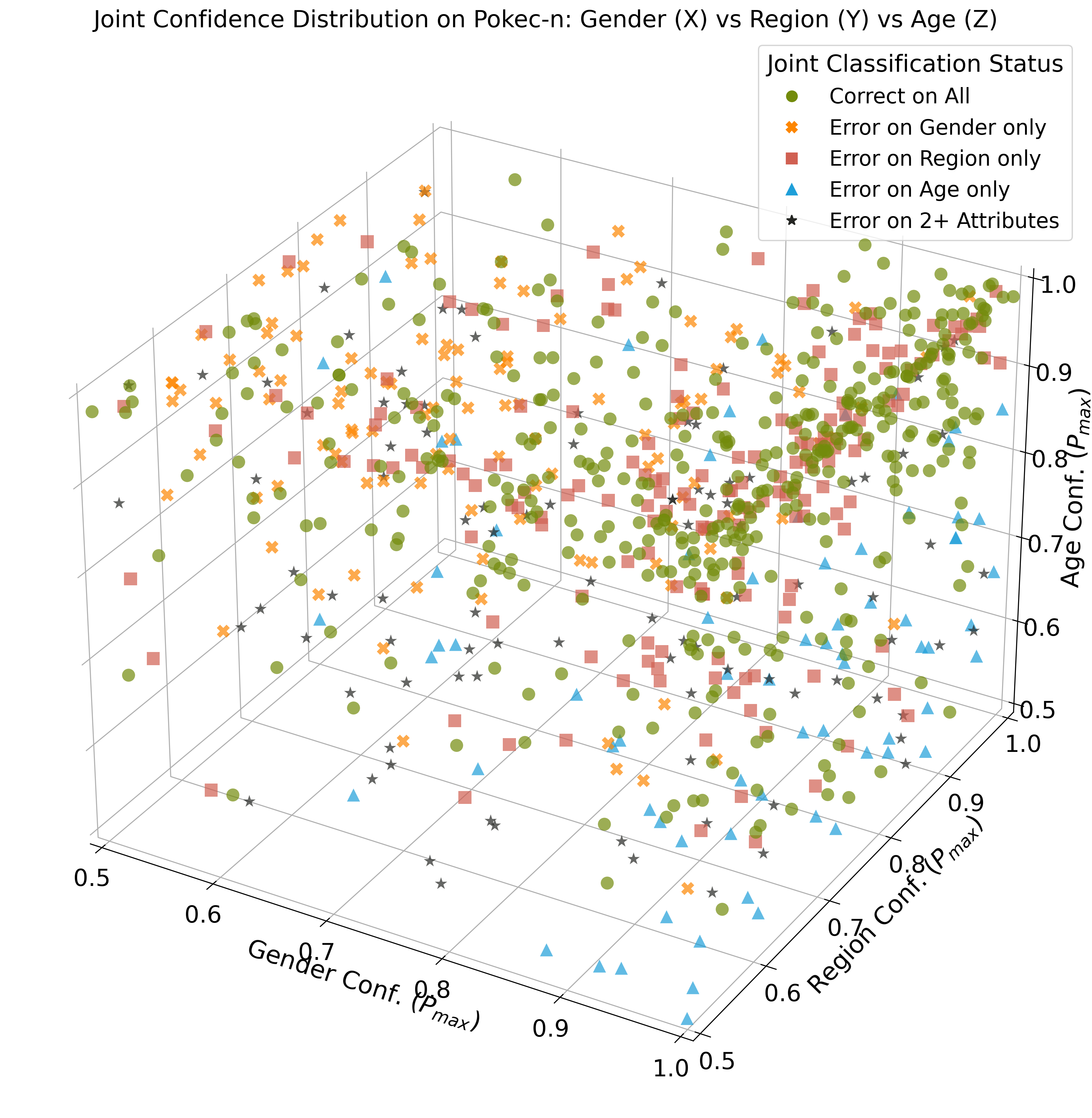} 
    \end{subfigure}\hfill
    \begin{subfigure}{0.5\textwidth}
        \centering
        \includegraphics[width=\linewidth]{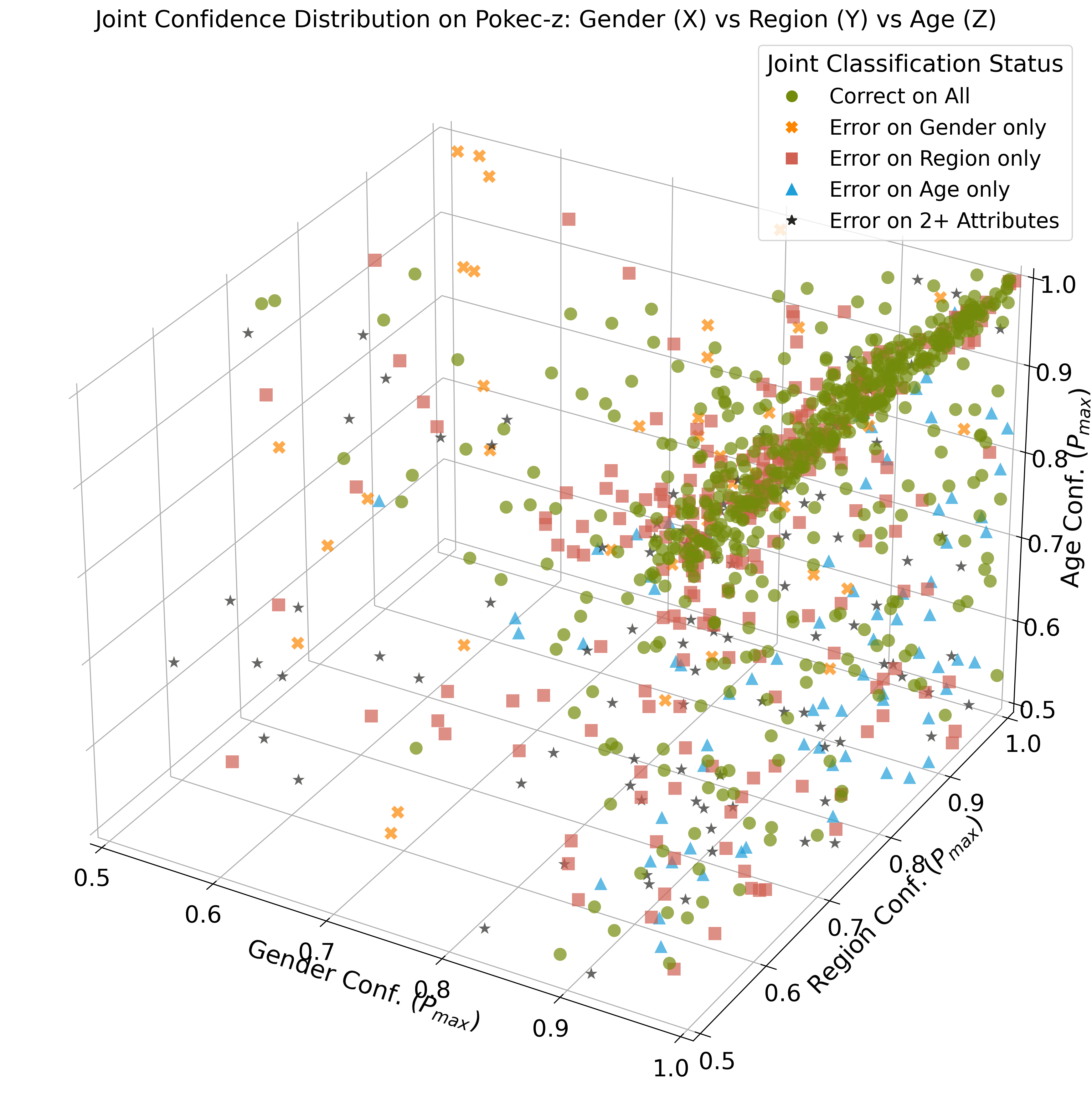} 
    \end{subfigure}
    
    \caption{Joint Confidence Distribution for Multiple Sensitive Attributes}
    \label{joint}
\end{figure*}

\noindent\textbf{Joint Confidence Distributions}
The joint confidence landscape of Taipan is further visualized in Figure \ref{joint}. We observe a clear and dense high-confidence region where all sensitive attributes are correctly inferred. In contrast, nodes with misclassified sensitive attributes are predominantly concentrated in low-confidence regions. Notably, nodes misclassified on only one sensitive attribute are not randomly distributed; instead, they cluster along the coordinate axes, where the attack model exhibits high confidence for one attribute but only moderate confidence for the other. This pattern reveals asymmetric and partial sensitive attribute leakage across different attack tasks. Overall, these observations indicate that Taipan’s internal confidence serves as a reliable proxy for G-MSAIAs.


\subsection{Adaptation Steps}
\label{sec_adpt}

\begin{figure*}[h] 
\centering
    
    

    \begin{subfigure}{0.5\textwidth}
        \centering
        \includegraphics[width=\linewidth]{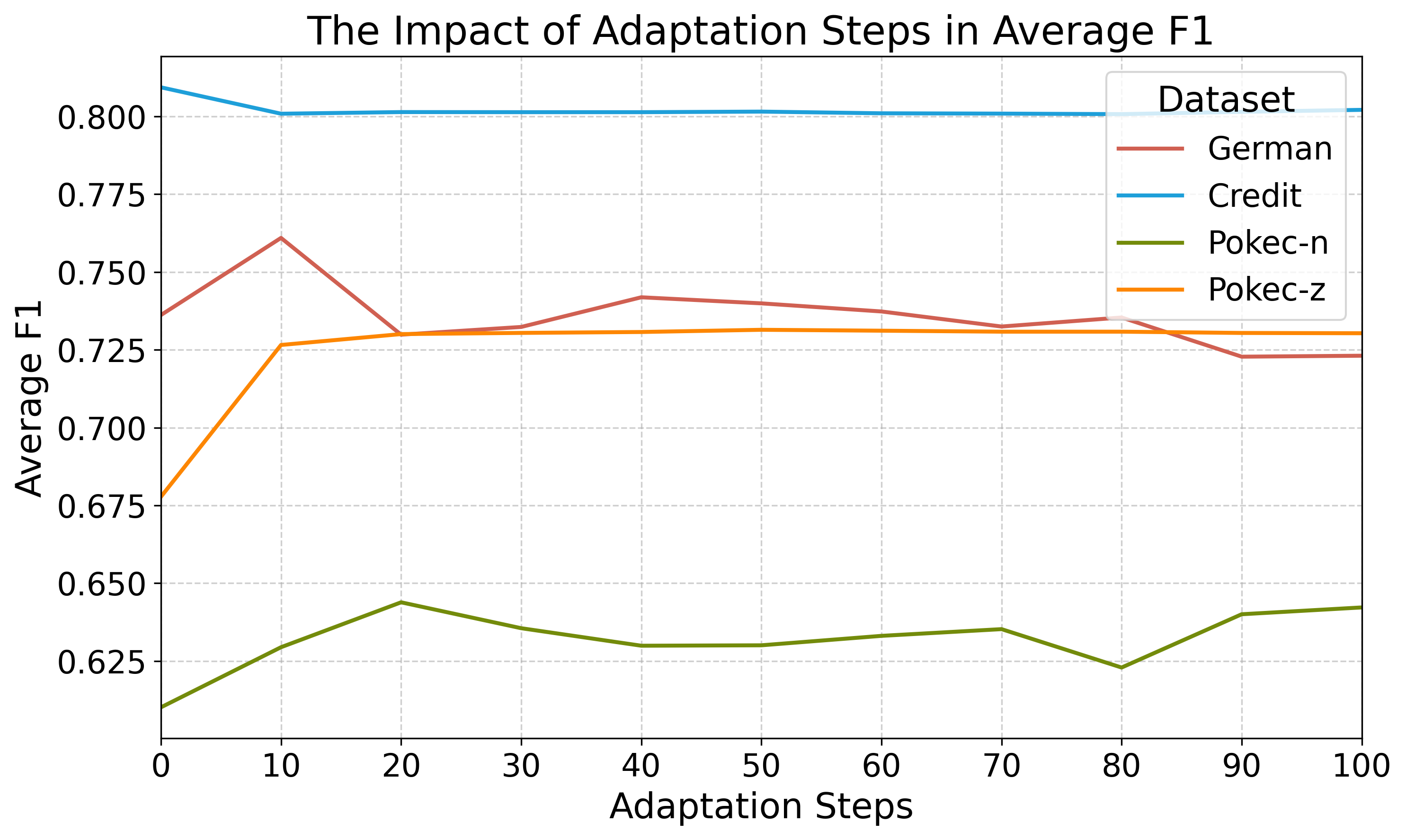} 
    \end{subfigure}\hfill
    \begin{subfigure}{0.5\textwidth}
        \centering
        \includegraphics[width=\linewidth]{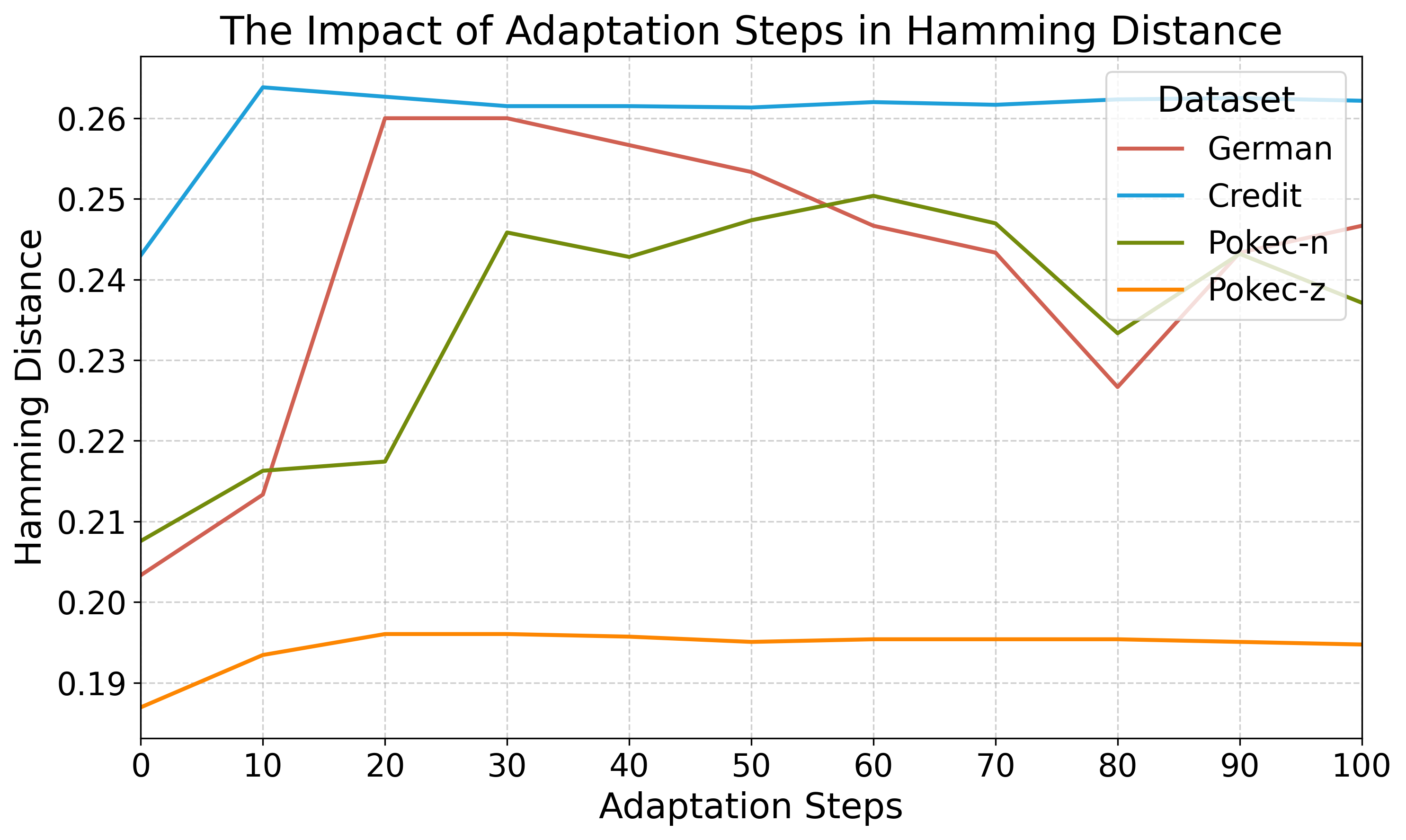} 
    \end{subfigure}

    \begin{subfigure}{0.5\textwidth}
        \centering
        \includegraphics[width=\linewidth]{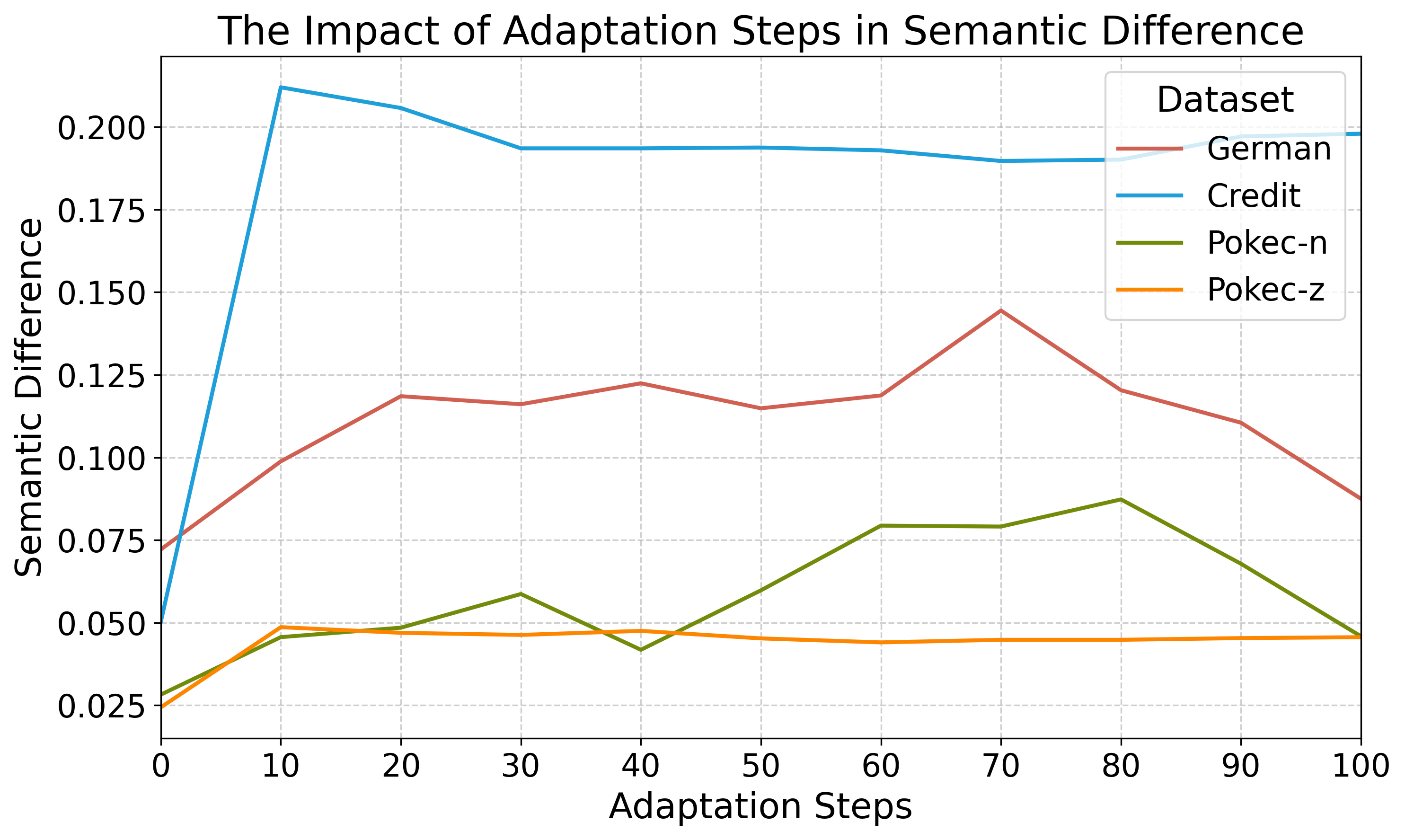} 
    \end{subfigure}\hfill
    \begin{subfigure}{0.5\textwidth}
        \centering
        \includegraphics[width=\linewidth]{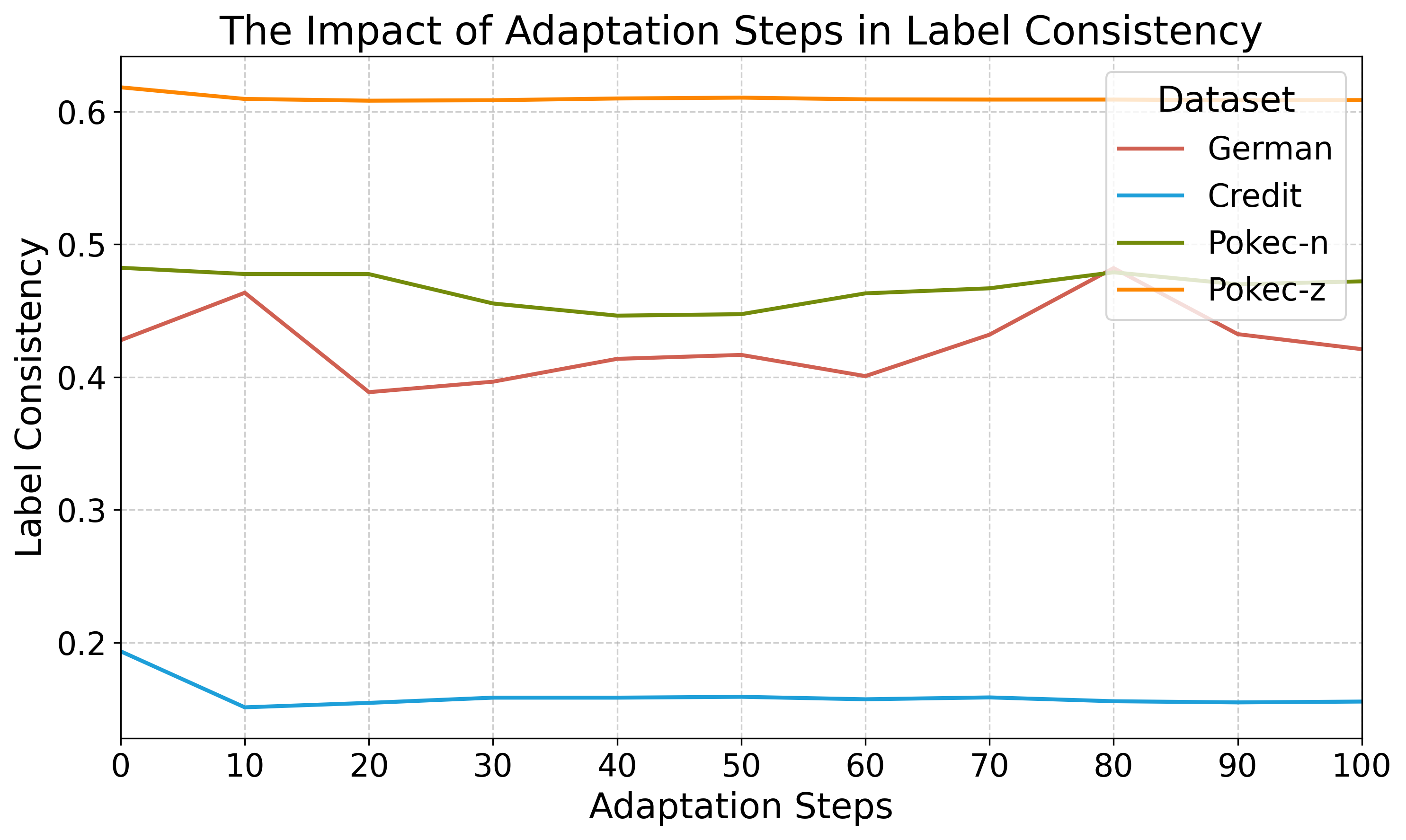} 
    \end{subfigure}
    
    \caption{Impact of Adaptation Steps on Semantic-based Metrics}
    \label{fig:adaptation_trends_rest}
\end{figure*}

From Figure \ref{fig:adaptation_trends_rest}, SD generally increases as the number of adaptation steps grows, indicating that excessive adaptation gradually introduces semantic drift from the auxiliary graph.
In contrast, within a small number of adaptation steps (e.g., fewer than $10$), SD remains relatively low, suggesting a favorable trade-off between semantic preservation and computational costs. LC exhibits a comparatively stable trend across adaptation steps for most datasets. Although German shows mild oscillations as the number of steps increases, its overall performance remains stable, and better LC is consistently achieved within the first $10$ adaptation steps. Taken together, these results suggest that a small number of adaptation steps is sufficient for G-MSAIAs.

\subsection{Node Degree and Node Homophily}
\label{sec_deg_homo}

\begin{figure*}[htbp]
    \centering
    
    
    \begin{subfigure}{0.5\textwidth}
        \includegraphics[width=\linewidth]{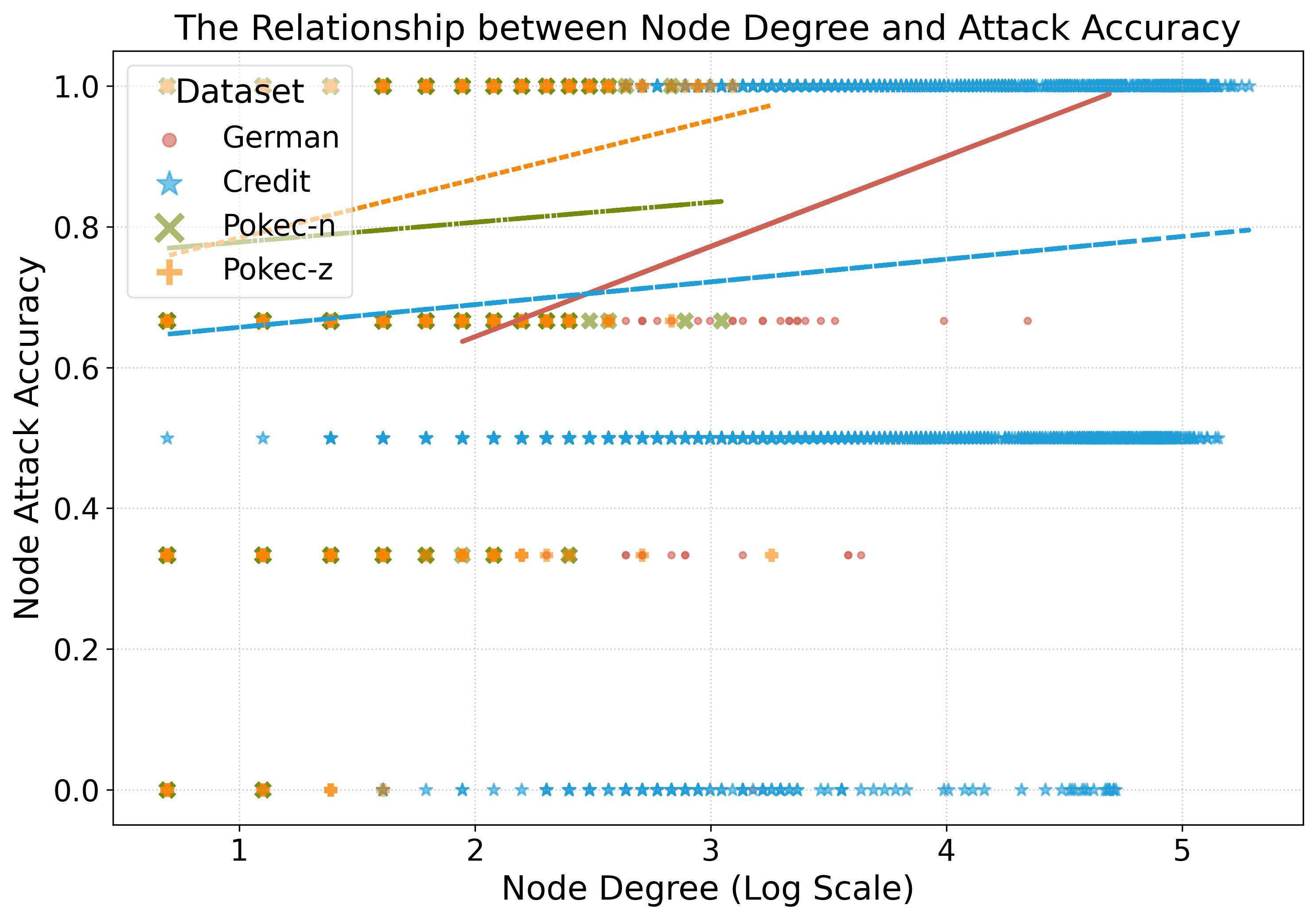}
    \end{subfigure}\hfill
    \begin{subfigure}{0.5\textwidth}
        \includegraphics[width=\linewidth]{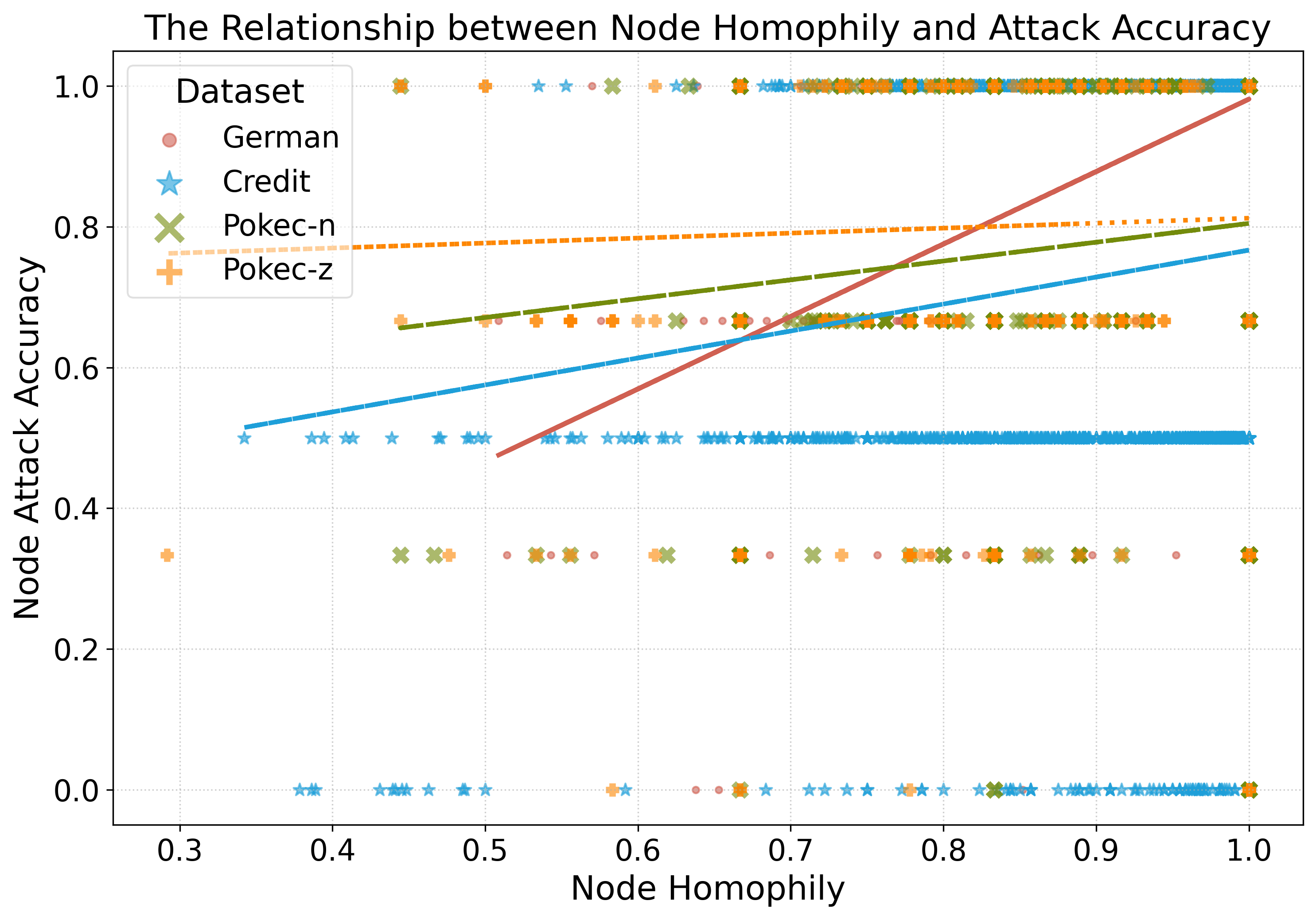}
    \end{subfigure}
    \caption{Impacts of Node Degree and Node Homophily on Attack Accuracy}
    \label{deg_homo_rest}
\end{figure*}

As shown in Figure \ref{deg_homo_rest}, both node degree and node homophily exhibit clear correlations with per-node attack accuracy, particularly on German. Compared to subset accuracy, node-level attack accuracy exhibits more stable patterns. This difference is expected, as subset accuracy is an all-or-nothing metric that drops to zero once any sensitive attribute is incorrectly inferred, leading to more abrupt variations. In contrast, node-level accuracy captures partial privacy leakage, thereby providing a smoother view of attack effectiveness. Importantly, although some nodes attain zero subset accuracy, the majority of points stays consistently above $0.5$, indicating that the adversary can still correctly infer a substantial fraction of sensitive attributes. This observation highlights that subset accuracy alone may underestimate privacy risks, as partial privacy leakage already constitutes a severe privacy threat.
 
\end{document}